%% file: mainv10.tex
\documentclass[american,reprint, floatfix,amsmath,amssymb, superscriptaddress,nofootinbib,prx,aps,longbibliography]{revtex4-2}

\usepackage{svg}
\usepackage{tablefootnote}
\usepackage{array}
\usepackage{tikz}
\usepackage{tikz-cd}
\usetikzlibrary{arrows.meta,positioning,calc}
\usetikzlibrary{calc,positioning,arrows.meta,decorations.pathmorphing}
\usetikzlibrary{calc,positioning,decorations.pathmorphing,decorations.pathreplacing}
\usepackage[colorlinks=true]{hyperref}
\hypersetup{allcolors=blue}
\usepackage[T1]{fontenc}

\usepackage{array}
\usepackage{varwidth}
\usepackage{stmaryrd}
\usepackage{graphicx} 	
\usepackage{amsmath}  	
\usepackage{enumitem}
\setitemize{noitemsep,topsep=0pt,parsep=0pt,partopsep=0pt,leftmargin=*}
\usepackage{amssymb}

\usepackage{enumitem}
\usepackage{tabularx}
\usepackage{ragged2e} 
\usepackage{algorithm}
\usepackage{algpseudocode}
\usepackage[caption=false]{subfig}

\makeatletter
\newcommand\listofappendices{
  \@starttoc{atoc}
}

\newif\ifrvx@inappendix
\rvx@inappendixfalse

\newif\ifrvx@incl 
\rvx@incltrue
\newcommand{\noappendixtoc}{\rvx@inclfalse}
\newcommand{\yesappendixtoc}{\rvx@incltrue}

\let\rvx@orig@appendix\appendix
\let\rvx@orig@addcontentsline\addcontentsline

\renewcommand{\appendix}{
  \rvx@inappendixtrue
  \rvx@orig@appendix
}

\renewcommand{\addcontentsline}[3]{
  \rvx@orig@addcontentsline{#1}{#2}{#3}
  \ifrvx@inappendix
    \ifrvx@incl
      \ifstrequal{#1}{toc}{
        \rvx@orig@addcontentsline{atoc}{#2}{#3}
      }{}
    \fi
  \fi
}
\makeatother

\newcommand{\nocontentsline}[3]{}
\let\origcontentsline\addcontentsline
\newcommand\stoptoc{\let\addcontentsline\nocontentsline}
\newcommand\resumetoc{\let\addcontentsline\origcontentsline}

\newtheorem{theorem}{Theorem}
\newtheorem{proposition}{Proposition}
\newtheorem{lemma}{Lemma}

\newtheorem{example}{Example}

\newtheorem{corollary}{Corollary}
\newtheorem{definition}{Definition}

\usepackage{enumitem}
\newcommand{\ignore}[1]{}

\newenvironment{proof}{\par\noindent\textbf{Proof:}\ }{\hfill$\square$\par}

\global\long\def\ket#1{\left|#1\right\rangle }

\global\long\def\sket#1{|#1\rangle\!\rangle}

\global\long\def\sbra#1{\left\langle\!\left\langle #1\right|\right.}

\global\long\def\sbraket#1#2{\langle\!\langle #1|#2\rangle\!\rangle }

\newcommand{\cT}{\mathcal{T}}
\newcommand{\invrej}{(\mathrm{Inv}_0,\mathrm{Rej}_1)}
\newcommand{\invinv}{(\mathrm{Inv}_0,\mathrm{Inv}_1)}
\newcommand{\invmsrej}{(\mathrm{Inv}_0,\mathrm{MS\_Rej}_1)}

\algrenewcommand\algorithmicrequire{\textbf{Input:}}
\algrenewcommand\algorithmicensure{\textbf{Output:}}
\algnewcommand\Input{\item[\algorithmicrequire]}
\algnewcommand\Output{\item[\algorithmicensure]}
\algrenewcommand\algorithmiccomment[1]{\hfill{\color{gray}\texttt{\textbackslash\textbackslash}~#1}}
\algnewcommand\LComment[1]{{\color{gray}\texttt{\textbackslash*}~#1~\texttt{*\textbackslash}}}

\begin{document}

\title{Syndrome aware mitigation of logical errors}

\author{Dorit Aharonov}
\affiliation{Qedma Quantum Computing, Tel Aviv, Israel}
\affiliation{The Benin School of Computer Science and Engineering, Hebrew University, Jerusalem, Israel}

\author{Yosi Atia}
\affiliation{Qedma Quantum Computing, Tel Aviv, Israel}

\author{Eyal Bairey}
\affiliation{Qedma Quantum Computing, Tel Aviv, Israel}

\author{Zvika Brakerski}
\affiliation{Qedma Quantum Computing, Tel Aviv, Israel}
\affiliation{Faculty of Mathematics and Computer Science,
Weizmann Institute of Science, Israel}

\author{Itsik Cohen}
\affiliation{Qedma Quantum Computing, Tel Aviv, Israel}

\author{Omri Golan}
\email{omri.golan@qedma.com}
\affiliation{Qedma Quantum Computing, Tel Aviv, Israel}

\author{Ilya Gurwich}
\affiliation{Qedma Quantum Computing, Tel Aviv, Israel}

\author{Netanel H.~Lindner}
\affiliation{Qedma Quantum Computing, Tel Aviv, Israel}
\affiliation{Department of Physics, Technion, Haifa, Israel}

\author{Maor Shutman}
\affiliation{Qedma Quantum Computing, Tel Aviv, Israel}

\begin{abstract}
Broad applications of quantum computers will require \textit{error correction} (EC). 
However, hardware roadmaps indicate that physical qubit numbers will 
remain limited in the foreseeable future, 
leading to residual logical errors that constrain the size and accuracy of 
achievable computations. 
Recent work suggested \textit{logical error mitigation} (LEM), 
which applies known \textit{error mitigation} (EM) methods 
to logical errors, eliminating their effect at the cost of a runtime overhead. 

We introduce \textit{syndrome-aware logical error mitigation} (SALEM), 
which mitigates logical errors conditioned on the error syndromes measured during error correction.  
The runtime overhead of SALEM is exponentially lower than that of LEM schemes which do not make use of syndrome data, 
enabling substantially larger circuit volumes that can be executed accurately.  
Compared to the routinely used combination of error correction and syndrome rejection 
(post-selection), 
SALEM increases the size of reliably executable computations by orders of magnitude. In 
the practical setting where space and time overheads are fixed and error reduction methods are compared by their resulting estimation errors, we observe a surprising phenomenon: SALEM, which tightly combines EC with EM, can outperform physical EM even above the standard fault-tolerance 
(pseudo) threshold.  
Thus, SALEM can make use of EC in regimes of physical error rates where EC is commonly deemed useless.
\end{abstract}

\maketitle

\begin{figure*}[ht!]
\begin{centering}
\includegraphics[width=1\textwidth]{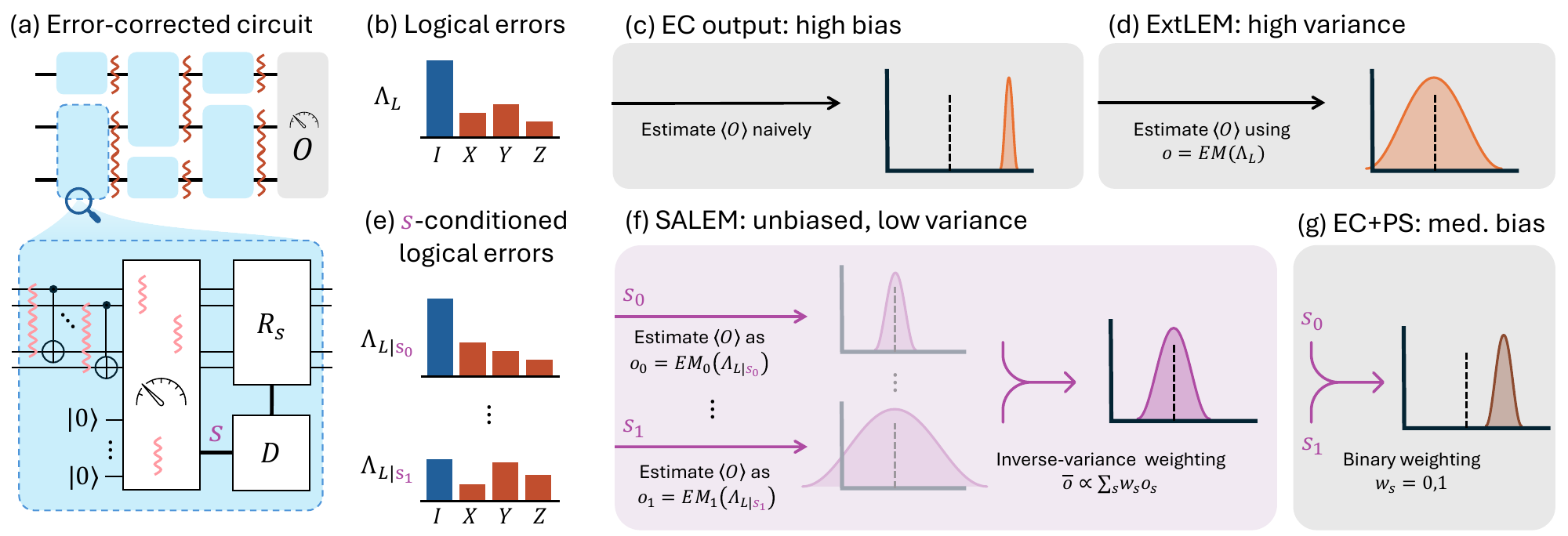}
\par\end{centering}
\caption{ 
Conceptual overview of syndrome-aware logical error mitigation (SALEM). (a) An error-corrected logical circuit, ending with the measurement of a logical observable $O$. Within each logical gate, physical errors occur (pink zigzags), and a syndrome bit-string $s$ is measured, decoded ($D$), and used to recover ($R_s$) from the most probable physical error processes given the syndrome. Red zigzags indicate logical errors due to less probable physical error processes. 
(b) The logical error channel $\Lambda_L$ corresponds to the (syndrome-averaged) distribution of logical errors at the output of a logical gate, see Sec.~\ref{Sec: setup}.  
(c) Direct estimation of the expectation value $\langle O\rangle$ is biased, due to the  non-trivial logical error channel, $\Lambda_L\neq I$.
(d) Previously proposed `external logical error mitigation' (ExtLEM) applies a fixed error mitigation (EM) protocol $EM$, irrespective of the measured syndrome $s$. The protocol is designed to mitigate the channel $\Lambda_L$, producing an unbiased estimator $o$ for $\langle O \rangle$, at the expense of an increased statistical error (for a fixed number of shots). (e) Restricting attention to shots in which a particular value of the syndrome $s$ was measured, the $s$-conditioned logical error channel $\Lambda_{L|s}$ appears as the relevant error channel. 
(f) `Syndrome-aware logical error mitigation' (SALEM), introduced here, applies a separate EM protocol $EM_s$, designed to mitigate the channel $\Lambda_{L|s}$, to shots in which the syndrome $s$ is measured. The resulting unbiased estimators $o_{s}$ for different syndromes are averaged with `inverse-variance' weights $w_s$, producing an unbiased estimator $\overline{o}$ with a significantly reduced shot overhead. (g) A standard approach to mitigating logical errors using syndrome data is `post-selection' (EC+PS), which can be described as a simple averaging with binary weights, corresponding to `accepted' or `rejected' syndromes, without applying mitigation in the accepted case. This leads to a remaining bias due to logical errors in accepted syndromes (as well as a sub-optimal increase in variance due to rejected syndromes). 
\label{fig: mainV2}}
\end{figure*}

\section{Introduction}

Quantum computation promises  dramatic algorithmic speedups (‘quantum advantages’, QAs) over classical computation, for a variety of applications \cite{quantum_algorithm_zoo}.
Fulfilling this promise requires a solution to the problem of errors in quantum processing units (QPUs)
which quickly accumulate to render even small quantum computations useless. The detrimental effect of errors is commonly agreed to be the main bottleneck towards realizing QAs. 

The long-term solution for errors in QPUs is fault-tolerant (FT) quantum computation \cite{Aharonov_1997,Knill_1998,Kitaev_1997, preskillaliferis},
 where logical qubits used in quantum algorithms are  redundantly encoded onto physical qubits 
 by a quantum error correcting code \cite{Peres_1983,shor_1995,steane_1996,
 eczoo} such that dominant errors can be detected and inverted during computation. Recent experiments have made significant progress towards realizing error correction (EC), but are still focused on few-qubit logical circuits, and have so far demonstrated limited improvements in error rates due to EC \cite{1-bluvstein2024logical,3-da2404demonstration,
4-reichardt2024demonstration,5-acharya2024quantum,putterman2024hardware,v477-jw8l,
eickbusch2025dynamic_surface_codes,salesrodriguez2025logical_magic_state_distillation,
yamamoto2025qec_molecular_energies,Bluvstein2025}. This is due to the value of the physical error rates of current QPUs ($\epsilon\sim10^{-3}$, that is, a physical error every $\sim 10^3$ physical operations), as well as the limited number of physical qubits currently available, ranging from tens to hundreds. Recent theoretical advances have made significant strides towards more efficient EC schemes \cite{xu2024constant,
bravyi_2024, scruby2024high, gidney2024magic}, 
predicting logical error rates as low as $\epsilon_L\sim10^{-5}-10^{-8}$ (per logical qubit) with physical error rates of near-term hardware \cite{google_quantum_ai,
ibm_quantum_technology, quera_qec_2024, quantinuum_roadmap_2023,
ionq_roadmap_2024, iqm_roadmap_2024}. But even these logical error rates will not suffice to maintain the high output accuracies needed for many applications in and beyond the `MegaQuOp' regime (circuits with $\sim 10^6$ logical gates) \cite{WhyEM}, where the first QAs of significant industry relevance may be anticipated \cite{preskill2024megaquop}. More generally, for as long as high-fidelity physical qubit numbers remain limited, the logical error rates achievable with EC will restrict accessible logical circuit volumes (number of logical gates) and output accuracies, and thus the applicability of quantum computing \cite{WhyEM}.

An alternative near-term solution for hardware errors is termed `quantum error mitigation' (EM) \cite{bravyitemmegambetta, li2017efficient,
PhysRevX.8.031027, emreview,PhysRevA.104.052607,cai2021multi,
filippov2024scalability}
and is now standard practice when working with QPUs.
In EM, the execution of a given ideal (error-free) quantum circuit is replaced with multiple executions of (generally distinct) noisy circuits, the outcomes of which are post-processed to yield an estimate for the outcome of the ideal circuit. 
In terms of resources, EM therefore requires little or no overhead in physical qubit number, but instead implies an overhead in the number of circuit executions (`shots'), which translates to an overhead in QPU time. It is expected that QPUs paired with EM protocols will soon provide the first \textit{useful} finite  QAs \cite{WhyEM, zimboras2025myths, lanes2025frameworkquantumadvantage, QESEMpaper, eisert2025mind}. Accordingly, EM has already been used to estimate the output of ideal quantum circuits that are classically challenging, some of which are potentially impossible to simulate with super-computers in reasonable time \cite{kim2023evidence, haghshenas2025digitalquantummagnetismfrontier, Abanin2025ConstructiveInterference,  alam2025fermionicdynamicstrappedionquantum}. 
The required shot overhead 
for EM is generically exponential in the total infidelity, i.e., the product of the error rate per gate $\epsilon$ and the circuit volume $V$ \cite{takagi2022fundamental, Takagi2023universal, Tsubouchi2023universal, Quek_Eisert, Schuster_Yao, WhyEM}. This overhead still allows EM to accurately execute circuits  with volumes $V\sim(1-10)\epsilon^{-1}$, yielding a significant boost over execution without EM which is limited to circuits of volumes $V\sim\delta \epsilon^{-1}$, where $\delta$ is the required accuracy. With near-term error rates $\epsilon\sim 10^{-3}$, this limits EM to circuit volumes $V\sim10^3-10^4$, which is expected to suffice for surpassing classical simulation, but not for applications of major industry relevance. 

EM can similarly be used to boost the circuit volumes of error-corrected quantum circuits. The combination of the two approaches was recently proposed \cite{Suzuki2022, Piveteau2021, Lostaglio2021, Xiong2020, Tsubouchi2024} and very recently demonstrated \cite{zhang2025demonstrating}. We refer to this general approach as `logical error mitigation' (LEM). In terms of resources, LEM uses available QPU time \textit{and} physical qubit numbers, to maximize accessible circuit volume and output accuracy. The LEM methods described in the literature thus far may be summarized as follows: First, apply EC to replace physical gates by corresponding FT logical gates with a reduced error rate. Then apply EM to these logical gates, as if they are physical gates. We refer to this approach as ‘external LEM’ (ExtLEM), as it does not `look inside' the logical gates, and specifically, does not make use of the syndrome data measured within them. The volumes enabled by ExtLEM can be estimated by $V\sim(1-10)\epsilon_L^{-1}$ where $\epsilon_L$ is the logical error rate. These volumes are significantly larger than $V\sim\delta\epsilon_L^{-1}$ enabled by EC alone. Improving the runtime incurred by ExtLEM, and thus the enabled volumes, is critical in order to accelerate the path towards QAs of major industry relevance. 

\begin{figure*}[ht]
\begin{centering}
\includegraphics[width=1\textwidth]{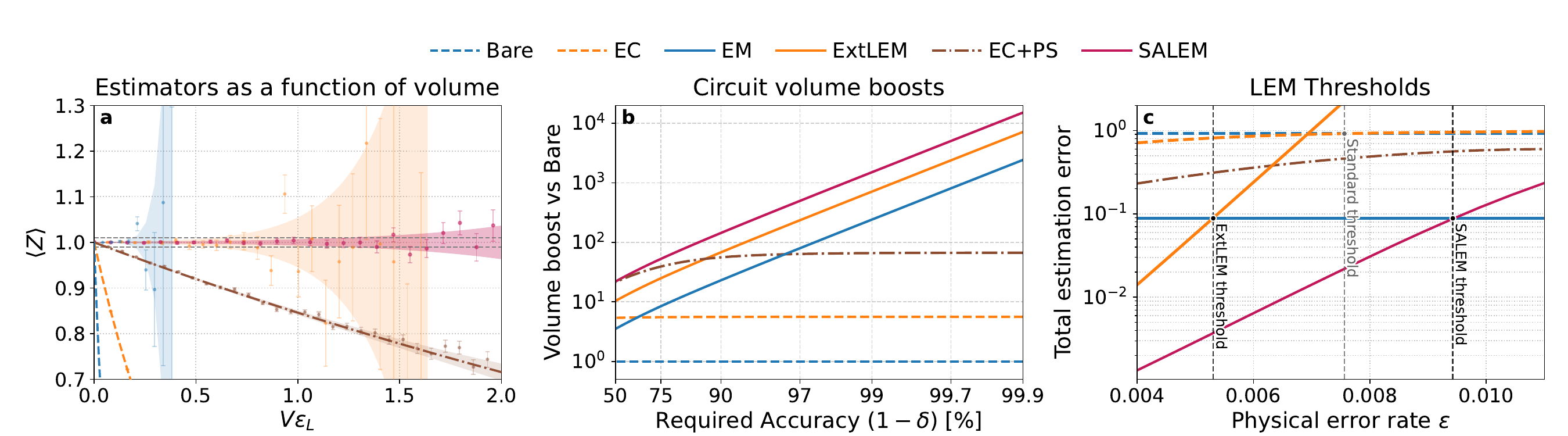} 
\par\end{centering}
\caption{Performance benefits of (coarse-grained) SALEM. 
(a) Comparison of estimators for an expectation value $\langle Z\rangle$ due to different error reduction methods, as a function of logical circuit volume $V$,  normalized with the logical error rate $\epsilon_L\approx10^{-4}$, for a fixed physical error $\epsilon=10^{-3}$. EC is based on a distance 4 surface code memory circuit subject to circuit-level physical errors, see Fig.~\ref{Fig: Bi-SALEM} and Appendix \ref{Appendix: numerical simulations} for technical details. All error reduction methods are allocated the same space and time overheads, relative to the number of qubits and the runtime needed to meet the allowed estimation error ($1\%$, gray dashed lines) in an idealized error-free QPU. Specifically, we allocate a space-time volume corresponding to one-hundred error-corrected shots per error-free shot. The physical methods Bare and EM exploit the given space overhead by parallelizing shots, and have a lower shot time than methods involving EC. Shaded bands show the expected statistical spread of estimators (mean $\pm$ one standard deviation), with mean curves shown only for biased methods and points illustrating a representative finite-shot realization. SALEM outperforms all methods, maintaining the required accuracy for significantly larger circuit volumes. 
(b) The `circuit volume boost' (CVB) is the multiplicative improvement in achievable volume for a given required accuracy  \cite{WhyEM}. Methods limited by bias (EC and EC+PS), can provide significant CVBs, which do not change much with allowed inaccuracy $\delta$. Methods limited by statistical errors (EM, ExtLEM and SALEM), provide a CVB $\sim \delta^{-1}$, and are therefore superior at small $\delta$ (high required accuracy). SALEM achieves the largest circuit volumes among all methods, at both high and low accuracy.
(c) The behavior of estimation errors (including both biases and statistical errors) for the different methods as a function of physical error rate reveals the standard `FT (pseudo) threshold' (gray dot), as well as two new types of (pseudo) thresholds, obtained by comparing ExtLEM and SALEM to physical EM (black dots). The logical circuit volume is scaled relative to the \textit{physical} error rate, $V=2.5/\epsilon$, such that the Bare and EM methods have constant estimation errors. At practical volumes $V=O(1/\epsilon_L)$, the SALEM threshold is higher than the standard FT threshold.  \label{fig: performance}}
\end{figure*}

In this work we follow up on an idea alluded to in our previous work \cite{WhyEM}, and introduce \textit{syndrome-aware logical error mitigation} (SALEM), which exploits the syndrome data measured within logical gates as part of EC (the approach works similarly also for error detection, but we focus in this exposition on EC). SALEM makes use of syndrome data in order to achieve a reduced shot overhead compared to 
ExtLEM, and consequently enables larger circuit volumes and output accuracies, given fixed resources. A `fine-grained' version of SALEM is depicted in Fig.~\ref{fig: mainV2}, and compared to EC (alone), ExtLEM and the standard combination of EC and post-selection (EC+PS) \cite{smith2024mitigating, PrabhuReichardt2024}. Figure~\ref{fig: performance} demonstrates the performance improvements of SALEM, achieved already with a simple `coarse-grained' version of SALEM, where syndromes are grouped to a small number of subsets (some of which may be rejected).

Figure~\ref{fig: performance}(a) demonstrates the behavior of estimators due to different error reduction methods as a function of circuit volume. In this example, SALEM maintains the required accuracy $1-\delta=99\%$ at volumes over $2\times$ larger than with ExtLEM, $20\times$ larger than with EC+PS, and $200\times$ larger than with EC. Figure~\ref{fig: performance}(b) shows how this boost in circuit volume (CVB) \cite{WhyEM}, measured relative to bare circuit execution (Bare),  changes with the required accuracy, showing that the advantage of SALEM over ExtLEM is essentially independent of $\delta$, while the advantage over  EC and EC+PS, which are bias-limited, grows with the required accuracy, as $\sim \delta^{-1}$.  

The performance of LEM relative to physical EM depends on the physical error rate $\epsilon$, which determines the logical error rate $\epsilon_L$. In Fig.~\ref{fig: performance}(a)-(b), $\epsilon$ is significantly below the `FT (pseudo) threshold', defined by the break-even condition $\epsilon_{L}(\epsilon)=\epsilon$; such that $\epsilon_L\ll \epsilon$, and EC is much better than bare circuit execution (Bare). Accordingly, ExtLEM, and more so SALEM, significantly outperform physical EM (with $3\times$ and $6\times$ larger accessible volumes in this example). 

Importantly, in the context of LEM, the FT threshold  no longer marks the point at which EC becomes beneficial. Instead, we regard new (pseudo) thresholds as more relevant: these are the physical error rate values at which ExtLEM and SALEM, respectively, become preferable to using physical EM. Figure \ref{fig: performance}(c) shows estimation errors for different methods at physical error rates around the FT threshold, revealing the new thresholds.  Due to the space-time overhead of EC, both the ExtLEM and SALEM thresholds are volume-dependent (see Appendix \ref{Appendix: thresholds}). The ExtLEM threshold is always lower than the FT threshold, and approaches it at large volumes. However, the SALEM threshold is always higher than the ExtLEM threshold. Importantly, for sufficiently large circuit volumes, it \textit{surpasses the FT threshold}, and continues to \textit{increase} with volume. As shown in Fig.~\ref{fig: performance}, the SALEM threshold exceeds the FT threshold already at practical volumes of $O(1/\epsilon_L)$. Thus, through SALEM, the use of EC can be beneficial even in a regime where EC is not expected to be useful, namely, {\it above} the standard FT threshold.

\section{Setup\label{Sec: setup}}

Consider an error-corrected logical circuit $C$, as shown in Fig.~\ref{fig: mainV2}(a). The circuit implements logical gates onto logical qubits, encoded in a larger number of physical qubits.
Within each logical gate, a syndrome bit-string $s$ (`local syndrome') is measured, and subsequently decoded to obtain a corresponding recovery operation. With high probability, the recovery is successful, and no logical error, corresponding to an unwanted logical gate, occurs. 
Logical gates are usually constructed to be `fault-tolerant' to some order $t$ (`$t$-FT'), such that events including up to $t$ physical errors (`faults') are properly recovered from, and logical errors are due to events including $\geq t+1$ physical errors. The probability distribution $\mathbb{P}(\sigma )$ of possible logical errors $\sigma$ defines the \textit{logical error channel} $\Lambda_L=\sum_\sigma \mathbb{P}(\sigma )\sigma$, see Fig.~\ref{fig: mainV2}(b). 

We stress that logical errors can generally occur due to events involving faults in \textit{multiple} logical gates. As a result, it is a priori unclear whether logical error channels $\Lambda_L$ associated with \textit{individual} gates, as depicted in Fig.~\ref{fig: mainV2}(a)-(b), can be usefully defined. 
In Appendix \ref{Appendix: predictive logical channels} we show that it is possible to define and, to leading order in the physical error rate, \textit{locally} characterize logical error channels associated with individual logical layers and gates, yielding a purely logical description of faulty error-corrected circuits that provides a basis for LEM. This is a central technical contribution of this work, which may be of independent interest.

Given a logical error channel $\Lambda_L$, the corresponding \textit{logical error rate} $\epsilon_L=1-\mathbb{P}(\sigma=I)$,  which is the probability for a non-trivial logical error in the corresponding logical gate, can be viewed as the limiting factor for the performance of EC. In particular, the expectation value $\langle O \rangle_{\boldsymbol{\Lambda}_L}$ of a logical operator $O$ measured at the end of the circuit will generically deviate from its ideal (error-free) value $\langle O \rangle= \langle O \rangle_{\boldsymbol{I}}$; where the `global channel' $\boldsymbol{\Lambda}_L$ denotes the list  of logical error channels associated with all logical gates in the circuit $C$.  Naive estimation of $\langle O \rangle$, by averaging over 
the measured values of $O$ in $N$ shots, will  produce the biased outcome $\langle O \rangle_{\boldsymbol{\Lambda}_L}\neq \langle O\rangle$ in expectation, with a variance $\leq 1/N$, see Fig.~\ref{fig: mainV2}(c). 

LEM protocols are designed to eliminate the bias $b=|\langle O\rangle_{\boldsymbol{\Lambda}_L}-\langle O\rangle|$ due to logical errors, at the expense of a shot overhead.  Proposed ExtLEM methods (Fig.~\ref{fig: mainV2}(d)) do this by applying a fixed EM  protocol $EM$ to all shots, designed to mitigate the global channel $\boldsymbol{\Lambda}_L$, producing an estimator  $o$ that depends on this channel \cite{Suzuki2022, Piveteau2021, Lostaglio2021, Xiong2020, Tsubouchi2024, zhang2025demonstrating}.
The protocol $EM$ is \textit{unbiased} (possibly, for a specific instance) if it reproduces the ideal result in expectation, $\mathbb{E}[{o}]=\langle O \rangle$. The cost of this  estimation is a \textit{shot overhead} $\Gamma$ --  defined through a given upper bound $\mathbb{V}[o]\leq \Gamma/N$ on the variance $\mathbb{V}[o]$  as a function of the number of shots $N$ used to produce $o$. 

\section{Fine-grained SALEM\label{Sec: SALEM}}

We start by discussing a  `fine-grained' version of SALEM (FG-SALEM), in which each syndrome is treated separately. FG-SALEM reuses the syndrome data generated within logical gates, by (i) separately mitigating shots in which distinct syndromes were measured, and (ii) properly averaging over the resulting per-syndrome estimates. 

Consider first a specific logical gate. Restricting attention to shots in which a particular syndrome $s$ is measured, the relevant error channel for that logical gate is the $s$-conditioned logical error channel, $\Lambda_{L|s}=\sum_\sigma \mathbb{P}(\sigma|s)\sigma$ (Fig.~\ref{fig: mainV2}(e)).
In fact, logical error channels in error-corrected circuits can generally depend also on syndromes measured in past logical gates (Appendix \ref{Appendix: syndrome-conditioned exact channels}). Importantly, we find that, at leading order in the physical error rate, this dependence is limited to syndrome data measured in a few recent causally-connected logical gates (see  Appendix~\ref{sec:approx logical channels}). We keep the notation $\Lambda_{L|s}$ in the main text for simplicity, though $s$ here should be understood as including past syndrome data relevant to the particular logical gate. 
 We collect the local syndromes  measured in all logical gates in the circuit  to a `global syndrome' $\boldsymbol{s}$, and  refer to the list of all conditioned logical channels in the circuit as an `$\boldsymbol{s}$-conditioned  global logical error channel' $\boldsymbol{\Lambda}_{L|\boldsymbol{s}}$. We discuss the challenging computational task of mapping $s\mapsto\Lambda_{L|s}$ (or $\boldsymbol{s}\mapsto \boldsymbol{\Lambda}_{L|\boldsymbol{s}}$) in Sec.~\ref{Sec: fine-grained SALEM and ML decoding}. 

Given $N_{\boldsymbol{s}}$ shots (out of a total of $N$) in which a particular $\boldsymbol{s}$ is observed, FG-SALEM applies an EM protocol $EM_{\boldsymbol{s}}$ designed to mitigate $\boldsymbol{\Lambda}_{L|\boldsymbol{s}}$, and produces an estimator $o_{\boldsymbol{s}}$ (Fig.~\ref{fig: mainV2}(f)). Note that a non-zero $N_{\boldsymbol{s}}$ may be as low as 1. In this section we thus restrict attention to `single-shot' protocols $EM_{\boldsymbol{s}}$, that can produce an estimator $o_{\boldsymbol{s}}$ using even a single shot (Appendix \ref{Appendix: single shot}). This restriction is not required for `coarse-grained' SALEM (CG-SALEM), discussed below.

In particular, one may consider single-shot protocols $EM_{\boldsymbol{s}}$ that (in expectation) invert $\boldsymbol{\Lambda}_{L|\boldsymbol{s}}$ locally, by sampling from quasi-probability (QP) distributions \cite{bravyitemmegambetta, 
PhysRevX.8.031027, van2023probabilistic, QESEMpaper} (see Appendix \ref{Appendix: FG-SALEM-QP}); or that sample informationally-complete measurement bases and invert $\boldsymbol{\Lambda}_{L|\boldsymbol{s}}$ globally, in (extensive) post-processing \cite{filippov2023scalable, qiskit_pna}. Note that FG-SALEM with local inversion is adaptive, modifying the executed circuit in real time based
on the already-measured part of $\boldsymbol{s}$, generalizing the
syndrome-dependent recovery operation of  EC (Appendix \ref{Appendix: implementation}).  Both local and global inversion protocols require characterization of  the channels $\boldsymbol{\Lambda}_{L|\boldsymbol{s}}$, and are in principle unbiased if this characterization is exact.
The primary examples we study in this paper are based on local inversion protocols, as their shot overhead is well understood, they do not inherently require extensive classical computation, and they apply straightforwardly even to local channels $\Lambda_{L|s}$ with a large (order-1) infidelity $\epsilon_{L|s}$ - which generically appear in EC and are important for SALEM.  

If all protocols $EM_{\boldsymbol{s}}$ are unbiased, each $o_{\boldsymbol{s}}$ reproduces $\langle O\rangle$ in expectation, $\mathbb{E}[o_{\boldsymbol{s}}|N_{\boldsymbol{s}}]=\langle O \rangle$, but with a different shot overhead $\Gamma_{\boldsymbol{s}}$, defined by $\mathbb{V}[o_{\boldsymbol{s}}|N_{\boldsymbol{s}}]\leq \Gamma_{\boldsymbol{s}}/N_{\boldsymbol{s}}$ (Fig.~\ref{fig: mainV2}(f)). An aggregated estimator $\overline{o}\propto\sum_{\boldsymbol{s}} w_{\boldsymbol{s}} o_{\boldsymbol{s}}$ is then constructed as a weighted average over syndrome subsets, with weights $w_{\boldsymbol{s}}$. The estimator $\overline{o}$ is unbiased for any choice of weights, which may therefore be chosen to minimize the FG-SALEM shot overhead $\Gamma^{FG}_{SALEM}$, defined by  $\mathbb{V}[\overline{o}]\leq\Gamma^{FG}_{SALEM}/N$. As shown in  Appendix~\ref{Appendix: FG-SALEM gen. exp and var},  the optimal choice is given by `inverse-variance' (IV) weights $w_{\boldsymbol{s}}=N_{\boldsymbol{s}}/\Gamma_{\boldsymbol{s}}$, and the shot overhead for the resulting $\overline{o}$ is given (up to negligible corrections) by the `harmonic expectation' of $\Gamma_{\boldsymbol{s}}$ over $\boldsymbol{s}$,
\begin{align}
    \Gamma^{FG}_{SALEM}=\mathbb{H}[\Gamma_{\boldsymbol{s}}]:=\mathbb{E}[\Gamma_{\boldsymbol{s}}^{-1}]^{-1}. \label{Eq: Gamma_SALEM}
\end{align}

The FG-SALEM shot overhead in Eq.~\eqref{Eq: Gamma_SALEM} is significantly lower than that of ExtLEM, assuming both are based on the same underlying EM protocol. To gain intuition for this statement, assume the underlying EM protocol satisfies the simplified but representative functional relation $\Gamma(\epsilon)=e^{\lambda V\epsilon}$, between the infidelity $\epsilon$ of the mitigated error channel (averaged over all $V$ gates in the circuit),  and the corresponding shot overhead $\Gamma$. The parameter $\lambda$ appearing in the exponent (and thus, strongly affecting the shot overhead), is an order-1 coefficient that may be referred to as the ‘blowup rate’ of the EM protocol.  Thus, the shot overhead for ExtLEM is $\Gamma_{ExtLEM}=e^{\lambda V\epsilon_L}$, and the shot overheads for FG-SALEM are $\Gamma_{\boldsymbol{s}}=e^{\lambda V \epsilon_{L|\boldsymbol{s}}}$. Noting that $\epsilon_{L}=\mathbb{E}[\epsilon_{L|\boldsymbol{s}}]$, we have $\Gamma_{ExtLEM}=e^{\lambda V \mathbb{E}[\epsilon_{L|\boldsymbol{s}}]}=e^{ \mathbb{E}[\log\Gamma_{\boldsymbol{s}}]}=\mathbb{G}[\Gamma_{\boldsymbol{s}}]$, which is the `geometric expectation' of $\Gamma_{\boldsymbol{s}}$. The well-known HM-GM inequality then shows that 
\begin{equation}
\Gamma^{FG}_{SALEM}=\mathbb{H}[\Gamma_{\boldsymbol{s}}]\leq \mathbb{G}[\Gamma_{\boldsymbol{s}}]=\Gamma_{ExtLEM},\label{Eq: H<G}
\end{equation}
with the gap determined by the non-uniformity of $\Gamma_{\boldsymbol{s}}$, and hence of $\epsilon_{L|\boldsymbol{s}}$. These are generically highly non-uniform in error-correcting circuits, since the appearance of different syndromes requires different minimal numbers of faults $m_{\boldsymbol{s}}\leq t+1$, such that $\mathbb{P}(\boldsymbol{s})\sim \epsilon^{m_{\boldsymbol{s}}}$, and  $\epsilon_{L|\boldsymbol{s}}\sim\epsilon^{t+1-m_{\boldsymbol{s}}}$ are separated into different orders in $\epsilon$. In particular, the `worst' syndromes, with $m_{\boldsymbol{s}}=t+1$, are responsible for a significant part of the gap between $\Gamma_{SALEM}^{FG}$ and $\Gamma_{ExtLEM}$, see Appendix \ref{Appendix: bi-SALEM} for a toy model. Generalizing   Eq.~\eqref{Eq: H<G}, Appendix \ref{Appendix: advantage of SALEM over ExtLEM} shows that $\Gamma_{SALEM}^{FG}\leq \Gamma_{ExtLEM}$ corresponds to a generalized mean inequality, which holds under mild convexity assumptions satisfied by a wide variety of EM protocols.

Many EM protocols show an approximately-exponential behavior, with blowup rate  $\lambda=\lim_{\epsilon\rightarrow0}(\epsilon V)^{-1}\log\Gamma(\epsilon)$. Table~\ref{Tab: per-syndrome lambda} shows numerical estimates for the FG-SALEM blowup rate $\lambda^{FG}_{SALEM}=\lim_{\epsilon_L\rightarrow0}(\epsilon_LV)^{-1}\log \Gamma^{FG}_{SALEM}$ in a number of examples based on QP distributions, exhibiting a significant improvement over $\lambda_{ExtLEM}=\lambda=4$ \cite{van2023probabilistic, QESEMpaper}. For each code in Tab.~\ref{Tab: per-syndrome lambda} we consider a fixed EC circuit, augmented with either a  `realistic' decoder, considered fast enough to run in real time, or the optimal \textit{degenerate maximum likelihood} (ML) decoder, which gives the lowest possible logical error $\epsilon_L$ for the chosen EC circuit \cite{deMartiiOlius2024decodingalgorithms}. We see that FG-SALEM has a lower blowup rate for realistic decoders. This seemingly positive result implies a subtle interplay between FG-SALEM and optimal decoding, which we now discuss.

\renewcommand{\arraystretch}{1.2}  
\begin{table}[t]
\caption{Blowup rates for FG-SALEM based on QP distributions, in distance-3 codes. For each code we consider a FT logical memory circuit, equipped with a realistic sub-optimal decoder (minimal-weight perfect matching, MWPM \cite{pymatching}, or a minimal-weight lookup table, LUT \cite{Chao_2018}) or the optimal ML decoder, and subjected to a circuit-level noise model (see Appendix \ref{Appendix: numerical simulations}). Values for $\lambda^{FG}_{SALEM}$ are to be contrasted with the blowup rate for ExtLEM, $\lambda=4$. To make this comparison explicit, we also report the shot overhead improvement factor
$\Gamma_{ExtLEM}/\Gamma^{FG}_{SALEM}=\exp\!\big((4-\lambda^{FG}_{SALEM})\epsilon_L V \big)$, for $V=3/\epsilon_L$.\label{Tab: per-syndrome lambda}}
\label{tab:thresholds}
\centering
\begin{ruledtabular}
\begin{tabular}{lcccc}
Code & \multicolumn{2}{c}{Surface} & \multicolumn{2}{c}{Steane} \\
\hline
Decoder & MWPM & ML & LUT & ML \\
\hline
$\lambda^{FG}_{SALEM}$ & 2.3 & 3.3 & 2.5 & 3.0 \\
$\Gamma_{ExtLEM}/\Gamma^{FG}_{SALEM}$  
  & $164$ & $8.2$ & $90.0$ & $20.1$ \\
\end{tabular}
\end{ruledtabular}
\end{table}

\section{Fine-grained SALEM as an extension of optimal decoding\label{Sec: fine-grained SALEM and ML decoding}}

The implementation of FG-SALEM requires solving the `syndrome-conditioned logical characterization' problem: compute $\Lambda_{L|s}$ given $s$.
Doing this to high accuracy is feasible for codes with a small number of possible syndromes, like the distance-3 codes in Table~\ref{Tab: per-syndrome lambda}. However, for large codes, this is generally intractable, as it is at least as hard as ML decoding, which is a $\#P$-hard computational task \cite{iyer2013hardness}. Indeed, given the channel $\Lambda_{L|s}=\sum_\sigma \mathbb{P}(\sigma|s)\sigma$ one can perform ML decoding by extracting the most likely logical error $\sigma_{s}=\text{argmax}_\sigma \mathbb{P}(\sigma|s)$, and applying the recovery $\sigma_s^{-1}$.  Note that $\Lambda_{L|s}$ is the logical error channel after a logical gate, including the  decoder used within it (Fig.~\ref{fig: mainV2}(a)). Accordingly, $\sigma_s$ is the most likely logical error after decoding, and $\sigma_s\neq I$ represents a `decoding error'.

We can now explain why $\lambda^{FG}_{SALEM}$ is lower for the realistic decoders in Tab.~\ref{Tab: per-syndrome lambda}: The local inversion of $\Lambda_{L|s}$ can be viewed as first implementing the ML recovery $\sigma_{s}^{-1}$ (without a shot overhead), and subsequently inverting the residual channel $\sigma_{s}^{-1}\Lambda_{L|s}$ obtained after ML decoding (with an overhead which is independent of the decoder actually used). Thus, the shot overhead $\Gamma^{FG}_{SALEM}$ can be independent of the decoder used, implicitly improving it to ML.  However, the logical error $\epsilon_L$ is larger for realistic decoders, leading to a smaller blowup rate $\lambda_{SALEM}^{FG}$ in the realistic case (see Appendix \ref{Appendix: bi-SALEM} for a toy model).

Conceptually, we see that FG-SALEM (with exact characterization) is a generalization of ML decoding - inverting the entire error channel $\Lambda_{L|s}$ as opposed to `just' the most likely error $\sigma_s$. Just as ML decoding is used to define the optimal performance to which realistic decoders aspire to, and which can be approached in exchange for classical complexity, we view  FG-SALEM as marking an extreme point in the tradeoff between quantum complexity (shot overhead) and classical complexity. 

FG-SALEM may be practical for certain large codes by leveraging TN decoders, which perform an approximate  syndrome-conditioned logical characterization, see Sec.~\ref{Sec: discussion}.  In the following section we take an alternative approach towards a scalable implementation of SALEM.

\section{Coarse-grained SALEM\label{Sec: coarse-grained SALEM main text}} 

\begin{figure*}[ht!]
\begin{centering}
\includegraphics[width=1\textwidth]{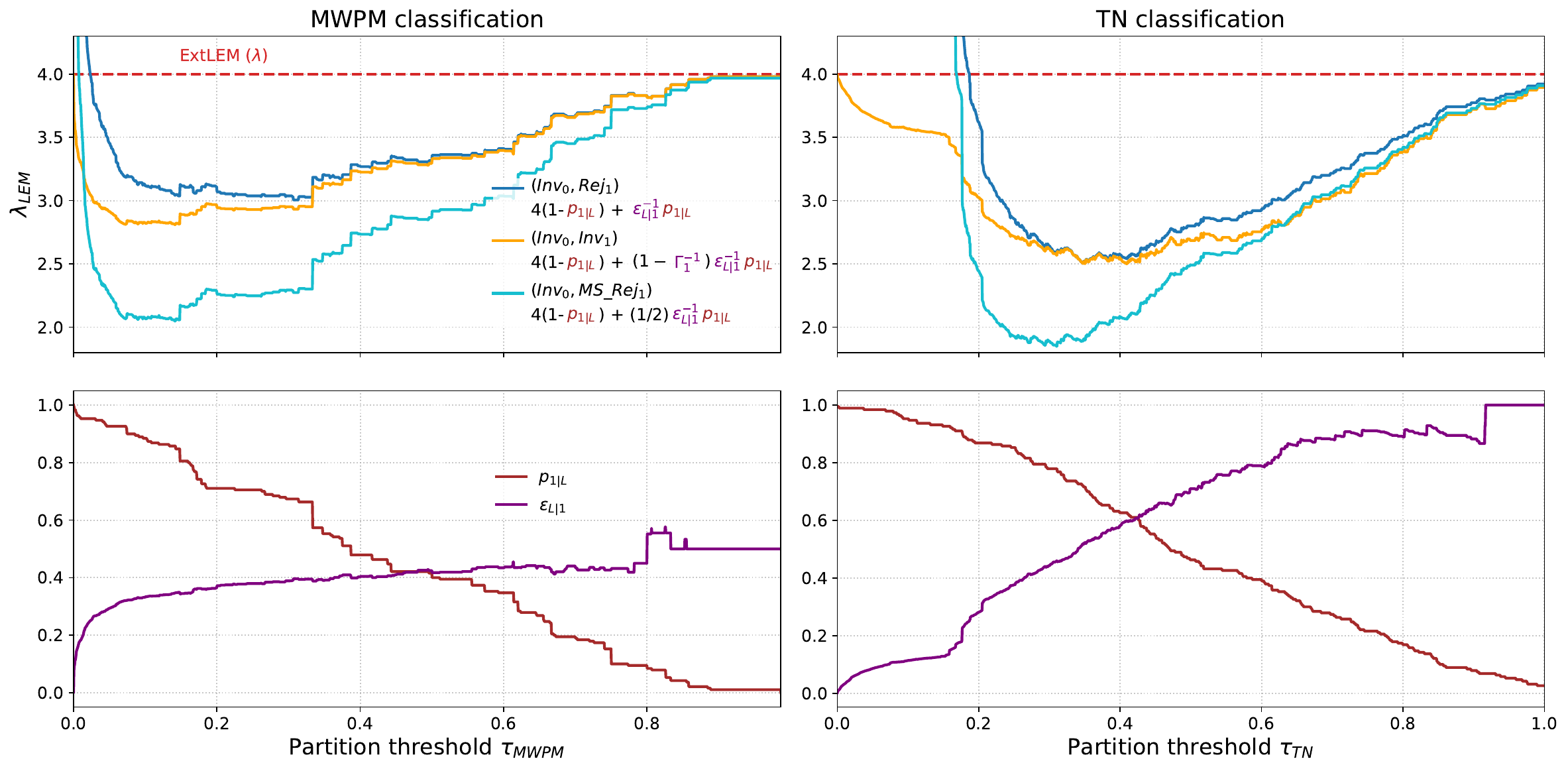}
\par\end{centering}
\caption{
Performance of coarse-grained (binary) SALEM. Top panels: Blowup rates $\lambda^{\{S_0,S_1\}}_{SALEM}$  for different variants of CG-SALEM based on binary partitions $S=S_0\cup S_1$ of syndromes (`binary SALEM'). The `good' set $S_0$ should have a small conditioned logical error $\epsilon_{L|0}$, and is mitigated using local inversion ($\text{Inv}_0$). The `bad' set $S_1$ should have a large conditioned logical error $\epsilon_{L|1}$, and is mitigated by rejection (blue), local inversion (orange) or mid-shot rejection (cyan). In each panel, the bad set $S_1$ shrinks from $S$ to $\emptyset$ along the x-axis. Minimal values of $\lambda_{LEM}$ for each binary SALEM variant indicate the optimal partition for that variant. The $(\text{Inv}_0,\text{Inv}_1)$ variant always improves upon both ExtLEM (based on local inversion) and $(\text{Inv}_0,\text{Rej}_1)$, but requires that classification be performed in real-time and does not significantly improve upon $(\text{Inv}_0,\text{Rej}_1)$ when using (near-) optimal partitions. Mid-shot rejection also requires a real-time classifier, and $(\text{Inv}_0,\text{MS\_Rej}_1)$ can provide the best blowup rate (of QPU time) among the three variants, depending on the depth of the mitigated logical circuit (data shown corresponds to depth $D=\epsilon_L^{-1}$). The data is obtained from simulations of logical memory circuits for the $d=4$ surface code, decoded using the MWPM algorithm \cite{pymatching}, representing a sub-optimal `realistic' decoder. We consider two families of classifiers: one based on the MWPM decoder itself (Left Panels) \cite{smith2024mitigating, meister2024efficientsoftoutputdecoderssurface, gidney2023yokedsurfacecodes}, which can therefore run in real-time; and a stronger but slower family of classifiers based on a TN decoder (Right Panels) \cite{piveteau2023tensornetworkdecoding}, which may only be possible to run in post-processing. The x-axes correspond to a threshold that defines a specific classifier within each family, setting a cutoff on a decoder-derived proxy for the conditional logical error (see Appendix~\ref{Appendix: surface code simulations}). 
Numerically obtained blowup rates are well approximated by simple expressions in terms of $\epsilon_{L|1}=\mathbb{P}(\text{logical error}|S_1)$ and 
$p_{1|L}=\mathbb{P}(S_1|\text{logical error})$, the latter representing the fraction of logical errors `contained' in the set $S_1$ (see top legend and Appendix~\ref{Appendix: bi-SALEM}-\ref{Appendix: mid-shot reject}). These quantities are plotted in the Bottom Panels. Maximizing both $\epsilon_{L|1}$ and $p_{1|L}$ is desirable but impossible, and optimal partitions are obtained when the two are balanced.   
  \label{Fig: Bi-SALEM}
 }
\end{figure*}

We consider two modifications of the FG-SALEM protocol discussed so far. These modifications define `coarse-grained' SALEM (CG-SALEM) protocols, that can reach blowup rates close to those of FG-SALEM, without requiring significant real-time classical computation, beyond what is already used by realistic decoders, see Fig.~\ref{Fig: Bi-SALEM}. Moreover, CG-SALEM relaxes the limitations on the types of EM protocols which can be used in practice, and can even be made compatible with any EM protocol. 

Our first modification is a coarse-graining of the set of global syndromes. This can be achieved by partitioning each set of local syndromes $S=\{s\}$ (namely, syndromes associated with a specific logical gate) to  subsets $S_k,\ k=0,1\dots,K$, such that $S=\cup_{k=0}^K S_k$.  Collecting all local subset indices $k$ we obtain a global index $\boldsymbol{k}$, and global syndrome subsets $S_{\boldsymbol{k}}$. Given such a partition, CG-SALEM will apply an EM protocol $EM_{\boldsymbol{k}}$ designed to mitigate the channel $\boldsymbol{\Lambda}_{L|\boldsymbol{k}}$, to the $N_{\boldsymbol{k}}$ shots in which syndromes $\boldsymbol{s}\in S_{\boldsymbol{k}}$ are measured, obtaining an estimator $o_{\boldsymbol{k}}$. 
The CG-SALEM estimator $\overline{o}$ will then be obtained as a weighted average of $o_{\boldsymbol{k}}$ with IV weights $w_{\boldsymbol{k}}=N_{\boldsymbol{k}}/\Gamma_{\boldsymbol{k}}$, where $\Gamma_{\boldsymbol{k}}$ is a bound on the shot overhead of $EM_{\boldsymbol{k}}$. Note that more generally, one can directly partition the set of global syndromes.

The second modification is the possible rejection of $S_K$, implemented by assigning a weight $w_{\boldsymbol{k}}=0$ to global indices $\boldsymbol{k}$ including at least one instance of $K$. This may be viewed as a limiting case of IV weighting, with infinite shot overhead $\Gamma_{\boldsymbol{k}}\rightarrow\infty$ assigned to such global indices $\boldsymbol{k}$.

In analogy with Eq.~\eqref{Eq: Gamma_SALEM}, coarse-graining with IV weighting leads to the shot overhead $\Gamma_{SALEM}^{\{S_{\boldsymbol{k}}\}}=\mathbb{H}[\Gamma_{\boldsymbol{k}} ]$. In analogy with Eq.~\eqref{Eq: H<G}, this overhead is always better than ExtLEM, but worse than FG-SALEM,  
\begin{align}
    \Gamma_{SALEM}^{FG}\leq \Gamma_{SALEM}^{\{S_{\boldsymbol{k}}\}}\leq \Gamma_{ExtLEM},
\end{align}
with finer partitions providing an overhead which is closer to FG-SALEM. Rejecting $S_{K}$ further increases the shot overhead to 
\begin{align}
    \Gamma_{SALEM}^{\{S_{\boldsymbol{k}}\},Rej}=\mathbb{P}_{Acc}^{-1} \mathbb{H}\left[\Gamma_{\boldsymbol{k}}|Acc\right]\geq \Gamma_{SALEM}^{\{S_{\boldsymbol{k}}\}},
\end{align}
where $\mathbb{P}_{Acc}$ is the (global) acceptance probability. An equality occurs if $\Gamma_{\boldsymbol{k}}=\infty$ for all rejected $\boldsymbol{k}$. Note, however, that since final measurements from rejected shots are not used, such shots can in principle be rejected once the first rejected local syndrome is measured. Such `mid-shot rejection' does not change the shot overhead, but we find that it can significantly reduce QPU time overheads, for circuits with practically relevant volumes and aspect ratios (Appendix \ref{Appendix: mid-shot reject}). Thus, when minimizing QPU time as opposed to shot number, the optimal SALEM protocol would include a subset of syndromes that are mid-shot rejected, while accepted syndromes would be completely partitioned to singletons, as in FG-SALEM.

A main advantage of CG-SALEM over FG-SALEM is that mapping syndromes to corresponding logical error channels is computationally tractable for large codes. Indeed, instead of the syndrome-conditioned logical characterization ($s\mapsto\Lambda_{L|s}$) needed for FG-SALEM, which may be computationally hard, CG-SALEM requires two  computationally tractable tasks:

\textbf{Partition and classification:} Define a partition $S=\cup_kS_k$, and classify each measured syndrome $s$ to its containing subset $S_k$. Useful partitions should preserve a significant fraction of the shot overhead improvement of FG-SALEM over ExtLEM, and allow for computationally efficient classification, $s\mapsto k$. We find that such useful partitions can be constructed using `soft-output decoders' \cite{smith2024mitigating, meister2024efficientsoftoutputdecoderssurface, gidney2023yokedsurfacecodes, toshio2025decoderswitchingbreakingspeedaccuracy}. In addition to proposing a recovery operation, such decoders produce a metric $\tau$ for the probability of a logical error given the proposed recovery. Fig.~\ref{Fig: Bi-SALEM} shows the performance of CG-SALEM with partitions based on the value of $\tau$. 

\textbf{Subset-conditioned characterization:} Estimate the  channels $\Lambda_{L|k}$ for (accepted subsets) $k=0,\dots,K$. We focus on small numbers $K$ of subsets, such that all channels $\Lambda_{L|k}$ can be efficiently computed in pre-processing and stored in a short lookup table.  
We describe our `physical-to-logical' characterization (P2LC) protocol in Appendix \ref{sec:approx logical channels}. Appendix \ref{Sec: color code simulations}  demonstrates numerically  that the logical error channels produced by P2LC allow for  CG-SALEM which is essentially unbiased.

CG-SALEM offers several additional advantages over FG-SALEM. First,  
if the protocols $EM_{\boldsymbol{s}}$ require adaptive logic, the reduction to a smaller number of protocols $EM_{\boldsymbol{k}}$ simplifies this logic. For example, mapping $s$ to a sample from a corresponding QP distribution for $\Lambda_{L|k}^{-1}$ can be done efficiently for a small number of subsets. Note that mid-shot rejection is adaptive, as opposed to standard rejection. Second, coarse graining and rejection may be used to
ensure that all (accepted) subsets $S_{\boldsymbol{k}}$ are probable enough
to implement protocols $EM_{\boldsymbol{k}}$  requiring a large number
of shots in order to produce an estimator, such as exponential zero-noise extrapolation (ZNE) \cite{PhysRevX.8.031027}.

For $K=0$ (no partition) we reproduce ExtLEM. The case $K=1$ (binary partition) with a rejected subset $S_{1}$ and a complementary accepted subset $S_{0}$, to which a trivial protocol $EM_{\boldsymbol{0}}=\texttt{do\_nothing}$ is applied, corresponds to the commonly used  combination of EC and post-selection (EC+PS) \cite{smith2024mitigating, PrabhuReichardt2024} (Fig.~\ref{fig: mainV2}(g)). A significant drawback of EC+PS is that it leaves un-mitigated the accepted error channels $\boldsymbol{\Lambda}_{L|\boldsymbol{0}}$, leading to a bias $b=|\langle O\rangle - \langle O\rangle_{\boldsymbol{\Lambda}_{L|\boldsymbol{0}}}|$. An additional drawback is that the shot overhead $\mathbb{P}_{acc}^{-1}$ for mitigating the channels $\boldsymbol{\Lambda}_{L|\boldsymbol{k}\neq \boldsymbol{0}}$ by rejection is sub-optimal, unless $\Gamma_{\boldsymbol{k}}=\infty$ for all $\boldsymbol{k}\neq \boldsymbol{0}$. Such overhead divergences are unavoidable when the channels $\boldsymbol{\Lambda}_{L|\boldsymbol{k}\neq \boldsymbol{0}}$ are singular, see Appendix \ref{Appendix: bi-SALEM}.

The simplest novel instance of SALEM is obtained for $K=1$ by assigning a non-trivial protocol $EM_{\boldsymbol{0}}$ to the accepted set $S_0$ (and rejecting $S_1$). As demonstrated in Fig.~\ref{fig: performance}, this already suffices to achieve a significant advantage over both ExtLEM and EC+PS. With a single accepted set, this version of SALEM is compatible with \textit{any} protocol $EM_{\boldsymbol{0}}$, which can be performed non-adaptively in all shots, while  classification and rejection can be performed in post-processing. If classification is nevertheless performed in real-time, adaptivity can be used to reduce quantum runtime, by either (i) accepting and mitigating the set $S_1$, or (ii) rejecting it mid-shot, once a syndrome $s\in S_1$ is observed.  
In Fig.~\ref{Fig: Bi-SALEM} we compare these different variants of CG-SALEM based on binary partitions (`binary SALEM'), demonstrating the tradeoffs between classical real-time processing and quantum runtime (see Appendices~\ref{Appendix: bi-SALEM} and \ref{Appendix: surface code simulations} for details).

\section{Discussion and outlook\label{Sec: discussion}}

 EM on the physical level plays a key role in working with existing QPUs, extending the  circuit volumes that can be executed accurately by orders of magnitude. Likewise, 
LEM is expected to similarly play a key role in utilizing error-corrected QPUs for as long as physical qubit numbers remain limited  \cite{WhyEM, eisert2025mind}. 
In this work we demonstrated that LEM can be made significantly more efficient, by re-purposing the syndrome data generated as part of EC. Effectively using this data requires syndrome classifiers, which are closely related to a growing body of work on `soft-output decoders' \cite{smith2024mitigating, meister2024efficientsoftoutputdecoderssurface, gidney2023yokedsurfacecodes, toshio2025decoderswitchingbreakingspeedaccuracy}. Thus, SALEM is intimately connected with EC, and we believe the two should be viewed holistically, as a framework that maximizes output accuracy given an allowance of physical qubits, quantum runtime, and classical compute resources for real-time processing, as well as for pre- and post-processing.

Rapid progress has been made recently in experimental demonstrations of EC \cite{1-bluvstein2024logical,2-mayer2024benchmarking,
4-reichardt2024demonstration}. We anticipate the focus of such experiments to gradually shift from demonstrating an improvement over noisy circuit execution, to demonstrating the more practically relevant improvement over physical EM using SALEM. The `SALEM threshold' we defined corresponds precisely to this comparison. We find, perhaps counterintuitively, that despite the large space-time overhead of EC, the efficiency of SALEM can make EC useful  even in noise rate regimes where EC (alone) does not improve upon bare circuit execution. Moreover, SALEM can provide practical value in the very near term, when applied to QPUs supporting tens of \textit{error detected} logical qubits, which are already becoming available to end users \cite{1-bluvstein2024logical, ransford2025helios_98qubit}.

Our work identifies several directions for further research. First, in our numerical simulations, we focused on well-studied EC codes, namely color and surface codes, where SALEM could be simulated on top of well-known fault-tolerance constructions and decoders, with publicly accessible implementations. It would be interesting to estimate the performance of SALEM on top of more recently proposed EC schemes \cite{xu2024constant,
bravyi_2024, scruby2024high}. 

Second, we found that in non-adaptive instances of SALEM, a `strong but slow' TN decoder operating in post-processing can be useful for syndrome classification. We relied on the recent work of Ref.~\cite{piveteau2023tensornetworkdecoding}, which extended TN decoders to the circuit-level setting. In doing so we found that runtimes for TN decoders must be significantly improved if they are to be practically used, even in post-processing. Reference~\cite{kaufmann2024blockbp} demonstrated a significant speedup for TN decoders in the code-capacity setting, and it would be interesting to extend this approach to the circuit level. 
With fast and accurate enough TN decoders, it may even be possible to perform FG-SALEM, since these decoders can be used to approximate the required channels $\Lambda_{L|s}$. 

Third, in Appendix~\ref{Appendix: lower bounds} we show that SALEM can have a shot overhead that is lower than that of any possible LEM protocol that does not make use of syndrome data (i.e., any ExtLEM protocol). Whether or not SALEM makes optimal use of syndrome data, and accordingly, whether syndrome-based LEM methods that are more efficient than SALEM exist, is an interesting and practically relevant open question. 

Last but not least, it would be very interesting to establish experimentally that SALEM can indeed be more efficient than physical EM even above the standard FT threshold, establishing that EC can in fact be useful in this regime.

\vspace{10pt}

\textbf{Acknowledgments} We thank  our colleagues at Qedma for invaluable discussions and helpful feedback during this project. We also thank Jay Gambetta and John Preskill for insightful discussions and remarks. 

\vspace{10pt}

\textbf{Competing interests}  This work describes methods that are the subject of U.S. and international patent applications filed by Qedma Quantum Computing Ltd. \cite{qedma_logical_errors_2024}.

\vspace{10pt} 

\textbf{Note added} While finalizing this manuscript, References~\cite{Zhou2025ErrorMitigation,Dinca2025ErrorMitigation} appeared on the arXiv, discussing LEM based on soft-output decoders. Ref.~\cite{Zhou2025ErrorMitigation} describes the special case of binary SALEM with a rejected subset, in the code-capacity setting, where no faults occur during error correction.
Ref.~\cite{Dinca2025ErrorMitigation} describes EC+PS, as well as a LEM protocol based on heuristic post-processing of syndrome-resolved data. 

\textbf{Note added in v2} Several related works \cite{xiao2026,zheng2026,yuan2026,kumar2026,tsubouchi2026} appeared after posting the first version of this paper.   Ref.~\cite{xiao2026,zheng2026} discuss logical characterization based on syndrome data, finding an exponential advantage in QPU time over external logical characterization. This is closely related to the advantage of physical-to-logical characterization (P2LC)
over external characterization reported on in Appendix \ref{sec:approx logical channels}.
Indeed, syndrome-based characterization can be understood as first mapping syndrome data to a physical error model, which is then mapped to a logical error model \cite{xiao2026,zheng2026}. Ref.~\cite{yuan2026,kumar2026} study binary SALEM with a rejected subset, discussed in Sec.~\ref{Sec: coarse-grained SALEM main text}. They  base their protocol  on error-detection codes, and describe logical characterization methods that share similarities with P2LC. Ref.~\cite{tsubouchi2026} performs an information-theoretic study of syndrome-based LEM, in a simplified setting with one round of EC and code-capacity errors,  establishing that FG-SALEM, described in Sec.~\ref{Sec: SALEM}, is the shot-optimal LEM method among those that measure only the target observable. Ref.~\cite{tsubouchi2026} proves a lower bound on the FG-SALEM overhead in this setting, distinct from the lower bounds we obtain in Appendix \ref{Appendix: lower bounds}. They further suggest measuring observables correlated with the target observable as a means to overcome this bound.

\clearpage

\appendix

\numberwithin{equation}{subsection}
\renewcommand{\theequation}{\thesubsection.\arabic{equation}}

\noappendixtoc

\section*{Appendix}
\yesappendixtoc

\vspace{1em}
\listofappendices 

\subsection{Setup\label{Appendix: setup}}

\subsubsection{Error-corrected logical gates and layers}

Let $C=G_D\cdots G_1$ denote a faulty error-corrected quantum circuit, comprised of $D$ error-corrected logical layers $G_j$. Each $G_{j}$ is an adaptive physical circuit, including gadgets implementing a desired logical layer $g_j$, as well as syndrome measurement, decoding and recovery. Note that we use the term `syndrome' to refer to the bit-string including the results of \textit{all} mid-circuit measurements performed within  $G_{j}$, including e.g., repeated or flagged stabilizer measurements. We consider realistic `circuit-level' noise models, where each physical operation in $G_{j}$, including state preparations, measurements and unitary gates, carries a physical error (or `fault') channel, with infidelity at most $\epsilon$.  

We start our analysis by considering each layer $G_j$ as a whole, ignoring any local structure until Appendix \ref{sec:approx logical channels}. Accordingly, we consider the stabilizer code $[\![n,k,d]\!]$ involving all $k$ logical qubits in the circuit \cite{gottesman1997}, ignoring for now any internal block structure that may be present. The parameter $n$ represents the total number of physical `data qubits' in the circuit, that are passed between layers, and excludes auxiliary qubits that are used within each layer for e.g., syndrome measurement or resource-state preparation, and are reset between layers. We denote the Hilbert space of physical data qubits by $H$ and the code subspace of logical qubits by $K$, such that
\begin{align}
    (\mathbb{C}^2)^{\otimes k}\cong K\subset H=(\mathbb{C}^2)^{\otimes n}.
\end{align}
We use calligraphic letters for the corresponding spaces of Hermitian operators, 
\begin{align}
    \mathcal{K}=Herm(K)\subset Herm(H)=\mathcal{H},
\end{align}
such that physical input and output states for each layer are  elements of $\mathcal{H}$, while logical states are  elements of $\mathcal{K}$. Using the `super-bra-ket' notation, we denote by $\sket{c}\in \mathcal{K}$ a general (possibly mixed) code state. By convention, any statement involving $\sket{c}$ should be understood as valid for all code states in $\mathcal{K}$. The layers 
\begin{align}
    G_j:\mathcal{H}\rightarrow\mathcal{H}
\end{align}
are trace-preserving (TP) and completely-positive (CP) super-operators. In the absence of faults, each error-corrected layer $G_j$  implements a desired unitary logical layer 
\begin{align}
    g_j:\mathcal{K}\rightarrow\mathcal{K}
\end{align}
on the code space. Denoting the fault-free, or `ideal', version of $G_j$ by $G_{j,\mathrm{ideal}}$, this may be expressed as 
\begin{align}
    G_{j,\mathrm{ideal}}\sket{c}=g_j\sket{c}.
\end{align} 
The syndrome measurements, decoding and recovery in each $G_{j}$ are constructed such that $G_{j,\mathrm{ideal}}$ can be written as
\begin{align}
    G_{j,\mathrm{ideal}}=g_j\mathrm{EC}_j, \label{Eq: ideal EC}
\end{align}
where $\mathrm{EC}_j:\mathcal{H}\rightarrow\mathcal{K}$ is an ideal round of error correction, which maps any physical state to a code state, and acts trivially on the code. We therefore have
\begin{align}
    G_{j,\mathrm{ideal}}:\mathcal{H}\rightarrow\mathcal{K}.
\end{align}

\subsubsection{Logical and correctable errors}

We denote by $\mathcal{P}_n$ the group of $4^n$ physical Pauli super-operators $\sigma:\mathcal{H}\rightarrow\mathcal{H}$, acting by conjugation. The stabilizer group $\mathcal{S}\subset \mathcal{P}_n$ includes $2^{n-k}$ physical Paulis that act trivially on $\mathcal{K}$. The group of logical Paulis $\mathcal{L}\subset \mathcal{P}_n$ includes the physical Paulis $\sigma_{\mathrm{log}}\in \mathcal{P}_n$ that map $\mathcal{K}$ to itself. The restrictions $\sigma_{\mathrm{log}}|_{\mathcal{K}}$ of logical Paulis to $\mathcal{K}$ correspond to the cosets $\{\sigma_{\mathrm{log}}\mathcal{S}\}=\mathcal{L}/\mathcal{S}\cong\mathcal{P}_k$, isomorphic to the $4^k$ $k$-qubit Paulis $\mathcal{P}_k$ acting on logical qubits. We use an arbitrary list of representatives $\{\sigma_{\mathrm{log}}^b\}_{b=1}^{4^k}$ for these logical cosets of the stabilizer group. 

The syndrome cosets $\mathcal{P}_n/\mathcal{L}\cong (\mathcal{P}_n/\mathcal{S})/(\mathcal{L}/\mathcal{S})$ are in one-to-one correspondence with the $2^{n-k}$ possible syndromes of $G_{j,\mathrm{ideal}}$ (`ideal syndromes'),\footnote{Note, however, that additional syndromes are possible in the presence of faults. The syndrome bit-string of $G_j$ (and therefore also of $G_{j,\mathrm{ideal}}$) includes the outcomes of all physical measurements performed in $G_{j}$, including e.g., repeated or flagged stabilizer measurements, and generally contains more than $n-k$ bits.} and $2^{n-k}$ corresponding orthogonal subspaces of the form $\sigma(\mathcal{K})\subset \mathcal{H}$. Choosing an arbitrary set of representatives $\{\sigma^a\}_{a=1}^{2^{n-k}}$ for the syndrome cosets, we write these orthogonal subspaces as $\mathcal{K}^a=\sigma^a(\mathcal{K})$, with $\mathcal{K}^1=\mathcal{K}$. The representatives $\sigma^a$ are often chosen to be algebraically-simple `destabilizers' \cite{AaronsonGottesman2004}, or `pure errors' \cite{Poulin2006}. 

We instead represent the syndrome cosets operationally, using \textit{correctable errors} $\sigma_{\mathrm{corr}}\in \mathcal{P}_n$. These are defined by the requirement 
\begin{align}
    G_{j,\mathrm{ideal}}\sigma_{\mathrm{corr}}\sket{c}=g_j\sket{c}. \label{Eq: corr def}
\end{align}
That is, $G_{j,\mathrm{ideal}}$ produces the desired output state $g_j\sket{c}$, even if the input state $\sket{c}$ is corrupted by the physical Pauli error $\sigma_{\mathrm{corr}}$. Note that the set of correctable errors depends on the decoder in $G_{j,\mathrm{ideal}}$ (inherited from $G_{j}$), and can therefore be layer-dependent.\footnote{Correctable errors typically include the set of Paulis $\sigma\in \mathcal{P}_n$ with weight (or support) $w_\sigma\leq \lfloor (d-1)/2\rfloor$, where $d=\min_{\sigma_{\mathrm{log}}\in\mathcal{L}-\mathcal{S}}w_{\sigma_{\mathrm{log}}}$ is the code distance, as well as some but not all Paulis with a higher weight. However, even the inclusion of \textit{all} $w_\sigma\leq \lfloor (d-1)/2\rfloor$ Paulis depends on the decoder, which may be designed differently, depending on the circuit structure and circuit-level noise model. More generally, the code distance $d$ does not directly enter our analysis. We introduce the related `fault-tolerance level' $t$ in Appendix \ref{sec:approx logical channels}.  } Explicitly, the ideal EC cycle in Eq.~\eqref{Eq: ideal EC} can be written as 
\begin{align}
    \mathrm{EC}_j=\sum_{a=1}^{2^{n-k}} R^a_j\Pi^a, \label{Eq: explicit ideal EC}
\end{align}
where $\Pi^a=R^a_j\Pi^1 R^a_j$ is the orthogonal projector on $\mathcal{K}^a$, $R^a_j\in \mathcal{P}_n$ is the corresponding recovery operation, and the ideal decoder $a\mapsto R_j^a$ is fixed by the decoder in $G_j$. Equations \eqref{Eq: explicit ideal EC}, \eqref{Eq: ideal EC} and \eqref{Eq: corr def} show that every correctable error $\sigma_{\mathrm{corr}}$ is given by one of the recoveries $R^a_j$, modulo $\mathcal{S}$. 

It follows that correctable cosets $\sigma_{\mathrm{corr}}\mathcal{S}$ of the stabilizer group, or equivalently, restrictions $\sigma_{\mathrm{corr}}|_{\mathcal{K}}:\mathcal{K}\rightarrow \mathcal{H}$, are in one-to-one correspondence with ideal syndromes, and we use an arbitrary list of representatives $\{\sigma_{\mathrm{corr}}^a\}_{a=1}^{2^{n-k}}$ (e.g., the recoveries $R_j^a$). It is then possible to uniquely decompose every physical Pauli error $\sigma$ on the code space as a product of a correctable representative and a logical representative
\begin{align}
\sigma|_{\mathcal{K}}=(\sigma_{\mathrm{corr}}^a\sigma_{\mathrm{log}}^b)|_{\mathcal{K}}:\mathcal{K}\rightarrow \mathcal{H}.\label{Eq: sigma=sigma_c sigma_l}  
\end{align}
The action of  $G_{j,\mathrm{ideal}}$ on a Pauli-erred input state $\sigma \sket{c}=\sigma_{\mathrm{corr}}^a\sigma_{\mathrm{log}}^b\sket{c}$ then produces the code state
\begin{align}
    g_j\sigma_{\mathrm{log}}^b\sket{c}=G_{j,\mathrm{ideal}}\sigma \sket{c}, \label{Eq: logical error}
\end{align}
with a possibly non-trivial logical error $\sigma_{\mathrm{log}}^b$.

\subsubsection{Input and output error channels}

As opposed to the fault-free layer $G_{j,\mathrm{ideal}}$, the output of the faulty layer $G_j$ is not purely logical, and involves also correctable errors. This is due to `late faults' within $G_j$, which are not equivalent to any input error. Examples include physical  measurement errors, or physical errors occurring after the syndrome measurement circuit in $G_j$. Given an input error channel 
\begin{align}
    \Lambda_{\mathrm{in}}^{(j)}:\mathcal{K}\rightarrow \mathcal{H},
\end{align}
mapping code states to physical states, we can define an output error channel 
\begin{align}
    \Lambda_{\mathrm{out}}^{(j)}:\mathcal{K}\rightarrow \mathcal{H},
\end{align}
via
\begin{align}
\Lambda_{\mathrm{out}}^{(j)}g_j=G_j\Lambda_{\mathrm{in}}^{(j)}.\label{Eq: in->out}
\end{align}
This can be expressed as a commutative diagram,  
\begin{equation}
\begin{tikzcd}[column sep=large, row sep=large]
\mathcal{K}
  \arrow{r}{g_j}
  \arrow{d}[swap]{\Lambda_{\mathrm{in}}^{(j)}}
&
\mathcal{K}
  \arrow{d}{\Lambda_{\mathrm{out}}^{(j)}}
\\
\mathcal{H}
  \arrow{r}[swap]{G_j}
&
\mathcal{H}
\end{tikzcd}
\label{Eq:in-out-diagram}\ .
\end{equation}
Below we use the term `error channel' or `channel' to refer to any hermiticity-preserving (HP) and TP super-operator associated with errors of any kind (input, output or logical). We do not require that error channels be CP, and comment on this property when relevant.

\begin{example} {\bf (Adaptive Clifford circuits)} \label{example CAP}
    As an important example, let $G_j$ be a circuit comprised of  unitary Clifford gates, ancilla qubits prepared and mid-circuit-measured in the  standard basis, and  Pauli gates conditioned on mid-circuit measurement results; with Pauli fault channels before or after every physical operation. If $\Lambda_{\mathrm{in}}^{(j)}$ is a Pauli channel, the output  $\Lambda_{\mathrm{out}}^{(j)}$ is also a Pauli channel, and can be written as 
\begin{align}
    \Lambda_{\mathrm{out}}^{(j)}=\sum_{a,b}p_{a,b}^{(j)}(\sigma_{\mathrm{corr}}^{a} \sigma_{\mathrm{log}}^{b})|_{\mathcal{K}},\label{Eq: out = joint corr*log}
\end{align}
using Eq.~\eqref{Eq: sigma=sigma_c sigma_l}. Note, however, that the joint distribution $p_{a,b}^{(j)}$ does not generally decompose as a product of its marginals, and accordingly, $\Lambda_{\mathrm{out}}^{(j)}$ can't be written as a product of a correctable Pauli channel and a logical Pauli channel. 

Though Eq.~\eqref{Eq: out = joint corr*log} is written in terms of representatives $\sigma_{\mathrm{corr}}^{a}, \sigma_{\mathrm{log}}^{b}$, the resulting channel $\Lambda_{\mathrm{out}}^{(j)}$ is only defined on  $\mathcal{K}$, and therefore independent of the choice of representatives. Each such choice defines a distinct extension $\hat{\Lambda}_{\mathrm{out}}^{(j)}:\mathcal{H}\rightarrow \mathcal{H}$ of the output channel to a Pauli channel on the physical space. Such extensions will be used in Appendix \ref{Appendix: Implementable QP distributions}
\end{example}

\subsubsection{Logical state preparation and  measurement}

Each error-corrected shot starts with a faulty error-corrected state preparation $\sket{G_\psi}\in \mathcal{H}$. This is the output of a faulty adaptive circuit, acting on a fixed initial physical state, and aimed at producing a \textit{fixed} initial logical state $\sket{\psi}\in \mathcal{K}$. In the absence of faults, $\sket{G_{\psi,\mathrm{ideal}}}=\sket{\psi}$. The collective effect of the fault channels within $\sket{G_\psi}$ can be described by an output error channel $\Lambda_{\mathrm{out}}^{\psi}:\mathcal{K}\rightarrow\mathcal{H}$ such that
\begin{align}
    \Lambda_{\mathrm{out}}^{\psi}\sket{\psi}=\sket{G_\psi}. \label{Eq: prep out}
\end{align}
This is analogous to  Eq.~\eqref{Eq: in->out}, where there is no input channel, and $\sket{G_\psi}$, $\sket{\psi}$ and $\Lambda_{\mathrm{out}}^{\psi}$ play the roles of $G_j,g_j$ and $\Lambda_{\mathrm{out}}^{(j)}$, respectively. Note that $\Lambda_{\mathrm{out}}^{\psi}$ is not unique, since Eq.~\eqref{Eq: prep out} only specifies its action on a single logical state $\sket{\psi}$. 

\vspace{5pt}

A logical observable is an element $O\in \mathcal{K}$. At the end of each error-corrected shot, a faulty error-corrected logical
measurement $G_O$, corresponding to a logical observable $O$, is performed. $G_O$ is a faulty adaptive circuit that maps physical states to classical distributions over the spectrum of $O$. Restricting attention to the resulting expectation values, we write the error-corrected measurement as a linear functional $\sbra{G_O}:\mathcal{H}\rightarrow \mathbb{R}$. In the absence of faults and input errors, it satisfies
$\sbraket{G_{O,\mathrm{ideal}}}{c}=\sbraket{O}{c}$  for any code state $\sket{c}\in \mathcal{K}$.  In the presence of correctable input errors, we obtain the same output,
\begin{align}
    \sbra{G_{O,\mathrm{ideal}}}\!\sigma_{\mathrm{corr}}\sket{c}=\sbraket{O}{c},
\end{align}
in analogy with Eq.~\eqref{Eq: corr def}. For a general input error
\(\sigma=\sigma_{\mathrm{corr}}\sigma_{\mathrm{log}}\),
\begin{align}
   \sbra{G_{O,\mathrm{ideal}}}\!\sigma\sket{c}=\sbra{O}\!\sigma_{\mathrm{log}}\sket{c},
\end{align}
in analogy with Eq.~\eqref{Eq: logical error}, where $\sbra{O}$ plays the role of $g_j$. This is still a logical expectation value, i.e., the expectation value of the logical operator in a logical state.

Given an input error channel
\(\Lambda_{\mathrm{in}}^O:\mathcal{K}\to\mathcal{H}\), we can define a logical error channel
\(\Lambda_L^O:\mathcal{K}\to\mathcal{K}\)
such that
\begin{align}
    \langle\!\langle O|\Lambda_L^O=\sbra{G_O}\!\Lambda_{\mathrm{in}}^O.  \label{Eq: in->L O}
\end{align}
Indeed, by the standard isomorphism between linear functionals and vectors, the functional 
$\sbra{G_O}\!\Lambda_{\mathrm{in}}^O:\mathcal{K}\to\mathbb{R}$ corresponds to a unique operator \(O'\in\mathcal{K}\) such that
\(\sbra{G_O}\!\Lambda_{\mathrm{in}}^O=\sbra{O'}\).
The difference between \(O'\) and \(O\) is due to the combined effect of input errors and faults. We may then define
\(\Lambda_L^O:\mathcal{K}\to\mathcal{K}\) to be any logical channel satisfying
\(\sbra{O'}=\sbra{O}\Lambda_L^O\), such that Eq.~\eqref{Eq: in->L O} holds. Comparing to Eq.~\eqref{Eq: in->out}, we can view Eq.~\eqref{Eq: in->L O} as stating that the output error channel associated with $G_O$ is logical.  

As an example, if $G_O$ is an adaptive Clifford circuit (Example \ref{example CAP}) corresponding to a logical Pauli observable $O$, with a Pauli input error channel, the effect of errors and faults reduces to a probability \(P_{\mathrm{flip}}\) that the output of $G_O$, valued in $\{\pm1\}$, is flipped. In this case
\(O'=(1-2P_{\mathrm{flip}})O\),
so the logical channel $\Lambda_L^O$ reduces to a scalar
\((1-2P_{\mathrm{flip}})\), which can be inverted in post-processing.

The faulty error-corrected expectation value $\mathbb{E}_\psi[O]$ of a logical observable $O$ in the initial logical state $\psi$, evolved under the circuit $C$, is defined as
\begin{align}
\mathbb{E}_\psi[O]:=\sbra{G_O}\!G_D\cdots G_1\sket{G_\psi}.\label{Eq: exp val}
\end{align}

\subsubsection{Syndrome-conditioning \label{Appendix: synd cond}}

In order to discuss syndrome-conditioned quantities, we represent each layer 
 $G_j=\sum_{s_j} G_{j,s_j}$ as a `quantum instrument' $\{G_{j,s_j}\}_{s_j}$, with CP but trace-non-increasing maps $G_{j,s_j}$, corresponding to the different outcomes $s_j$ of the syndrome measurement within $G_j$ \cite{Watrous2018}. Given an input channel $\Lambda_{\mathrm{in}}^{(j)}$, the probability to observe the syndrome measurement result $s_j$ is 
\begin{align}
     p_{s_j}^{(j)}=\sbra{ I}\!G_{j,s_j}\Lambda_{\mathrm{in}}^{(j)}\sket{c}, \label{Eq: prob}
\end{align}
which depends on $\Lambda_{\mathrm{in}}^{(j)}$ but must be independent of the logical state $\sket{c}$.\footnote{Otherwise, the measured syndromes can be used to distinguish between different logical states, and syndrome measurement will generally result in the loss of logical information.} The syndrome-conditioned output state is given by 
$G_{j,s_{j}}\Lambda_{\mathrm{in}}^{(j)}\sket{c}/p_{s_j}^{(j)}$, and can be used to define a syndrome-conditioned output channel $\Lambda_{\mathrm{out}|s_j}^{(j)}:\mathcal{K}\rightarrow \mathcal{H}$ via
\begin{align}
\Lambda_{\mathrm{out}|s_j}^{(j)}g_j\sket{c}=G_{j,s_j}\Lambda_{\mathrm{in}}^{(j)}\sket{c}/p_{s_j}^{(j)},\label{Eq: in->out|s}
\end{align}
such that $\Lambda_{\mathrm{out}}^{(j)}=\sum_{s_j}p_{s_j}^{(j)}\Lambda_{\mathrm{out}|s_j}^{(j)}$. As in the syndrome-averaged case, if $G_j$ is an adaptive Clifford circuit (Example \ref{example CAP}), and $\Lambda_{\mathrm{in}}^{(j)}$ is a Pauli channel, the conditioned output $\Lambda_{\mathrm{out}|s_j}^{(j)}$ is also a Pauli channel. 

For simplicity, we do not condition on the syndromes measured within the state preparation and measurement, $\sket{G_\psi}$ and $\sbra{G_O}$, though this is certainly possible.  

For the entire circuit, the joint probability to observe a sequence of syndromes $\boldsymbol{s}=(s_1,\dots,s_D)$ (a `global syndrome') is 
\begin{align}
    p_{\boldsymbol{s}}=\sbra{I}\!G_{D,s_D}\cdots G_{1,s_1}\sket{G_\psi},\label{Eq: p_s}
\end{align}
and the syndrome-conditioned final state is
\begin{align}
   \sket{\boldsymbol{s}}_\psi=G_{D,s_D}\cdots G_{1,s_1}\sket{G_\psi}/p_{\boldsymbol{s}}.\label{Eq: |s>>}    
\end{align}
The corresponding syndrome-conditioned expectation value of the logical observable $O$ is 
\begin{align}
    \mathbb{E}_\psi[O|\boldsymbol{s}]:=&\sbraket{G_O}{\boldsymbol{s}}_\psi.\label{Eq: s-cond exp val 1}
\end{align}

\subsection{Predictive logical error channels \label{Appendix: predictive logical channels}}

\subsubsection{Syndrome-averaged  logical error channels\label{Sec: logical channels}}

Our goal in this subsection is to define logical error channels $\Lambda_{L}^{(j)}:\mathcal{K}\rightarrow\mathcal{K}$ that are `predictive', providing a purely logical description of the faulty error-corrected expectation value of Eq.~\eqref{Eq: exp val},
\begin{align}
\mathbb{E}_\psi[O]=\sbra{O}\!\Lambda_{L}^{(D)}g_D\cdots \Lambda_{L}^{(1)}g_1\sket{\psi}.\label{Eq: C=prod(Lambda g)}
\end{align}
We additionally require that each $\Lambda_{L}^{(j)}$ captures, in a precise sense clarified below, the logical errors that appear immediately after $G_j$, but not before it. Such logical error channels can then be characterized locally (Appendix \ref{sec:approx logical channels}) and subsequently targeted with LEM protocols, in order to reproduce the ideal logical expectation value 
\begin{align}
    \langle O\rangle:=\sbra{O}\!g_D\cdots g_1\sket{\psi}.
\end{align}

Properly defining the channels  $\Lambda_L^{(j)}$ is subtle, because logical errors after $G_j$ are due to combinations of (i) internal faults within $G_j$, and (ii) certain faults in earlier layers that were not corrected.   We would like to describe the effect of these earlier faults as a correctable input channel $\Lambda_{\mathrm{in}}^{(j)}$. However, as discussed below Eq.~\eqref{Eq: out = joint corr*log}, it is unclear how to separate  correctable errors from past logical errors. We address these challenges by iteratively defining input, output and logical channels for the layers $G_j$:  
\begin{equation}   
\Lambda_{\mathrm{in}}^{(j)} \mapsto \Lambda_{\mathrm{out}}^{(j)} \mapsto  \Lambda_{L}^{(j)}\mapsto \Lambda_{\mathrm{in}}^{(j+1)},
\label{eq:Lambda_full_recursion}
\end{equation}
see Fig.~\ref{fig:faulty_gate_decomposition} for an overview.

\begin{figure}[H]
\centering
\input{fig_decomposition_of_G_j3.tikz.tex}
\caption{Iterative construction of logical error channels using Eq.~\eqref{Eq: in->out}, \eqref{Eq: out->L} and \eqref{Eq: def of next in}. Faulty error-corrected layers (pink) are gradually replaced by desired logical layers and logical error channels (white), until a fully logical description is attained (Eq.~\eqref{Eq: C=prod(Lambda g)}). At intermediate steps, input (or output) errors channels map the logical space to the physical space, connecting the two descriptions (Eq.~\eqref{eq:G_to_g}).  
}
\label{fig:faulty_gate_decomposition}
\end{figure}

Given an input channel $\Lambda_{\mathrm{in}}^{(j)}$, we obtain the output channel $\Lambda_{\mathrm{out}}^{(j)}:\mathcal{K}\rightarrow \mathcal{H}$ via Eq.~\eqref{Eq: in->out}. We then define the logical error channel $\Lambda_{L}^{(j)}:\mathcal{K}\rightarrow\mathcal{K}$ using the ideal version of the \textit{next} layer $G_{j+1,\mathrm{ideal}}:\mathcal{H}\rightarrow \mathcal{K}$, such that
\begin{align}
g_{j+1}\Lambda_L^{(j)}=G_{j+1,\mathrm{ideal}}\Lambda_{\mathrm{out}}^{(j)},\label{Eq: out->L}
\end{align}
\vspace{-20pt}
\begin{equation}
\begin{tikzcd}[column sep=large, row sep=large]
\mathcal{K}
  \arrow{r}{\Lambda_L^{(j)}}
  \arrow{dr}[swap]{\Lambda_{\mathrm{out}}^{(j)}}
&
\mathcal{K}
  \arrow{r}{g_{j+1}}
&
\mathcal{K}
\\
&
\mathcal{H}
  \arrow{ur}[swap]{G_{j+1,\mathrm{ideal}}}
&
\end{tikzcd}\ .\nonumber
\end{equation}
This is a channel version of Eq.~\eqref{Eq: logical error}. Note that  $G_{j+1,\mathrm{ideal}}$ is a fictitious object used here to define logical error channels -- we do not assume that any physical operations in the circuit are fault-free. This usage of $G_{j+1,\mathrm{ideal}}$ allows us to avoid an unrealistic  assumption commonly made in the literature, involving a final round of fault-free stabilizer measurements within $G_j$, see e.g.,  \cite{Chubb2018StatMechCorrelated}. 

Given $\Lambda_{\mathrm{out}}^{(j)}$ and $\Lambda_{L}^{(j)}$, we define the input to the next faulty layer $G_{j+1}$ as\footnote{In any practical setting, syndrome-averaged logical error channels are close to the identity, and therefore invertible.}
\begin{align}
\Lambda_{\mathrm{in}}^{(j+1)}=\Lambda_{\mathrm{out}}^{(j)}(\Lambda_{L}^{(j)})^{-1},
\label{Eq: def of next in}
\end{align} 
\vspace{-20pt}
\begin{equation}
\begin{tikzcd}[column sep=large, row sep=large]
\mathcal{K}
  \arrow{rr}{(\Lambda_L^{(j)})^{-1}}
  \arrow{dr}[swap]{\Lambda_{\mathrm{in}}^{(j+1)}}
&&
\mathcal{K}
  \arrow{dl}{\Lambda_{\mathrm{out}}^{(j)}}
\\
&
\mathcal{H}
&
\end{tikzcd}\ . \nonumber
\end{equation}
such that $\Lambda_{\mathrm{out}}^{(j)}=\Lambda_{\mathrm{in}}^{(j+1)}\Lambda_{L}^{(j)}$. We view this relation as a channel version of the correctable-logical decomposition in Eq.~\eqref{Eq: sigma=sigma_c sigma_l}. Indeed, it follows from Eq.~\eqref{Eq: out->L}-\eqref{Eq: def of next in} that  $\Lambda_{\mathrm{in}}^{(j+1)}$ satisfies the defining property of correctable errors (Eq.~\eqref{Eq: corr def}), 
\begin{align}
    G_{j+1,\mathrm{ideal}}\Lambda_{\mathrm{in}}^{(j+1)}=g_{j+1}, \label{Eq: corr def channel}
\end{align}
and may therefore be viewed as `correctable'.\footnote{However, as discussed below Eq.~\eqref{Eq: out = joint corr*log}, the output channel does not generally admit a decomposition as the product of a correctable channel and a logical channel. Accordingly, 
$\Lambda_{\mathrm{in}}^{(j+1)}$  may include non-correctable  errors, and may not be CP.
In Appendix~\ref{sec:approx logical channels} we show that, at leading order in the physical error rate, $\Lambda_{\mathrm{in}}^{(j+1)}$ can be replaced by a CP map containing only correctable errors. 
This subtlety passes on to our output and logical error channels, which are only CP at leading order. We stress that complete-positivity is important mathematically and conceptually, but practically irrelevant for predicting and mitigating faulty expectation values.\label{Foot: CP?}}  

We use Eq.~\eqref{Eq: prep out} to define the first input channel, such that $\Lambda^{(1)}_{\mathrm{in}}:\mathcal{K}\rightarrow\mathcal{H}$ is any channel satisfying 
\begin{align}
    \Lambda_{\mathrm{in}}^{(1)}\sket{\psi}=\sket{G_\psi}. \label{Eq: initial 0}
\end{align}
We then use Eq.~\eqref{Eq: in->out}, \eqref{Eq: out->L} and \eqref{Eq: def of next in} iteratively, obtaining input, output and logical error channels, such that: 
\begin{align}
    G_{j} \cdots G_1\sket{G_\psi} 
    \label{eq:G_to_g}=& G_{j}\Lambda_{\mathrm{in}}^{(j)}\Lambda^{(j-1)}_Lg_{j-1}\cdots \Lambda_L^{(1)}g_1\sket{\psi}\\
   =&\Lambda_{\mathrm{out}}^{(j)}g_j\Lambda^{(j-1)}_Lg_{j-1}\cdots \Lambda_L^{(1)}g_1\sket{\psi} \nonumber,
\end{align}
for all $j=1,\dots,D$ (see Fig.~\ref{fig:faulty_gate_decomposition}). The final logical channel is obtained using Eq.~\eqref{Eq: in->L O}, such that $\Lambda_L^{(D)}:\mathcal{K}\rightarrow\mathcal{K}$ is any channel satisfying 
\begin{align}
    \sbra{O}\!\Lambda_L^{(D)}=\sbra{G_O}\!\Lambda_{\mathrm{out}}^{(D)},  \label{Eq: out->L last}
\end{align}
and the desired Eq.~\eqref{Eq: C=prod(Lambda g)} holds. Note that our definitions of the first and last logical error channels $\Lambda_L^{(1)},\Lambda_L^{(D)}$ incorporate all faults occurring within state preparation and measurement, $\sket{G_\psi},\sbra{G_O}$, respectively.

\subsubsection{Syndrome-conditioned logical error channels \label{Appendix: syndrome-conditioned exact channels}}

We now generalize the analysis in   Appendix \ref{Sec: logical channels} to the syndrome-conditioned case, resulting in Definition \ref{def channels} and Theorem \ref{thm channels} below.

Since we do not condition on the syndrome measured within state preparation, the first input channel $\Lambda_{\mathrm{in}}^{(1)}$ is unchanged from the syndrome-averaged case (Eq.~\eqref{Eq: initial 0}). The probability to obtain the syndrome measurement result $s_1$ within the first layer $G_1=\sum_{s_1} G_{1,s_1}$ is then
\begin{align}
    p_{s_1}^{(1)} = \sbra{I}\!G_{1,s_{1}}\Lambda_{\mathrm{in}}^{(1)}\sket{c}, \label{Eq: prob|s 1}
\end{align}
independent of the choice of $\sket{c}\in \mathcal{K}$, and the syndrome-conditioned output channel is given by
\begin{align}
\Lambda_{\mathrm{out}|s_1}^{(1)}g_1=G_{1,s_1}\Lambda_{\mathrm{in}}^{(1)}/p_{s_1}^{(1)}.\label{Eq: in->out|s 1}
\end{align}
We also define the `joint output channel' $\Lambda_{\mathrm{out},s_1}^{(1)}=p_{s_1}^{{(1)}}\Lambda_{\mathrm{out}|s_1}^{(1)}$, which corresponds to the joint distribution of output errors and syndromes. Following Eq.~\eqref{Eq: out->L}, the conditioned logical channel is then defined through\footnote{Note that we do not condition on the syndrome $s_{2,\mathrm{ideal}}$ measured in $G_{2,\mathrm{ideal}}$ -- we go back to this point at the end of this section.} 
\begin{align}
g_{2}\Lambda_{L|s_1}^{(1)}=G_{2,\mathrm{ideal}}\Lambda_{\mathrm{out}|s_1}^{(1)},\label{Eq: out->L|s 1}
\end{align}
and we also define the joint logical channel $\Lambda^{(1)}_{L,s_1}=p^{(1)}_{s_1}\Lambda^{(1)}_{L|s_1}$. Clearly, we have
\begin{align}
    \Lambda_{\mathrm{out}}^{(1)}&=\sum_{s_1} p_{s_1}^{(1)}\Lambda_{\mathrm{out}|s_1}^{(1)}=\sum_{s_1} \Lambda_{\mathrm{out},s_1}^{(1)},\label{Eq: standard relations}\\
    \Lambda_L^{(1)}&=\sum_{s_1} p_{s_1}^{(1)}\Lambda_{L|s_1}^{(1)}=\sum_{s_1} \Lambda_{L,s_1}^{(1)}.\nonumber
\end{align}

Following Eq.~\eqref{Eq: def of next in}, we define the syndrome-conditioned input to $G_2$ as\footnote{As opposed to the syndrome-averaged case, syndrome-conditioned logical error channels need not be close to the identity, and may not be invertible. In the context of SALEM, non-invertible error channels lead to vanishing inverse-variance weights (see Sec.~\ref{Sec: coarse-grained SALEM main text}), corresponding to an effective rejection of shots in which the corresponding syndromes are observed. We therefore restrict attention to global syndromes $\boldsymbol{s}=(s_1,\dots,s_D)$ for which all logical error channels are invertible.}  
\begin{align}
\Lambda_{\mathrm{in}|s_1}^{(2)}:=\Lambda_{\mathrm{out}|s_1}^{(1)}(\Lambda_{L|s_1}^{(1)})^{-1}=\Lambda_{\mathrm{out},s_1}^{(1)}(\Lambda_{L,s_1}^{(1)})^{-1}. \label{Eq: L->next in|s 1}
\end{align}
As opposed to Eq.~\eqref{Eq: standard relations}, the naive relation $\Lambda_{\mathrm{in}}^{(2)}=\sum_{s_1} p_{s_1}^{(1)}\Lambda_{\mathrm{in}|s_1}^{(2)}$ does not hold. Instead, we have 
\begin{align}
   \Lambda_{\mathrm{in}}^{(2)}\Lambda_{L}^{(1)} =\Lambda_{\mathrm{out}}^{(1)}=\sum_{s_1} \Lambda_{\mathrm{out},s_1}^{(1)}=\sum_{s_1} \Lambda_{\mathrm{in}|s_1}^{(2)}\Lambda_{L,s_1}^{(1)},
\end{align}
where the first equality follows from Eq.~\eqref{Eq: def of next in}, and the last from Eq.~\eqref{Eq: L->next in|s 1}. Thus,
\begin{align}
\Lambda_{\mathrm{in}}^{(2)}
=
\left(
\sum_{s_1}\Lambda_{\mathrm{in}|s_1}^{(2)}\Lambda_{L,s_1}^{(1)}
\right)
\left(
\sum_{s_1}\Lambda_{L,s_1}^{(1)}
\right)^{-1},
\label{Eq: generalized conditioning}
\end{align}
which reduces to the naive relation if each joint channel $\Lambda_{L,s_1}^{(1)}$ is replaced by its total probability $p_{s_1}^{(1)}=\sbra{I}\!\Lambda_{L,s_1}^{(1)}\sket{c}$. Equation~\eqref{Eq: generalized conditioning} is therefore a generalized form of conditioning, where $\Lambda_{\mathrm{in}|s_1}^{(2)}$ is not just conditioned on the past syndrome $s_1$ (as our simple notation suggests), but on the joint distribution of past logical errors with that syndrome. To provide context for Eq.~\eqref{Eq: generalized conditioning}, note that $\{\Lambda^{(1)}_{L,s_1}\}_{s_1}$ is itself a quantum instrument, and can therefore be understood as an `operation-valued measure', generalizing the classical probability measure $p_{s_1}^{(1)}$ \cite{Gudder2023QuantumInstruments}. 

Given $\Lambda_{\mathrm{in}|s_1}^{(2)}$, the application of $G_2=\sum_{s_2}G_{2,s_2}$ defines the next conditioned output channel 
\begin{align}
    \Lambda_{\mathrm{out}|s_1,s_2}^{(2)}g_2=G_{2,s_2}\Lambda_{\mathrm{in}|s_1}^{(2)}/p_{s_2|s_1}^{(2)} \label{Eq: in|s1->out|s12},
\end{align}
occurring with probability 
\begin{align}
    p_{s_2|s_1}^{(2)}=\sbra{I}\!G_{2,s_2}\Lambda_{\mathrm{in}|s_1}^{(2)}\sket{c}, \label{Eq: p_s2|s1}
\end{align}
 which depends on the input channel  $\Lambda_{\mathrm{in}|s_1}^{(2)}$, but not on the logical state $\sket{c}$. It follows from Eq.~\eqref{Eq: generalized conditioning}-\eqref{Eq: in|s1->out|s12} that $\Lambda_{\mathrm{out}|s_1,s_2}^{(2)}$ is related to its syndrome-averaged counterpart via a naive averaging over $s_2$ with $p_{s_2|s_1}^{(2)}$, while  averaging over $s_1$ again involves the `operation-valued measure' $\Lambda_{L,s_1}^{(1)}$ as opposed to the naive $p_{s_1}^{(1)}$,
 \begin{align}
   \Lambda_{\mathrm{out}}^{(2)}g_2=\left(\sum_{s_1,s_2}p_{s_2|s_1}^{(2)}\Lambda_{\mathrm{out}|s_1,s_2}^{(2)}g_2 \Lambda_{L,s_1}^{(1)}\right)\left(\sum_{s_1}\Lambda_{L,s_1}^{(1)}\right)^{-1}.\label{Eq: generalized conditioning L 3}
\end{align}
 
The fault-free version $G_{3,\mathrm{ideal}}$ defines the next logical error channel,
 \begin{align}
  g_3\Lambda_{L|s_1,s_2}^{(2)} =G_{3,\mathrm{ideal}}\Lambda_{\mathrm{out}|s_1,s_2}^{(2)},  
 \end{align}
which satisfies an averaging relation as in Eq.~\eqref{Eq: generalized conditioning L 3}, 
\begin{align}
   \Lambda_{L}^{(2)}g_2=\left(\sum_{s_1,s_2}p_{s_2|s_1}^{(2)}\Lambda_{L|s_1,s_2}^{(2)}g_2 \Lambda_{L,s_1}^{(1)}\right)\left(\sum_{s_1}\Lambda_{L,s_1}^{(1)}\right)^{-1}.\label{Eq: generalized conditioning L 2}
\end{align}
The latter is equivalent to a simple averaging relation of `logical prefix channels',
\begin{align}
    \Lambda_{L}^{(2)}g_2\Lambda_{L}^{(1)}g_1=\sum_{s_1,s_2}p_{s_2|s_1}^{(2)}p_{s_1}^{(1)}\Lambda_{L|s_1,s_2}^{(2)}g_2 \Lambda_{L|s_1}^{(1)}g_1. \label{Eq: averaging}
\end{align}

The syndrome-conditioned quantities we defined reproduce the correct joint distribution of syndromes $p_{s_1,s_2}$, and the correct syndrome-conditioned state $\sket{s_1,s_2}_c$: 
\begin{align}
    p_{s_1,s_2}:=&\sbra{I}\!G_{2,s_2}G_{1,s_1}\sket{G_\psi}\\
    =&\sbra{I}\!G_{2,s_2}G_{1,s_1}\Lambda_{\mathrm{in}}^{(1)}\sket{\psi}\nonumber\\
    =&\sbra{I}\!G_{2,s_2}\Lambda_{\mathrm{out}|s_1}^{(1)}g_1\sket{\psi}p_{s_1}^{(1)}\nonumber\\
    =&\sbra{I}\!G_{2,s_2}\Lambda_{\mathrm{in}|s_1}^{(2)}\left(\Lambda_{L|s_1}^{(1)}g_1\sket{\psi}\right)p_{s_1}^{(1)}\nonumber\\
    =&p_{s_2|s_1}^{(2)}p_{s_1}^{(1)},\nonumber
\end{align} 
where we used Eq.~\eqref{Eq: initial 0}, \eqref{Eq: in->out|s 1}, \eqref{Eq: L->next in|s 1} and \eqref{Eq: p_s2|s1} with $\sket{c}=\Lambda_{L|s_1}^{(1)}g_1\sket{\psi}$. Similarly, 
\begin{align}
    \sket{s_1,s_2}_c:=&G_{2,s_2}G_{1,s_1}\sket{G_\psi}/p_{s_1,s_2}\\
    =&G_{2,s_2}G_{1,s_1}\Lambda_{\mathrm{in}}^{(1)}\sket{\psi}/p_{s_2|s_1}^{(2)}p_{s_1}^{(1)}\nonumber\\
    =&G_{2,s_2}\Lambda_{\mathrm{out}|s_1}^{(1)}g_1\sket{\psi}/p_{s_2|s_1}^{(2)}\nonumber\\
    =&G_{2,s_2}\Lambda_{\mathrm{in}|s_1}^{(2)}\left(\Lambda_{L|s_1}^{(1)}g_1\sket{\psi}\right)/p_{s_2|s_1}^{(2)}\nonumber\\
    =&\Lambda_{\mathrm{out}|s_1,s_2}^{(1)}g_2\Lambda_{L|s_1}^{(1)}g_1\sket{\psi}.\nonumber
\end{align}

Continuing in this manner, we obtain logical error channels 
$\Lambda^{(j)}_{L|s_1,\dots,s_j}=\Lambda^{(j)}_{L|s_{1:j}}$, and similarly $\Lambda^{(j)}_{\mathrm{out}|s_{1:j}}$ and $\Lambda^{(j)}_{\mathrm{in}|s_{1:j-1}}$, occurring with probabilities $p_{s_j|s_{1:j-1}}^{(j)}$: 
\begin{definition}\label{def channels}{\bf (Syndrome-conditioned logical channels)}
    Let  $C=\sum_{s_D}G_{D,s_D}\cdots \sum_{s_1}G_{1,s_1}$ be a faulty error-corrected circuit, written in terms of quantum instruments $\{G_{1,s_1}\}_{s_1},\dots, \{G_{D,s_D}\}_{s_D}$, with faulty error-corrected state preparation $\sket{G_\psi}$ and observable measurement $\sbra{G_O}$. We iteratively define conditioned syndrome probabilities $p_{s_j|s_{1:j-1}}^{(j)}$, and corresponding input, output and logical  error channels 
    \begin{align}
        \Lambda_{{\mathrm{in}}|s_{1:j}}^{(j+1)},\Lambda_{\mathrm{out}|s_{1:j}}^{(j)}&:\mathcal{K}\rightarrow\mathcal{H},\\ \Lambda_{L|s_{1:j}}^{(j)}&:\mathcal{K}\rightarrow\mathcal{K}, \nonumber
    \end{align}
by 
\begin{align}
    p_{s_j|s_{1:j-1}}^{(j)}&=\sbra{I}\!G_{j,s_j}\Lambda_{\mathrm{in}|s_{1:j-1}}^{(j)}\sket{c},\\
    \Lambda_{\mathrm{out}|s_{1:j}}^{(j)}g_j&=G_{_j,s_j}\Lambda_{\mathrm{in}|s_{1:j-1}}^{(j)}/p_{s_j|s_{1:j-1}}^{(j)},
\end{align}
for $j=1,\dots,D$, where $\sket{c}\in \mathcal{K}$ is an arbitrary code state, and by
\begin{align}
    g_{j+1}\Lambda_{L|s_{1:j}}^{(j)} &=G_{j+1,\mathrm{ideal}}\Lambda_{\mathrm{out}|s_{1:j}}^{(j)},\\
    \Lambda_{\mathrm{in}|s_{1:j}}^{(j+1)}&=\Lambda_{\mathrm{out}|s_{1:j}}^{(j)}(\Lambda_{L|s_{1:j}}^{(j)})^{-1},\label{Eq: in deff}
\end{align}
for $j=1,\dots,D-1$. The first input channel is any channel $\Lambda_{\mathrm{in}}^{(1)}:\mathcal{K}\rightarrow\mathcal{H}$  satisfying 
\begin{align}
    \Lambda_{\mathrm{in}}^{(1)}\sket{\psi}=\sket{G_\psi}, \label{Eq: initial}
\end{align}
while the last logical channel is any channel  $\Lambda_{L|s_{1:D}}^{(D)}:\mathcal{K}\rightarrow\mathcal{K}$ satisfying
\begin{align}
    \sbra{O}\!\Lambda_{L|s_{1:D}}^{(D)} &=\sbra{G_O}\!\Lambda_{\mathrm{out}|s_{1:D}}^{(D)}.\label{Eq: final}
\end{align} 
\end{definition}

It follows from the above discussion that the resulting syndrome-conditioned probabilities and logical channels give a purely logical description for syndrome-conditioned expectation values in the faulty error-corrected circuit: 
\begin{theorem}\label{thm channels}{\bf (Syndrome-conditioned logical error channels are predictive)} Let $C=\sum_{s_D}G_{D,s_D}\cdots \sum_{s_1}G_{1,s_1}$ be a faulty error-corrected circuit, with joint syndrome distribution 
\begin{align}
    p_{\boldsymbol{s}}:=\sbra{I}\!G_{D,s_D}\cdots G_{1,s_1}\sket{G_\psi},
\end{align}
syndrome-conditioned final state 
\begin{align}
   \sket{\boldsymbol{s}}_\psi:=G_{D,s_D}\cdots G_{1,s_1}\sket{G_\psi}/p_{\boldsymbol{s}}, \label{Eq: s-cond state}   
\end{align}
and syndrome-conditioned expectation value \begin{align}
    \mathbb{E}_\psi[O|\boldsymbol{s}]:=\sbraket{G_O}{\boldsymbol{s}}_\psi,\label{Eq: s-cond exp val}
\end{align}
for a logical observable $O$. Then, the probabilities  $p_{s_j|s_{1:j-1}}^{(j)}$ given by Def~\ref{def channels} accurately reproduce the joint distribution of  syndromes,
\begin{align}
    p_{\boldsymbol{s}}=p_{s_D|s_{1:D-1}}^{(D)}\cdots p_{s_2|s_1}^{(2)}p_{s_1}^{(1)},\label{Eq: probs}
\end{align}
and the logical channels $\Lambda_{L|s_{1:j}}^{(j)}$ given by Def~\ref{def channels} accurately predict the faulty syndrome-conditioned expectation value, 
\begin{align}
\mathbb{E}_\psi[O|\boldsymbol{s}]=\sbra{O}\Lambda_{L|s_{1:D}}^{(D)}g_D\cdots \Lambda_{L|s_{1:2}}^{(2)}g_2\Lambda_{L|s_1}^{(1)}g_1\sket{\psi}.\label{Eq: C=prod(Lambda g)|s}
\end{align}
\end{theorem}

Note that our construction of $\Lambda_{L|s_{1:j}}^{(j)}$ in Def~\ref{def channels} involves the fault-free layers $G_{i,\mathrm{ideal}}$ with $i\leq j+1$, which produce corresponding syndromes $s_{i,\mathrm{ideal}}$ that one may condition on. As described above,  these ideal layers are fictitious objects, introduced in order to define logical error channels. They differ from the faulty layers $G_{i}$  that actually appear in the circuit. Accordingly, conditioning on $s_{i,\mathrm{ideal}}$ does not uniquely determine the value of $s_{i}$. Different treatments of $s_{i,\mathrm{ideal}}$, such as averaging, conditioning on $0$, or conditioning on a subset, simply correspond to different valid conventions, that differ by passing certain logical errors between consecutive channels $\Lambda_{L|s_{1:i}}^{(i)},\Lambda_{L|s_{1:i+1}}^{(i+1)}$. Different conventions lead to equivalent sets of logical channels, satisfying Theorem \ref{thm channels}. 

\vspace{5pt}

The following corollary is an immediate extension of our analysis to logical channels conditioned on subsets of syndromes, and captures the relation between syndrome- and subset-conditioned logical error channels (see Eq.~\eqref{Eq: averaging}). 
\begin{corollary}\label{corr subset channels}{\bf (Subset-conditioned logical channels)} 
Given a partition $S^{(j)}=\cup_{k=1}^{K_j} S_{k}^{(j)}$ of the syndromes in each layer, and a global index $\boldsymbol{k}=(k_1,\dots,k_D)$ with $k_j=1,\dots,K_j$, define subset-conditioned probabilities $p^{(j)}_{k_j|k_{1:j-1}}$ and  channels $\Lambda_{\mathrm{out}|k_{1:j}}^{(j)}, \Lambda_{L|k_{1:j}}^{(j)}$ and $\Lambda_{\mathrm{in}|k_{1:j}}^{(j+1)}$ in analogy with Def~\ref{def channels}, using $G_{j,k_j}=\sum_{s_j\in S_{k_j}^{(j)}}G_{j,s_j}$ in place of $G_{j,s_j}$. These channels predict subset-conditioned expectation values, 
\begin{align}
\mathbb{E}_\psi[O|\boldsymbol{k}]=\sbra{O}\Lambda_{L|k_{1:D}}^{(D)}g_D\cdots \Lambda_{L|k_{1:2}}^{(2)}g_2\Lambda_{L|k_1}^{(1)}g_1\sket{\psi},\label{Eq: C=prod(Lambda g)|k}
\end{align}
and are related to the syndrome-conditioned logical error channels via 
\begin{align}
    &(p_{k_j|k_{1:j-1}}^{(j)}\cdots p_{k_1}^{(1)})\Lambda_{L|k_{1:j}}^{(j)}g_j\cdots\Lambda_{L|k_1}^{(1)}g_1  \label{Eq: syndrome averaging of channels}\\
    =&\sum_{s_1\in S^{(1)}_{k_1},\dots,s_j\in S^{(j)}_{k_j}}(p_{s_j|s_{1:j-1}}^{(j)}\cdots p_{s_1}^{(1)})\Lambda_{L|s_{1:j}}^{(j)}g_j\cdots  \Lambda_{L|s_1}^{(1)}g_1,\nonumber
\end{align}
for all $j=1,\dots,D$. In particular, 
\begin{align}
    &\Lambda_{L}^{(j)}g_j\cdots\Lambda_{L}^{(1)}g_1  \label{Eq: syndrome averaging of channels 2}\\
    =&\sum_{s_{1:j}}(p_{s_j|s_{1:j-1}}^{(j)}\cdots p_{s_1}^{(1)})\Lambda_{L|s_{1:j}}^{(j)}g_j\cdots  \Lambda_{L|s_1}^{(1)}g_1,\nonumber
\end{align}
\end{corollary}

When discussing the generalities of SALEM in Appendix \ref{Appendix: optimal SALEM} and in the main text, we sometimes find it convenient to treat all logical error channels in the circuit as a single object: 
\begin{definition} {\bf (Global error channel)}
    We refer to the list of conditioned logical error channels associated with all layers $j=1,\dots,D$ as a `global syndrome-conditioned logical error channel' 
    \begin{align}
        \boldsymbol{\Lambda}_{L|\boldsymbol{s}}=(\Lambda^{(j)}_{L|s_{1:j}})_{j=1}^D,
    \end{align}
    with entry-wise inverse $\boldsymbol{\Lambda}_{L|\boldsymbol{s}}^{-1}=((\Lambda^{(j)}_{L|s_{1:j}})^{-1})_{j}$.  
    Given a partition $S^{(j)}=\cup_k S_{k}^{(j)}$ of the syndromes in each layer, and a global index $\boldsymbol{k}=(k^{(j)})_{j}$, the global subset-conditioned logical error channel    $\boldsymbol{\Lambda}_{L|\boldsymbol{k}}$ is defined in the same manner.
\end{definition}

\subsubsection{Locality and  characterization}\label{sec:approx logical channels}

The logical error channels $\Lambda_{L}^{(j)}$ constructed above are exactly predictive, satisfying Eq.~\eqref{Eq: C=prod(Lambda g)} (or Eq.~\eqref{Eq: C=prod(Lambda g)|s} or \eqref{Eq: C=prod(Lambda g)|k}, in the syndrome- or subset-conditioned cases). However, they are non-local in time, in the sense that $\Lambda_{L}^{(j)}$ depends on all layers $G_i$ with $i\leq j$. Moreover, the input channels we defined can involve non-correctable errors, and may not be CP (Footnote \ref{Foot: CP?}). The latter implies that the resulting logical channels may not be CP. 

In this section we show that in a standard fault-tolerance setting, and at a leading order in the physical error rate, the exact logical channel $\Lambda_L^{(j)}$ can be approximated by a CP channel $\tilde{\Lambda}_L^{(j)}$ that depends only on the previous layer $G_{j-1}$, the current layer $G_j$, and the ideal version of the next layer,  
$G_{j+1,\mathrm{ideal}}$. The construction involves the approximation of $\Lambda_{\mathrm{in}}^{(j)}$ by a CP channel that involves only correctable errors. This forms the basis for a logical characterization protocol that estimates $\tilde{\Lambda}_L^{(j)}$ efficiently in both quantum and classical runtime, discussed below.

\vspace{5pt}

In order to define fault-tolerance in a broad but convenient setting, we restrict attention to circuit-level noise models given by probabilistic distributions of faults.
\begin{definition} {\bf (Probabilistic faults, fault-paths, error-fault combinations)}
    We say that the layer $G_j$ has probabilistic faults with rate $\epsilon$, if each fault channel within it is given by a convex combination of CPTP maps (`faults'), where the identity map (`no fault') has coefficient $\geq1-\epsilon$. Labeling all possible subsets of faults (`fault-sets'), including at most one fault from each fault channel in $G_j$, by $f_j$, we then have the probabilistic decomposition
    \begin{align}
        G_j=\sum_{f_j} p_{f_j} G_{j}^{(f_j)},
    \end{align}
where $G_j^{(f_j)}:\mathcal{H}\rightarrow\mathcal{H}$ are CPTP maps, occurring with probabilities $p_{f_j}$. The 
empty fault-set $f_j=\emptyset$ satisfies 
\begin{align}
    G_j^{(\emptyset)}=G_{j,\mathrm{ideal}}.
\end{align}
The weight $w_{f_j}=|f_j|$ of a fault-set $f_j$ is the number of faults within it, such that $p_{f_j}\leq\epsilon^{w_{f_j}}$. We write bounds such as  
\begin{align}
    \sum_{f_j:\, w_{f_j}=m} p_{f_j}=O(\epsilon^m),
\end{align}
which explicitly keep track of the $\epsilon$ dependence of the total probability of a given class of fault-sets, while ignoring combinatorial factors corresponding to the number of fault-sets in the class. Such probability bounds can be translated to upper bounds on standard distance measures for corresponding super-operators, such as (entanglement) infidelity, Frobenius distance or diamond distance.  

When all layers in the circuit, including $G_0=\sket{G_\psi}$ and $G_{D+1}=\sbra{G_O}$, have probabilistic faults, we can assign a probability $p_{\boldsymbol{f}}=\prod_{j=0}^{D+1}p_{f_j}$ to each `fault-path' $\boldsymbol{f}=(f_j)_{j=0}^{D+1}$. 

An `error-fault combination' (EFC) is a pair $(\mathsf{e},f_j)$, including a CPTP input error $\mathsf{e}:\mathcal{K}\rightarrow\mathcal{H}$ and a fault-set $f_j$ in $G_j$. Each EFC produces a CPTP output error $\mathsf{e}':\mathcal{K}\rightarrow\mathcal{H}$, given by $\mathsf{e}'g_j=G_j^{(f_j)}\mathsf{e}$. Note that a probabilistic input error channel $\Lambda_{\mathrm{in}}=\sum_{\mathsf{e}}p_{\mathsf{e}}\mathsf{e}$ leads to a probabilistic output channel $\Lambda_{\mathrm{out}}=\sum_{\mathsf{e}'}p_{\mathsf{e}'}'\mathsf{e}'$, with 
\begin{align}
    p_{\mathsf{e}'}'=\sum_{(\mathsf{e},f_j):\ \mathsf{e}'g_j=G_j^{(f_j)}\mathsf{e}}p_{f_{j}}p_{\mathsf{e}}.    
\end{align}
An error $\mathsf{e}$ before layer $G_{j+1}$ has a corresponding CPTP logical error $\mathsf{e}_L:\mathcal{K}\rightarrow\mathcal{K}$, defined by $g_{j+1}\mathsf{e}_L =G_{j+1,\mathrm{ideal}} \mathsf{e}$. The error $\mathsf{e}$ is   correctable if $\mathsf{e}_L=I$. For a Pauli error $\mathsf{e}=\sigma|_{\mathcal{K}}=\sigma_{\mathrm{corr}}^a\sigma_{\mathrm{log}}^b|_{\mathcal{K}}$, we reproduce $\mathsf{e}_L=\sigma_{\mathrm{log}}^b|_{\mathcal{K}}$. 

The weight of an error $\mathsf{e}$ is defined as the number of physical data qubits on which an extension $\hat{\mathsf{e}}:\mathcal{H}\rightarrow\mathcal{H}$,  $\hat{\mathsf{e}}|_{\mathcal{K}}=\mathsf{e}$ acts non-trivially, minimized over such extensions. For a Pauli error $\mathsf{e}=\sigma|_{\mathcal{K}}$, this reduces to the minimal support over the corresponding stabilizer coset, $w_\sigma=\min_{\sigma'\in\sigma\mathcal{S}}|\sigma'|$. 
\end{definition}

With the above definitions, we assume that each layer $G_j$ is fault-tolerant in the following standard sense: 
\begin{definition}\label{def1}{\bf ($t$-fault-tolerance)}
    An error-corrected logical layer  is $t$-fault-tolerant ($t$-FT) if any combination of a weight $w_{\mathrm{in}}$ input error and $w_{f}$ internal faults, with $w_{\mathrm{in}}+w_{f}\leq t$, leads to a weight $w_{\mathrm{out}}\leq w_f$ output error \cite{Aharonov_1997,Gottesman2009QuantumErrorCorrection}. Similarly, an error-corrected state preparation is $t$-FT if $w_f\leq t$   implies $w_{\mathrm{out}}\leq w_f$. An error-corrected final measurement  is $t$-FT if the distribution of measurement outcomes is correct for $w_{\mathrm{in}}+w_{f}\leq t$.
\end{definition}

\begin{figure}[H]
\centering
\input{fig_proof_prop_1.tikz.tex}
\caption{Intuition for Lemma \ref{lemma1} and Proposition \ref{thm0} with $t=2$. Each panel shows three consecutive gates, with faults marked by $\times$'s. (a) The $2$-FT property implies at most one output error after $G_{j-1}$, so the additional fault in $G_j$ is not enough to cause a logical error. (b) $G_j$ has at most one input error  and two faults, so a logical error after $G_j$ can occur. However this is a sub-leading weight-4 fault-path. (c)  A leading order logical error may occur  after $G_{j-1}$, but is included in the approximate logical channel of $G_{j-1}$, not of $G_j$. (d) A potential leading order logical error after $G_j$.  
\label{fig:fault_accumulation}}
\end{figure}
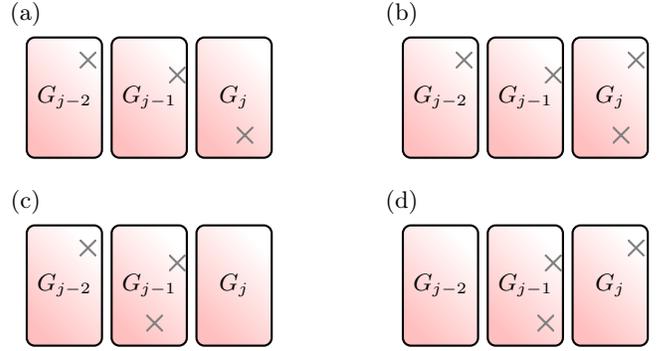

\begin{lemma}\label{lemma1}{\bf (Implications of $t$-fault-tolerance)} 
$t$-FT implies:  (i) Input errors with weight $\leq t$ are corrected in the absence of internal faults. (ii) The output error due to any combination of an input error and faults with $w_{\mathrm{in}}+w_{f}\leq t$ is correctable. (iii) If the number of faults in all pairs of consecutive logical layers $G_jG_{j-1}$ is at most $t$, there is no logical error in the final logical measurement \cite{preskillaliferis}.
(iv) Logical errors occur with probability $O(\epsilon^{t+1})$, due to fault-paths  involving at least $t+1$ faults in two consecutive layers $G_jG_{j-1}$. 
\end{lemma} 
\begin{proof}
    Setting $w_{f}=0$ gives (i). Noting that $w_{\mathrm{in}}+w_f\leq t$ and $w_{\mathrm{out}}\leq w_f$ imply $w_{\mathrm{out}}\leq t$, and using (i) for the next layer, where $w_{\mathrm{in}}^{(\mathrm{next})}=w_{\mathrm{out}}\leq t$ and $w_f^{(\mathrm{next})}=0$, implies (ii). To prove (iii), consider a fault-path starting at layer $G_i$ and ending at $G_{i+m}$, such that $w_{f}^{(j-1)}+w_{f}^{(j)}\leq t$ for all $j=i+1,\dots,i+m$ (with the convention $G_0=\sket{G_\psi}$ for fault paths involving faults within the logical state preparation). For $G_i$ we have $w_{\mathrm{in}}^{(i)}=0$ and $w_f^{(i)}\leq t$, so $w_{\mathrm{out}}^{(i)}\leq w_f^{(i)}$. Setting $w_{\mathrm{in}}^{(i+1)}=w_{\mathrm{out}}^{(i)}$, and using $w_{f}^{(i)}+w_{f}^{(i+1)}\leq t$ then gives $w_{\mathrm{out}}^{(i+1)}\leq w_{f}^{(i+1)}$. Proceeding in this manner shows that  $w_{\mathrm{out}}^{(j)}\leq w_{f}^{(j)}$ for all $j=i,\dots,i+m$. Using (i) then  shows that the final output error of the path, with weight $w_{\mathrm{in}}^{(i+m+1)}=w_{\mathrm{out}}^{(i+m)}\leq w_{f}^{(i+m)}\leq t$,  will be corrected by $G_{i+m+1}$, since $w_{f}^{(i+m+1)}=0$. If the path ends within the final logical measurement $\sbra{G_O}$, the fact that $w_{\mathrm{out}}^{(D)}\leq t$ implies that the classical measurement output is correct. Finally, statement (iv) is just  the contrapositive of (iii).
\end{proof}

\vspace{5pt}

Intuition for Lemma \ref{lemma1} is given in Fig.~\ref{fig:fault_accumulation}, which depicts a number of fault-paths that can and cannot generate logical errors at leading order $\epsilon^{t+1}$.

\begin{figure}[h]
\centering
\input{fig_logical_channel_construction.tikz.tex}
\caption{Computation of $\tilde{\Lambda}_L^{(j)}$ in Algorithm \ref{Algo: P2LC}. The channel $\Lambda_{\mathrm{in}}^{(j-1)}$ is replaced with $\tilde{\Lambda}_{\mathrm{in},0}^{(j-1)}=I$ and serves as input for $G_{j-1}$, to obtain $\tilde{\Lambda}_{\mathrm{out},0}^{(j-1)}$ using Eq.~\eqref{Eq: in->out}. Non-correctable errors in $\tilde{\Lambda}_{\mathrm{out},0}^{(j-1)}$ are replaced by $I$ to obtain the correctable channel $\tilde{\Lambda}_c^{(j-1)}$. Using Eq.~\eqref{Eq: in->out} again, $\tilde{\Lambda}_c^{(j-1)}$ serves as the input channel to $G_{j}$, resulting in $\tilde{\Lambda}_{\mathrm{out}}^{(j)}$. Finally, $G_{j+1,\mathrm{ideal}}$ is used to obtain  ${\tilde{\Lambda}_L^{(j)}}$  via Eq.~\eqref{Eq: out->L}.}
\label{fig:estimating_logical_channel_tilde}
\end{figure}
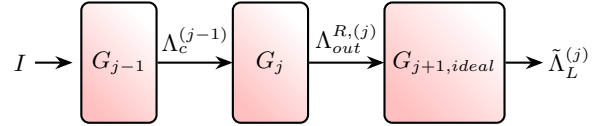

Motivated by Lemma \ref{lemma1}, we define an approximate logical channel $\tilde{\Lambda}_L^{(j)}$ which is temporally-local, see Algorithm~\ref{Algo: P2LC} and Fig.~\ref{fig:estimating_logical_channel_tilde}. We refer to this procedure as `physical-to-logical characterization' (P2LC), since logical error channels are obtained from (characterized) physical error channels through classical simulation, as opposed to a direct `external' logical characterization (ExtLC) \cite{combes2017, Suzuki2022, Piveteau2021}. P2LC has a number of advantages over ExtLC:

{\bf Context:}  ExtLC utilizes logical characterization circuits, which are typically periodic and often easy to classically simulate \cite{combes2017, Suzuki2022, Piveteau2021}. These differ significantly from the logical circuits used in quantum algorithms, and in which the logical layers (or gates) of interest appear. Thus, ExtLC characterizes logical layers in an incorrect context. In contrast, P2LC explicitly accounts for the correct context, by computing input error channels based on past algorithmic layers, and defining logical channels based on future algorithmic layers. Appendix \ref{Sec: color code simulations} demonstrates numerically that an incorrect treatment of input errors can lead to significant prediction biases. 

{\bf Runtime:} The QPU time needed for external characterization scales as $\epsilon_L^{-1}$, (typically due to a depth $\sim \epsilon_L^{-1}$ of logical characterization circuits \cite{combes2017, Suzuki2022, Piveteau2021}), which is exponential in the FT level $t$. In contrast, the QPU time for the physical characterization needed for P2LC scales as $\epsilon^{-1}$, with an additional polynomial dependence on $t$ \cite{qedma_logical_errors_2024}. Moreover, the number of fault-paths that need to be sampled in order to perform the classical simulations within P2LC, which naively scales as $\epsilon_L^{-1}$, can be made essentially independent of $\epsilon_L^{-1}$ (and hence of $t$) using an importance sampling of fault-paths with weight $\geq t+1$ \cite{qedma_logical_errors_2024}. 

{\bf Subset-conditioning:} Finally, as opposed to external characterization,\footnote{The approach of Ref.~\cite{girling2025} may allow for subset-conditioned external characterization.} P2LC can characterize subset-conditioned logical channels, as discussed below. Indeed, P2LC is used in the simulations described in Appendix \ref{Sec: color code simulations}, confirming its ability to accurately predict subset-conditioned faulty expectation values, and serve as a basis for (approximately) unbiased CG-SALEM. 

\vspace{5pt}

Within Algorithm \ref{Algo: P2LC}, we use the following additive decomposition of output error channels into correctable and non-correctable channels. While different decompositions may be used, this particular choice is useful since it leads to approximate channels $\tilde{\Lambda}_L^{(j)}$ that are directly related to the fault-paths appearing in Lemma \ref{lemma1}. We use this relation in the proof of Proposition \ref{thm0} below. 
\begin{definition} {\bf (Additive correctable and non-correctable parts)} \label{def: out=c+nc}
Given a probabilistic output error channel $\Lambda_{\mathrm{out}}=\sum_{\mathsf{e}}p_{\mathsf{e}}\mathsf{e}$, we define its additive correctable and non-correctable parts by
\begin{align}
    \Lambda_{c}=&I+\sum_{\mathsf{e}:\ \mathsf{e}_L=I} p_{\mathsf{e}}(\mathsf{e} -I),\\
    \Lambda_{nc}=&I+\sum_{\mathsf{e}:\ \mathsf{e}_L\neq I} p_{\mathsf{e}}(\mathsf{e} -I),\label{Eq: nc}
\end{align}
such that 
\begin{align}
   \Lambda_{\mathrm{out}}-I=(\Lambda_{c}-I)+(\Lambda_{nc}-I). \label{Eq: out=c+nc}
\end{align}
The correctable part $\Lambda_c$ (non-correctable part $\Lambda_{nc}$) can be viewed as obtained from $\Lambda$ by replacing non-correctable (correctable) errors by the identity. Both parts are CPTP, since they are convex combinations of CPTP errors.  The non-correctable part $\Lambda_{nc}$ is closely related to the logical error channel, 
\begin{align}
    \Lambda_L=\sum_{\mathsf{e}}p_{\mathsf{e}}\mathsf{e}_L=I+\sum_{\mathsf{e}:\ \mathsf{e}_L\neq I} p_{\mathsf{e}}(\mathsf{e}_L -I). \label{Eq: L~nc}
\end{align}

For a Pauli output channel $\Lambda_{\mathrm{out}}=\sum_{a,b}p_{a,b}\sigma_{\mathrm{corr}}^{a} \sigma_{\mathrm{log}}^{b}$ (Eq.~\eqref{Eq: out = joint corr*log}), we have the more explicit expressions,  
\begin{align}
    \Lambda_{c}=&I+\sum_a p_{a,1}(\sigma_{\mathrm{corr}}^{a} -I),\\
    \Lambda_{nc}=&I+\sum_a\sum_{b\neq 1}p_{a,b}(\sigma_{\mathrm{corr}}^{a} \sigma_{\mathrm{log}}^{b}-I),\nonumber\\
    \Lambda_L=&\sum_bp_b\sigma_{\mathrm{log}}^b
    =I+\sum_a\sum_{b\neq 1}p_{a,b}( \sigma_{\mathrm{log}}^{b}-I), \nonumber
\end{align}
with the conventions $\sigma_{\mathrm{log}}^{1}=I=\sigma_{\mathrm{corr}}^{1}$, and where $p_b=\sum_a p_{a,b}$ is the logical marginal.
\end{definition}

\begin{algorithm}[H]
  \caption{Physical-to-logical characterization (P2LC, channel version)}
  \label{Algo: P2LC}

  \begin{algorithmic}[1]
    \Require A $t$-FT error-corrected circuit $C=G_D\cdots G_1$; a physical
    error model for faults occurring in each error-corrected logical layer $G_j$.
    \Ensure An approximate logical error channel $\tilde{\Lambda}_{L}^{(j)}$
    for each layer $G_j$.

    \State Set $\tilde{\Lambda}_{L}^{(1)}=\Lambda_{L}^{(1)}$.
    \Comment{No approximations for the first layer.}

    \ForAll{$j \in \{2,\dots,D\}$ \textbf{in parallel}}
      \State Set $\tilde{\Lambda}_{\mathrm{in},0}^{(j-1)}
      = I$.
      \Comment{First approximation}

      \State Obtain $\tilde{\Lambda}_{\mathrm{out},0}^{(j-1)}$
      from
      $\tilde{\Lambda}_{\mathrm{out},0}^{(j-1)} g_j
      = G_j \tilde{\Lambda}_{\mathrm{in},0}^{(j-1)}$. 
      
      \Comment{Eq.~\eqref{Eq: in->out}}

      \State Extract $\tilde{\Lambda}_{c}^{(j-1)}$, the additive correctable
      part of $\tilde{\Lambda}_{\mathrm{out},0}^{(j-1)}$.
      
      \Comment{Def.~\ref{def: out=c+nc}}

      \State Set $\tilde{\Lambda}_{\mathrm{in}}^{(j)}
      = \tilde{\Lambda}_{c}^{(j-1)}$.
      \Comment{Second approximation}

      \State Obtain $\tilde{\Lambda}_{\mathrm{out}}^{(j)}$ from
      $\tilde{\Lambda}_{\mathrm{out}}^{(j)} g_j
      = G_j\tilde{\Lambda}_{\mathrm{in}}^{(j)}$.
      \Comment{Eq.~\eqref{Eq: in->out}}

      \If{$j\neq D$}

      \State Obtain $\tilde{\Lambda}_{L}^{(j)}$ from
      $g_{j+1}\tilde{\Lambda}_{L}^{(j)}
      = G_{j+1,\mathrm{ideal}}\tilde{\Lambda}_{\mathrm{out}}^{(j)}$.
      
      \Comment{Eq.~\eqref{Eq: out->L}}
      
      \Else $\ $ Obtain $\tilde{\Lambda}_{L}^{(D)}$ from $\sbra{O}\!\tilde{\Lambda}_L^{(D)}=\sbra{G_O}\!\tilde{\Lambda}_{\mathrm{out}}^{(D)}$.
      
      \Comment{Eq.~\eqref{Eq: out->L last}}
      \EndIf
    \EndFor

    \State \Return $\tilde{\Lambda}_{L}^{(1)},\dots,\tilde{\Lambda}_{L}^{(D)}$.
  \end{algorithmic}
\end{algorithm}
We stress that Algorithm \ref{Algo: P2LC} computes the channels $\tilde{\Lambda}_L^{(j)}$ in parallel, as opposed to our definition of the exact channels $\Lambda_L^{(j)}$, which is serial. 

The following result quantifies the difference between the exact channel $\Lambda_L^{(j)}$ and its approximate version $\tilde{\Lambda}_L^{(j)}$, which is temporally local and CP by construction. 
\begin{proposition}\label{thm0}{\bf (Logical error channels are temporally-local and CP at leading order)}
The approximate logical error channel $\tilde{\Lambda}_L^{(j)}$ is CP and depends only on the faulty layer $G_{j}$, the previous faulty layer $G_{j-1}$, and the fault-free version of the next layer $G_{j+1,\mathrm{ideal}}$. Assuming all layers are $t$-FT, the approximate channel captures leading-order logical errors, 
\begin{align}
    \tilde{\Lambda}_L^{(j)}=\Lambda_L^{(j)}+O(\epsilon^{(j)}_L\epsilon),\label{Eq: Lambda~ approx Lambda}
\end{align}
where $\epsilon_L^{(j)}=\Theta(\epsilon^{t+1})$ is the generic scaling of the infidelity of $\Lambda_L^{(j)}$.
\end{proposition}

\begin{proof}
Algorithm \ref{Algo: P2LC} differs from the exact construction of $\Lambda_{L}^{(j)}$ in Appendix \ref{Sec: logical channels} by modifying the input channels $\Lambda_{\mathrm{in}}^{(j-1)}\mapsto I$ and $\Lambda_{\mathrm{in}}^{(j)}\mapsto \Lambda_{c}^{(j-1)}$, so our task is to show that each of these changes modifies $\Lambda_{L}^{(j)}$ by at most $O(\epsilon_L^{(j)}\epsilon)=O(\epsilon^{t+2})$. We show this inductively over the layers $G_j$. For $j=1$ there's nothing to prove. 

For $j=2$, we start by considering the second modification,  $\Lambda_{\mathrm{in}}^{(2)}\mapsto \Lambda_{c}^{(1)}$. Since each fault in the state preparation $\sket{G_\psi}$ and in $G_1$ has probability $\leq\epsilon$, we have $\Lambda_{\mathrm{out}}^{(1)}=I+O(\epsilon)$, while  $\Lambda_{L}^{(1)}=I+O(\epsilon_L^{(1)})$, with $\epsilon_L^{(1)}=O(\epsilon^{t+1})$ (Lemma \ref{lemma1}). It follows that $\Lambda_{\mathrm{out}}^{(1)}-\Lambda_L^{(1)}=O(\epsilon)$, such that      
\begin{align}
    \Lambda_{\mathrm{in}}^{(2)}=&I+(\Lambda_{\mathrm{out}}^{(1)}-\Lambda_L^{(1)})(\Lambda_L^{(1)})^{-1}\label{Eq: in apprx}\\
    =&I+(\Lambda_{\mathrm{out}}^{(1)}-\Lambda_L^{(1)})(I+[(\Lambda_L^{(1)})^{-1}-I])\nonumber\\
=&I+\Lambda_{\mathrm{out}}^{(1)}-\Lambda_L^{(1)}+O(\epsilon_L^{(1)}\epsilon).\nonumber
\end{align}
Writing $\Lambda_{\mathrm{out}}^{(1)}$ in terms of its correctable and non-correctable parts (Eq.~\eqref{Eq: out=c+nc}), 
\begin{align}
\Lambda_{\mathrm{in}}^{(2)}=\Lambda_{c}^{(1)}+(\Lambda_{nc}^{(1)}-\Lambda_L^{(1)})+O(\epsilon_L^{(1)}\epsilon).
\end{align}
Comparing Eq.~\eqref{Eq: nc} to Eq.~\eqref{Eq: L~nc}, we see that the difference $\Lambda_{nc}^{(1)}-\Lambda_L^{(1)}$ is $O(\epsilon_L^{(1)})$ and vanishes under $G_{2,\mathrm{ideal}}$. Using $G_2=G_{2,\mathrm{ideal}}+O(\epsilon)$,  we therefore have   
\begin{align}
    G_{2}(\Lambda_{nc}^{(1)}-\Lambda_L^{(1)})=O(\epsilon_L^{(1)}\epsilon).
\end{align}
Thus, 
\begin{align}
G_2\Lambda_{\mathrm{in}}^{(2)}=G_2\Lambda_{c}^{(1)}+O(\epsilon_L^{(1)}\epsilon),
\end{align}
showing that replacing $\Lambda_{\mathrm{in}}^{(2)}$ with $\Lambda_{c}^{(1)}$ in Eq.~(\ref{Eq: in->out}) can only modify $\Lambda_{\mathrm{out}}^{(2)}$, by $O(\epsilon_L^{(1)}\epsilon)$.  Plugging this into Eq.~(\ref{Eq: out->L}) we get that   $\Lambda_L^{(2)}$ is also changed by $O(\epsilon_L^{(1)}\epsilon)$.

Given the modification $\Lambda_{\mathrm{in}}^{(2)}\mapsto \Lambda_{c}^{(1)}$, the resulting approximate version of  $\Lambda_{L}^{(2)}$ can be written an expectation 
\begin{align}
    &\sum_{\boldsymbol{f}\subset G_2G_1\sket{G_\psi}}p_f \mathsf{e}_L(f)\label{Eq: sum_f}\\
    =&I+\sum_{\boldsymbol{f}\subset G_2G_1\sket{G_\psi}}p_f (\mathsf{e}_L(f)-I)\nonumber
\end{align}
over fault-paths $\boldsymbol{f}=(f_\psi,f_1,f_2)$ `contained' in $G_2G_1\sket{G_\psi}$, of resulting logical errors $\mathsf{e}_L(\boldsymbol{f})=\mathsf{e}_L(f_\psi,f_1,f_2)$ after $G_2$. Importantly, the usage of $\Lambda_{c}^{(1)}$ as an input to $G_2$ implies that fault-paths $\boldsymbol{f}$ that produce logical errors before $G_2$ (so $\mathsf{e}_L(f_\psi,f_1)\neq I$), are `reset', such that $\mathsf{e}_L(f_\psi,f_1,f_2)=\mathsf{e}_L(\emptyset,\emptyset,f_2)$ for such fault-paths. It follows that the sum in the second line of Eq.~\eqref{Eq: sum_f} can be restricted to fault-paths $\boldsymbol{f}$ that lead to a logical error after $G_2$ but not before it. According to Lemma \ref{lemma1}, such fault-paths can produce leading-order logical errors ($\mathsf{e}_L(\boldsymbol{f})\neq I$ and $p_{\boldsymbol{f}}=\Theta(\epsilon^{t+1})$) only if they are contained in $G_2G_1$ (so $\boldsymbol{f}=(\emptyset,f_1,f_2)$). We can therefore replace the input  $\Lambda_{\mathrm{in}}^{(1)}\mapsto I$, while incurring another sub-leading $O(\epsilon^{t+2})$ correction to $\Lambda_{L}^{(2)}$.  The resulting approximate channel $\tilde{\Lambda}_L^{(2)}$ is a probabilistic mixture of CPTP maps, and hence is CPTP. 

Analysis of subsequent layers $j\geq3$ proceeds in a similar manner. First, all input channels $\Lambda_{\mathrm{in}}^{(i)}$ with $i\leq j$ are (iteratively) replaced with $\Lambda_c^{(i-1)}$. The resulting approximate version of $\Lambda_L^{(j)}$ is then an expectation of fault-paths that lead to logical errors after $G_j$, but not before it. Lemma \ref{lemma1} then implies that fault-paths extending out of $G_{j}G_{j-1}$ can be discarded at leading order, by further replacing $\Lambda_{\mathrm{in}}^{(j-1)}\mapsto \Lambda_{c}^{(j-2)}\mapsto I$.
\end{proof}

\vspace{5pt}

We note that the approximation error in Eq.~\eqref{Eq: Lambda~ approx Lambda} can be improved by a constant factor by avoiding the second approximation in Algorithm \ref{Algo: P2LC} ($\Lambda_{\mathrm{in}}^{(j)}\mapsto \Lambda_c^{(j-1)}$). The approximation error can further be improved to $O(\epsilon^{t+h+1})$, using a longer history $G_j\cdots G_{j-h}$, with $h>1$, refining the first approximation in Algorithm \ref{Algo: P2LC} \cite{qedma_logical_errors_2024}.

So far we discussed P2LC in the syndrome-averaged case; we now consider the subset-conditioned case. First, Algorithm \ref{Algo: P2LC} can be generalized to the subset-conditioned case, by simply conditioning on the subset indices $k_{j-1},k_j$ measured within $G_{j-1},G_j$ (in steps 4 and 7 of the algorithm); producing an approximation $\tilde{\Lambda}^{(j)}_{L|k_{j-1},k_j}$ for $\Lambda^{(j)}_{L|k_{1:j}}$. 
An immediate generalization of Proposition \ref{thm0} applies if one is interested in conditioning on `good subsets', as in binary SALEM with rejection. The definition of `good subsets' is iterative: A subset $S_0^{(j)}$ of syndromes for the $j$'th layer is good if 
\begin{align}
    p_{k_j=0|k_{1:j-1}=\boldsymbol{0}}^{(j)}&=\mathbb{P}\left(s_j\in S_0^{(j)}\big |\ s_i\in S_0^{(i)},\ \forall i< j\right)\nonumber\\
    &=1-O(\epsilon^{t+1}), 
\end{align}
where $k_{1:j-1}=\boldsymbol{0}$ corresponds to good subsets $S^{(i)}_0$ for all previous layers $i<j$. In this case 
\begin{align}
    \tilde{\Lambda}^{(j)}_{L|k_{j-1}=0,k_j=0}=\Lambda^{(j)}_{L|k_{1:j}=\boldsymbol{0}}+O(\epsilon_{L|\boldsymbol{0}}^{(j)}\epsilon),
\end{align}
where $\epsilon_{L|\boldsymbol{0}}^{(j)}=O(\epsilon^{t+1})$ is the infidelity of  $\Lambda_{L|k_{1:j}=\boldsymbol{0}}^{(j)}$,  generalizing Eq.~\eqref{Eq: Lambda~ approx Lambda}. The general behavior of the approximation error when conditioning on `bad' subsets of syndromes is more involved, and is beyond the scope of this paper. In Appendix \ref{Sec: color code simulations} we numerically demonstrate that the logical error channels produced by P2LC allow for essentially unbiased binary SALEM, involving the inversion of logical channels conditioned on both `good' and `bad' subsets.

Finally, we note that P2LC can produce logical error channels that are local in both time and space \cite{qedma_logical_errors_2024}. Assume that each error-corrected logical layer has a block structure  $G_j=\otimes_\alpha G_{j,\alpha}$, where the error-corrected logical gates $G_{j,\alpha}$ involve non-overlapping subsets of physical (data and auxiliary) qubits $q_{j,\alpha}$ ($q_{j,\alpha}\cap q_{j,\beta}=\emptyset$ for $\alpha\neq \beta$), and carry circuit-level errors, such that physical errors supported on $q_{j,\alpha}$ can occur before/after every physical operation within each $G_{j,\alpha}$. Note, however, that physical cross-talk errors, supported on qubits in two (or more) subsets $q_{j,\alpha}\neq q_{j,\beta}$, are excluded. Let $Q_{j,\alpha}$ denote the non-overlapping subsets of logical qubits on which the corresponding desired logical gates $g_{j,\alpha}$ act, such that $k=\sum_\alpha |Q_{j,\alpha}|$ is the total number of logical qubits. 

\begin{figure}[t]
\centering
\input{fig_one_step_past_light_cone.tikz.tex}
\caption{Illustration of a one-step past and future light-cones required to compute the leading order logical error channel of a gate in layer $j$ acting on logical qubits $\{0,1\}$. It suffices to simulate this local environment in order to obtain the leading-order logical error channel $\tilde{\Lambda}_{L}^{(j,\{0,1\})}$.} 
\label{fig:fault_accumulation_lightcone}
\end{figure}
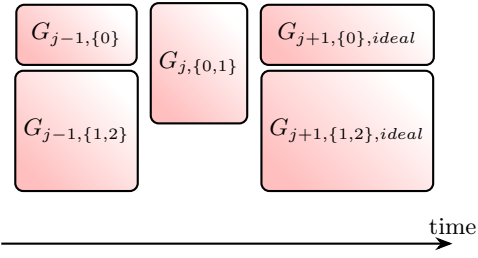

So far we assumed that each $G_j$ is $t$-FT as an operation on all physical data qubits. If each of the logical gates $G_{j,\alpha}$ is itself $t$-FT, we can obtain for each $G_{j,\alpha}$ an approximate space-time-local logical error channel $\tilde{\Lambda}_{L}^{(j,\alpha)}$, acting on  $Q_{j,\alpha}$, and which depends only on the gates $G_{j-1,\beta}$ in the previous layer that have an overlapping support with $G_{j,\alpha}$, $Q_{j-1,\beta}\cap Q_{j,\alpha}\neq\emptyset$ (the `one-step past light-cone'), as well as the ideal versions $G_{j+1,\beta,\mathrm{ideal}}$ of logical gates in the next layer with overlapping support (the `one-step future light-cone'), see Fig.~\ref{fig:fault_accumulation_lightcone}. Accordingly, $\tilde{\Lambda}_{L}^{(j,\alpha)}$ can be conditioned on subsets of syndromes in the one-step past light-cone. The relation between the approximate (syndrome-averaged) channels $\tilde{\Lambda}^{(j,\alpha)}_L$
 is given by 
 \begin{align}
     \Lambda^{(j)}_L=\otimes_\alpha \tilde{\Lambda}^{(j,\alpha)}_L + O(\epsilon_L^{(j)}\epsilon).
 \end{align}
 Thus, there are no logical cross-talk errors at leading order.

\subsection{FG-SALEM with QP distributions\label{Appendix: FG-SALEM-QP}}

\subsubsection{Implementable QP distributions \label{Appendix: Implementable QP distributions}}

Our primary example for the protocols $EM_{\boldsymbol{s}}$ used within FG-SALEM is a local inversion of $\boldsymbol{s}$-conditioned logical error channels $\Lambda_{L|s_{1:j}}^{(j)}$ (Def \ref{def channels}), based on quasi-probability (QP) distributions of the form
\begin{align}
(\Lambda^{(j)}_{L|s_{1:j}})^{-1}=\sum_b q^{(j)}_{b|s_{1:j}} \beta_{b,s_{1:j}}^{(j)}.\label{Eq: miti = QP log}
\end{align}
Here, $q^{(j)}_{b|s_{1:j}}\in \mathbb{R}$,   $\sum_b q^{(j)}_{b|s_{1:j}}=1$ are QPs, and $\beta_{b,s_{1:j}}^{(j)}:\mathcal{K}\rightarrow\mathcal{K}$ are logical basis operations. For Eq.~\eqref{Eq: miti = QP log} to be useful, the logical operations $\beta_{b,s_{1:j}}^{(j)}$ must correspond to physical operations $B_{b,s_{1:j}}^{(j)}:\mathcal{H}\rightarrow\mathcal{H}$ that can be implemented on the relevant QPU. This correspondence is generally defined through the output channels $\Lambda_{\mathrm{out}|s_{1:j}}^{(j)}:\mathcal{K}\rightarrow\mathcal{H}$, rather than the naive embedding $\mathcal{K}\hookrightarrow \mathcal{H}$.
\begin{definition} {\bf (Implementable logical basis operations)} A set of logical basis operations $\beta_{b,s_{1:j}}^{(j)}:\mathcal{K}\rightarrow\mathcal{K}$  is implementable in an error-corrected circuit $C$ run on a given QPU, if there exists a corresponding set  of physical operations $B_{b,s_{1:j}}^{(j)}:\mathcal{H}\rightarrow\mathcal{H}$ that can be implemented on the QPU, such that the following diagram is commutative, 
\begin{equation}
\begin{tikzcd}[column sep=large, row sep=large]
\mathcal{K}
  \arrow{r}{\beta_{b,s_{1:j}}^{(j)}}
  \arrow{d}[swap]{\Lambda_{\mathrm{out}\mid s_{1:j}}^{(j)}}
&
\mathcal{K}
  \arrow{d}{\Lambda_{\mathrm{out}\mid s_{1:j}}^{(j)}}
\\
\mathcal{H}
  \arrow{r}[swap]{B_{b,s_{1:j}}^{(j)}}
&
\mathcal{H}
\end{tikzcd}
\label{Eq: B diagram}
\end{equation}
or, 
\begin{align}
    B_{b,s_{1:j}}^{(j)}\Lambda_{\mathrm{out}|s_{1:j}}^{(j)}=\Lambda_{\mathrm{out}|s_{1:j}}^{(j)}\beta_{b,s_{1:j}}^{(j)}, \label{Eq: B~beta}
\end{align}
for all basis indices $b$, layers $j=1,\dots,D$ and global syndromes $\boldsymbol{s}=(s_1,\dots,s_D)$. Here, $\Lambda_{\mathrm{out}|s_{1:j}}^{(j)}:\mathcal{K}\rightarrow\mathcal{H}$ are the output channels in the circuit $C$ (Def~\ref{def channels}). 
\end{definition} 

Note that since Eq.~\eqref{Eq: B~beta}  only specifies the action of $B^{(j)}_{b,s_{1:j}}$ on the image of $\Lambda_{\mathrm{out}}^{(j)}$, the physical operations for a given set of logical basis operations are not unique. 

\begin{definition} {\bf (Mitigation super-operators)}
We refer to the super-operators generated by the physical QP distributions 
\begin{align}
\Lambda^{(j)}_{\mathrm{miti}|s_{1:j}}=\sum_b q^{(j)}_{b|s_{1:j}} B_b^{(j)}\label{Eq: miti = QP}
\end{align}
as `mitigation super-operators', $\Lambda^{(j)}_{\mathrm{miti}|s_{1:j}}:\mathcal{H}\rightarrow\mathcal{H}$. 
\end{definition}

\begin{lemma}\label{lemma miti channels}{\bf (Mitigation-inversion correspondence)} It follows from Eq.~\eqref{Eq: miti = QP log} and \eqref{Eq: B~beta} that mitigation super-operators satisfy the commutative diagram
\begin{equation}
\label{Eq: miti|s diagram}
\begin{aligned}
&
\vcenter{\hbox{
\begin{tikzcd}[
  column sep=huge,
  row sep=large
]
\mathcal{K}
  \arrow{r}{(\Lambda_{L\mid s_{1:j}}^{(j)})^{-1}}
  \arrow{d}[swap]{\Lambda_{\mathrm{out}\mid s_{1:j}}^{(j)}}
&
\mathcal{K}
  \arrow{d}{\Lambda_{\mathrm{out}\mid s_{1:j}}^{(j)}}
\\
\mathcal{H}
  \arrow{r}[swap]{\Lambda_{\mathrm{miti}\mid s_{1:j}}^{(j)}}
&
\mathcal{H}
\end{tikzcd}
}}
\ ,\\[1ex]
&
\Lambda_{\mathrm{miti}\mid s_{1:j}}^{(j)}
\Lambda_{\mathrm{out}\mid s_{1:j}}^{(j)}
=
\Lambda_{\mathrm{out}\mid s_{1:j}}^{(j)}
(\Lambda_{L\mid s_{1:j}}^{(j)})^{-1}.
\end{aligned}
\end{equation}
Thus, mitigation channels act on physical data qubits, but are equivalent to inverting logical channels on logical qubits. 

Using Eq.~\eqref{Eq: in deff}, mitigation super-operators also satisfy the above relation  with input in place of output error channels, 
\begin{align}
    \Lambda_{\mathrm{miti}|s_{1:j}}^{(j)}\Lambda_{\mathrm{in}|s_{1:j}}^{(j+1)}=\Lambda_{\mathrm{in}|s_{1:j}}^{(j+1)}(\Lambda_{L|s_{1:j}}^{(j)})^{-1}.
\end{align}
\end{lemma}

\vspace{5pt}
For adaptive Clifford circuits, the relation between logical and physical QP distributions reduces to a simple choice of representatives:
\begin{corollary} \label{cor Pauli miti channel}{\bf (QP distributions in adaptive Clifford circuits)} Assume that all error-corrected layers $G_j$ are adaptive Clifford circuits (Example \ref{example CAP}). In this case we can choose $B_{b,s_{1:j}}^{(j)}=\sigma_{\mathrm{log}}^b$ to be an arbitrary set of representatives for logical cosets of the stabilizer group, and $\beta_{b,s_{1:j}}^{(j)}=\sigma_{\mathrm{log}}^b|_{\mathcal{K}}$ to be the corresponding (representative-independent) restrictions to $\mathcal{K}$.  Accordingly,
\begin{align}
    \Lambda_{\mathrm{miti}|s_{1:j}}^{(j)}=\widehat{(\Lambda_{L|s_{1:j}}^{(j)})^{-1}}    
\end{align}
is an extension of the inverse logical channel $(\Lambda_{L|s_{1:j}}^{(j)})^{-1}$ from $\mathcal{K}$ to $\mathcal{H}$, obtained by choosing representatives for logical cosets. 
\end{corollary}
\begin{proof} 
To lighten notation, we do not indicate syndrome conditioning explicitly here. Using Example \ref{example CAP}, we can inductively show that all output error channels $\Lambda_{\mathrm{out}}^{(j)}$ are  Pauli error channels $\Lambda_{\mathrm{out}}^{(j)}=\sum_b p^{(j)}_{a,b} \sigma_{\mathrm{corr}}^a\sigma_{\mathrm{log}}^b|_{\mathcal{K}}$, and all logical channels are given by $\Lambda_L^{(j)}=\sum_b p^{(j)}_b \sigma_{\mathrm{log}}^b|_{\mathcal{K}}$, with the marginals $p_{b}^{(j)}=\sum_a p_{a,b}^{(j)}$. Noting that $\{\sigma_{\mathrm{log}}^b|_{\mathcal{K}}\}_{b=1}^{4^k}$ is isomorphic to the group of $k$-qubit Pauli super-operators, it follows that all inverse logical channels can be represented as in Eq.~\eqref{Eq: miti = QP log}, with $\beta_b^{(j)}=\sigma_{\mathrm{log}}^b|_{\mathcal{K}}$ (see Example \ref{Example: specific QP distributions} for specific constructions of $q^{(j)}_{b}$ from $p_{b}^{(j)}$). 

Let $\widehat{\Lambda_{\mathrm{out}}^{(j)}}:\mathcal{H}\rightarrow\mathcal{H}$ be an extension of $\Lambda_{\mathrm{out}}^{(j)}:\mathcal{K}\rightarrow\mathcal{H}$, obtained by choosing representatives,  $\widehat{\Lambda_{\mathrm{out}}^{(j)}}=\sum_b p^{(j)}_{a,b} \sigma_{\mathrm{corr}}^a\sigma_{\mathrm{log}}^b$. Setting $B_b=\sigma_{\mathrm{log}}^b$ and using the fact that Pauli super-operators commute shows that Eq.~\eqref{Eq: B~beta} holds:
\begin{align}
    &B_b^{(j)}\Lambda_{\mathrm{out}}^{(j)}\sket{c}\\
    =&B_b^{(j)}\widehat{\Lambda_{\mathrm{out}}^{(j)}}\sket{c}\nonumber\\
    =&\widehat{\Lambda_{\mathrm{out}}^{(j)}}B_b^{(j)}\sket{c}\nonumber\\
    =&\Lambda_{\mathrm{out}}^{(j)}\beta_b^{(j)}\sket{c}.\nonumber
\end{align}
It follows that $\Lambda^{(j)}_{\mathrm{miti}|s_{1:j}}=\sum_b q^{(j)}_{b|s_{1:j}} \sigma_{\mathrm{log}}^b=\widehat{(\Lambda_L^{(j)})^{-1}}$ is an extension of $(\Lambda_L^{(j)})^{-1}$.
\end{proof}

{~}

The analysis in this section and the subsequent results can be generalized to QP distributions that implement a reduction or amplification of logical error channels, such that 
\begin{align}
    \Lambda_{\mathrm{miti},\alpha}^{(j)}\Lambda_{\mathrm{out}}^{(j)}=\Lambda_{\mathrm{out}}^{(j)}(\Lambda_L^{(j)})^{\alpha-1},
\end{align}
with parameter $\alpha\geq 0$. For $\alpha\in (0,1)$ we obtain a partial reduction of the logical error channel, while $\alpha>1$ corresponds to amplification. Expectation values obtained with values $\alpha>0$ can then be used to extrapolate to $\alpha=0$, as part of a logical zero-noise extrapolation (ZNE). More generally, QP distributions that implement any  modification $\Lambda_L^{(j)}\mapsto \Lambda_L'^{(j)}$ of the logical error can be implemented by $\Lambda_{\mathrm{miti},\alpha}^{(j)}\Lambda_{\mathrm{out}}^{(j)}=\Lambda_{\mathrm{out}}^{(j)}(\Lambda_L^{(j)})^{-1}\Lambda_L'^{(j)}$.

\subsubsection{Implementation \label{Appendix: implementation}}

To implement the QP distribution in Eq.~\eqref{Eq: miti = QP}, we follow standard practice and write 
\begin{align}
q^{(j)}_{b|s_{1:j}}=W^{(j)}_{s_{1:j}} \mathsf{s}^{(j)}_{b|s_{1:j}}p^{(j)}_{b |s_{1:j}},    
\end{align}
with the `QP norm' $W^{(j)}_{s_{1:j}}=\sum_b |q^{(j)}_{b |s_{1:j}}|$ and signs $\mathsf{s}^{(j)}_{b|s_{1:j}}=\text{sign}(q^{(j)}_{b|s_{1:j}})$.  Assuming the basis $\beta_{b,s_{1:j}}^{(j)}$ is implementable, local inversion  of the logical channel can be achieved (in expectation) by applying the operation $B_b^{(j)}$ with probability $p^{(j)}_{b |s_{1:j}}$ (this is an $\boldsymbol{s}$-dependent circuit modification), and multiplying final measurement outcomes by the factor $f_{b,s_{1:j}}^{(j)}=\mathsf{s}_{b|s_{1:j}}^{(j)}W_{s_{1:j}}^{(j)}$. This single-shot procedure is detailed in Algorithm \ref{Algo: FG-SALEM-QP}. This can be understood as adding to 
the deterministic recovery $s_{1:j}\mapsto R_{s_{1:j}}$ of EC a (quasi) probabilistic recovery $s_{1:j}\mapsto B_b^{(j)}$.\footnote{As discussed in Sec.~\ref{Sec: fine-grained SALEM and ML decoding}, this (quasi-) probabilistic recovery can be viewed as including a deterministic correction of $R_{s_{1:j}}$ to the optimal ML recovery.}

\begin{algorithm}[H]
  \caption{Single-shot FG-SALEM with QP distributions}\label{Algo: FG-SALEM-QP}
  \begin{flushleft}
    \Input An error-corrected circuit  $C=\sum_{s_D}G_{D,s_D}\cdots \sum_{s_1}G_{1,s_1}$; a logical observable $\sbra{O}$; an implementable QP
  distribution Eq.~\eqref{Eq: miti = QP log} for each layer \(j = 1,\ldots,D\), with probabilities
  \(p_{b|s_{1:j}}^{(j)}\), basis operations \(B_{b}^{(j)}\),
  and factors \(f_{b,s_{1:j}}^{(j)}=\mathsf{s}_{b|s_{1:j}}^{(j)}W_{s_{1:j}}^{(j)}\). 
  \end{flushleft}

  \begin{algorithmic}[1]
  \State  \(f = 1\); \(s_{1:1} = ()\). \#Empty tuple
  \For{$j = 1$ \textbf{to} $D$}
\State Apply error corrected layer \(G_{j}\), and obtain a measured
  syndrome \(s_{j}\).
\State \(s_{1:j} = s_{1:j - 1} \cup (s_{j})\).
\State Sample \(b\) from the distribution
  \(p_{b|s_{1:j}}^{(j)}\), and apply \(B_b^{(j)}\). 
  \Comment{$\boldsymbol{s}$-dependent circuit modification}
\State \(f \mapsto f_{b,s_{1:j}}^{(j)} \times f\ \).
\EndFor
\State
  Measure $O$ to obtain a measurement outcome \(o'\).
\State \Return \(o = f \times o'\). 
\Comment{Post-processing}
    \Statex
  \end{algorithmic}
\end{algorithm}

The syndrome-conditioned variance of the single-shot estimator $o$ produced by Algorithm \ref{Algo: FG-SALEM-QP} satisfies the simple bound 
\begin{align}
    \Gamma_{\boldsymbol{s}}:=\mathbb{V}[o|\boldsymbol{s}]\leq W_{\boldsymbol{s}}^2, \label{Eq: Gamma_s<=W^2_s}
\end{align}
where 
\begin{align}
    W_{\boldsymbol{s}}=\prod_{j=1}^D W_{s_{1:j}}^{(j)},
\end{align}
and we assume WLOG that the observable $O$ is normalized in operator norm, $\|O\|_{\mathrm{op}}=1$. This motivates the $N$-shot estimator in Algorithm \ref{Algo: FG-SALEM-QP N-shot}. 

\begin{algorithm}[H]
  \caption{$N$-shot FG-SALEM with QP distributions}\label{Algo: FG-SALEM-QP N-shot}
  \begin{flushleft}
    \Input Number of shots \(N\); all input required for Algorithm \ref{Algo: FG-SALEM-QP}.
  \end{flushleft}

  \begin{algorithmic}[1]
  \State
  Run Algorithm \ref{Algo: FG-SALEM-QP} repeatedly \(N\) times to obtain the global syndromes \(\boldsymbol{s}_{i}\) and single-shot mitigated outcomes
  \(o_{i}\), \(i = 1,\ldots,N\).
\State \Return \(\overline{o} = \left( \sum_{i = 1}^{N}W_{\boldsymbol{s}_{i}}^{- 2} \right)^{- 1}\sum_{i = 1}^{N}{W_{\boldsymbol{s}_{i}}^{- 2}o_{i}}\).
    \Statex
  \end{algorithmic}
\end{algorithm}

\subsubsection{Expectation and shot overhead}

In this section we show that the $N$-shot estimator produced by Algorithm \ref{Algo: FG-SALEM-QP N-shot} is unbiased, and that its shot overhead is bounded by the Harmonic expectation of squared QP-norms $W_{\boldsymbol{s}}^2$. 

\begin{lemma}\label{lemma miti channels |s}{\bf (Mitigation super-operators reproduce ideal expectation values)} Adding $\Lambda_{\mathrm{miti}|s_{1:j}}^{(j)}$  after each $G_{j,s_j}$ in the $\boldsymbol{s}$-conditioned faulty expectation value (Eq.~\eqref{Eq: s-cond state}-\eqref{Eq: s-cond exp val}) produces the ideal expectation value, 
\begin{align}
    \langle O\rangle:=&\sbra{O}\!g_D\cdots g_1\sket{c}\\
    =&\sbra{G_O}\!\Lambda_{\mathrm{miti}|s_{1:D}}^{(D)}G_{D,s_D}\cdots \Lambda_{\mathrm{miti}|s_1}^{(1)}G_{1,s_1}\sket{G_\psi}/p_{\boldsymbol{s}}.\nonumber
\end{align}
for all $\boldsymbol{s}$. 
\end{lemma}
\begin{proof}
The result follows from Eq.~\eqref{Eq: probs}, the induction step
\begin{align}
    &\Lambda_{\mathrm{miti}|s_{1:j}}^{(j)}G_{j,s_j}\Lambda_{\mathrm{in}|s_{1:j-1}}^{(j)} /p_{s_j|s_{1:j-1}}^{(j)}\label{Eq: miti proof |s}\\
    =&\Lambda_{\mathrm{miti}|s_{1:j}}^{(j)}\Lambda_{\mathrm{out}|s_{1:j}}^{(j)}g_{j}\nonumber\\
  =& \Lambda_{\mathrm{out}|s_{1:j}}^{(j)} (\Lambda_{L|s_{1:j}}^{(j)})^{-1}g_{j}\nonumber\\
   =& \Lambda_{\mathrm{in}|s_{1:j}}^{(j+1)}g_j,\nonumber 
\end{align}
the initial step $\sket{G_\psi}=\Lambda_{\mathrm{in}}^{(1)}\sket{\psi}$ (Eq.~\eqref{Eq: initial}), and the final step $\sbra{G_O}\!\Lambda_{\mathrm{out}|s_{1:D}}^{(D)} (\Lambda_{L|s_{1:D}}^{(D)})^{-1}=\sbra{O}$ (Eq.~\eqref{Eq: final}).
\end{proof}

\vspace{5pt}

\begin{theorem}
    \label{thm0.5}{\bf (FG-SALEM with QP distributions is conditionally-unbiased)} 
    The output of Algorithm \ref{Algo: FG-SALEM-QP} is a `conditionally-unbiased' estimator for the ideal expectation value:
    \begin{align}
        \mathbb{E}[o|\boldsymbol{s}]=\langle O\rangle, \label{Eq: QP cond unbiased}
    \end{align}
    for all global syndromes $\boldsymbol{s}$.  
\end{theorem}

\begin{proof} 
We follow the discussion in Appendix \ref{Appendix: syndrome-conditioned exact channels}, adapting it to the presence of the basis operations $B^{(j)}_{b}$ and factors $f_{b,s_{1:j}}^{(j)}$, and then use Eq.~\eqref{Eq: miti = QP} and Lemma \ref{lemma miti channels |s}. 

Starting from the first layer, the $s_1$-conditioned state after $G_1$ is $G_{1,s_1}\sket{G_\psi}/p_{s_1}^{(1)}$. Given the measured $s_1$, Algorithm \ref{Algo: FG-SALEM-QP} samples a basis operation $B_{b_1}^{(1)}$ with probability $p_{b_1|s_1}^{(1)}$, producing the state 
\begin{align}
    \sket{s_1,b_1}=B_{b_1}^{(1)}G_{1,s_1}\sket{G_\psi}/p_{s_1}^{(1)},
\end{align}
which appears with probability 
\begin{align}
    p_{s_1,b_1}=p_{b_1|s_1}^{(1)}p_{s_1}^{(1)}.
\end{align}
The probability to obtain a syndrome $s_2$ in the second layer is then 
\begin{align}
    p_{s_2|s_1,b_1}^{(2)}=&\sbra{I}\!G_{2,s_2}\sket{s_1,b_1}\\
    =&\sbra{I}\!G_{2,s_2}B_{b_1}^{(1)}G_{1,s_1}\sket{G_\psi}/p_{s_1}^{(1)}\nonumber\\
    =&\sbra{I}\!G_{2,s_2}B_{b_1}^{(1)}\Lambda_{\mathrm{out}|s_1}^{(1)}\sket{\psi}\nonumber\\
    =&\sbra{I}\!G_{2,s_2}\Lambda_{\mathrm{out}|s_1}^{(1)}\beta_{b_1}^{(1)}\sket{\psi}\nonumber\\
    =&\sbra{I}\!G_{2,s_2}\Lambda_{\mathrm{in}|s_1}^{(2)}\left(\Lambda_{L|s_1}^{(1)}\beta_{b_1}^{(1)}\sket{\psi}\right)\nonumber\\
    =&p^{(2)}_{s_2|s_1},\nonumber
\end{align}
where we've used Def~\ref{def channels}, Eq.~\eqref{Eq: B diagram}, and the fact that $p^{(2)}_{s_2|s_1}=\sbra{I}\!G_{2,s_2}\Lambda_{\mathrm{in}|s_1}^{(2)}\sket{c}$ is independent of the logical state $\sket{c}$. We see that the distribution $p_{s_2|s_1,b_1}^{(2)}$ of $s_2$ within Algorithm \ref{Algo: FG-SALEM-QP} is independent of $b_1$, and identical to its distribution $p^{(2)}_{s_2|s_1}$ in the given faulty circuit. After the application of the second layer $G_2$, and the subsequent application of $B_{b_2}^{(2)}$ with probability $p_{b_2|s_{1:2}}^{(2)}$, the state is therefore 
\begin{align}
    \sket{s_1,b_1,s_2,b_2}=B_{b_2}^{(2)}G_{2,s_2}B_{b_1}^{(1)}G_{1,s_1}\sket{c}/p_{s_2|s_1}^{(2)}p_{s_1}^{(1)},
\end{align}
with probability 
\begin{align}
    p_{s_1,b_1,s_2,b_2}=&p^{(2)}_{b_1|s_{1:2}}p^{(2)}_{s_2|s_1}p_{b_1|s_1}^{(1)}p_{s_1}^{(1)}\\
    =&\left(p^{(2)}_{b_1|s_{1:2}}p_{b_1|s_1}^{(1)}\right)\left(p^{(2)}_{s_2|s_1}p_{s_1}^{(1)}\right).\nonumber
\end{align}
Continuing in this manner, we see that the iterative sampling procedure of $\boldsymbol{s}=(s_1,\dots,s_D)$ and $\boldsymbol{b}=(b_1,\dots ,b_D)$ can be understood as a two-step process, where first $\boldsymbol{s}$ is sampled from the joint distribution $p_{\boldsymbol{s}}=p_{s_D|s_{1:D-1}}^{(D)}\cdots p_{s_2|s_1}^{(2)}p_{s_1}^{(1)}$ (Eq.~\eqref{Eq: probs}), and then $\boldsymbol{b}$ is sampled from 
\begin{align}
    p_{\boldsymbol{b}|\boldsymbol{s}}=p_{b_D|s_{1:D}}^{(D)}\cdots p_{b_2|s_{1:2}}^{(2)}p_{b_1|s_1}^{(1)}.
\end{align}
The full output state is therefore
\begin{align}
    \sket{\boldsymbol{s},\boldsymbol{b}}=B_{b_D}^{(D)}G_{D,s_D}\cdots B_{b_1}^{(1)} G_{1,s_1}\sket{c}/p_{\boldsymbol{s}}
\end{align}
with probability $p_{\boldsymbol{s},\boldsymbol{b}}=p_{\boldsymbol{b}|\boldsymbol{s}}p_{\boldsymbol{s}}$, and the corresponding conditioned expectation value of the logical observable $O$ is $\mathbb{E}[O|\boldsymbol{s},\boldsymbol{b}]=\sbraket{O}{\boldsymbol{s},\boldsymbol{b}}$. Adding the factor $f_{\boldsymbol{b,\boldsymbol{s}}}=\prod_{j} f^{(j)}_{b_j,s_{1:j}}$ in post-processing corresponds to replacing $O$ with $f_{\boldsymbol{b,\boldsymbol{s}}}O$, such that 
\begin{align}
    \mathbb{E}[o|\boldsymbol{s},\boldsymbol{b}]=f_{\boldsymbol{b,\boldsymbol{s}}}\sbraket{O}{\boldsymbol{s},\boldsymbol{b}}.
\end{align}
We can now compute 
\begin{align}
    &\mathbb{E}[o|\boldsymbol{s}]\\
    =&\mathbb{E}\left[\mathbb{E}[o|\boldsymbol{s},\boldsymbol{b}]|\boldsymbol{s}\right]\nonumber\\
    =&\sum_{\boldsymbol{b}}p_{\boldsymbol{b}|\boldsymbol{s}}f_{\boldsymbol{b,\boldsymbol{s}}}\sbraket{O}{\boldsymbol{s},\boldsymbol{b}}\nonumber\\
    =&\sum_{\boldsymbol{b}}p_{\boldsymbol{b}|\boldsymbol{s}}f_{\boldsymbol{b,\boldsymbol{s}}}\sbra{O}\!B_{b_D}^{(D)}G_{D,s_D}\cdots B_{b_1}^{(1)} G_{1,s_1}\sket{c}/p_{\boldsymbol{s}}\nonumber\\
    =&\sum_{\boldsymbol{b}}\sbra{O}\!(q^{(D)}_{b_D,s_{1:D}}B_{b_D}^{(D)})G_{D,s_D}\cdots (q^{(1)}_{b_1,s_1}B_{b_1}^{(1)} )G_{1,s_1}\sket{c}/p_{\boldsymbol{s}}\nonumber\\
    =&\sum_{\boldsymbol{b}}\sbra{O}\!\Lambda_{\mathrm{miti}|s_{1:D}}^{(D)}G_{D,s_D}\cdots \Lambda_{\mathrm{miti}|s_{1}}^{(1)}G_{1,s_1}\sket{c}/p_{\boldsymbol{s}}\nonumber.
\end{align}
Lemma \ref{lemma miti channels |s} then implies that $\mathbb{E}[o|\boldsymbol{s}]=\langle O\rangle$.
\end{proof}

\vspace{5pt}

We now combine the QP-specific results above with the generalities of Appendix \ref{Appendix: optimal SALEM}, in order to describe the expectation and shot overhead of FG-SALEM based on QPs. 
\begin{corollary} {\bf (Expectation and shot overhead of FG-SALEM with QPs)} The SALEM estimator $\overline{o}$ produced by Algorithm \ref{Algo: FG-SALEM-QP N-shot} is unbiased, 
\begin{align}
\mathbb{E}[\overline{o}]=\langle O\rangle,  \label{Eq: SALEM QP unbiased}  
\end{align}
and its variance is bounded as $\mathbb{V}[\overline{o}]\leq \Gamma_{SALEM}^{FG}/N+O(1/N^2)$, with the shot overhead given by the harmonic expectation of squared QP norms, 
\begin{align}
    \Gamma_{SALEM}^{FG}=\mathbb{H}[W_{\boldsymbol{s}}^2]=\mathbb{H}\left[\prod_{j=1}^D (W_{s_{1:j}}^{(j)})^2\right]. \label{Eq: Gamma_SALEM QP}
\end{align}
\end{corollary}
\begin{proof}
Theorem \ref{thm0.5} and  Eq.~\eqref{Eq: Gamma_s<=W^2_s} state that $\mathbb{E}[o|\boldsymbol{s}]=\langle O\rangle$ and $\mathbb{V}[o|\boldsymbol{s}]\leq W_{\boldsymbol{s}}^2$, where $o$ is the single-shot estimator produced by Algorithm \ref{Algo: FG-SALEM-QP}. Using Corollary \ref{cor: single-shot}, these properties lead to $\mathbb{E}[o_{\boldsymbol{s}}|N_{\boldsymbol{s}}]=\langle O\rangle$ and $\mathbb{V}[o_{\boldsymbol{s}}|N_{\boldsymbol{s}}]\leq W_{\boldsymbol{s}}^2/N_{\boldsymbol{s}}$, where $o_{\boldsymbol{s}}$ is the corresponding $N_{\boldsymbol{s}}$-shot estimator (Eq.~\eqref{Eq: single -> multi}), and $N_{\boldsymbol{s}}$ is the number of shots (out of $N$) in which the global syndrome $\boldsymbol{s}$ was observed. We then have from Corollary \ref{cor: unbiased SALEM} and Theorem \ref{thm: shot overhead of FG-SALEM}, that the output $\overline{o}$ of Algorithm \ref{Algo: FG-SALEM-QP N-shot} satisfies Eq.~\eqref{Eq: SALEM QP unbiased} and \eqref{Eq: Gamma_SALEM QP}, respectively.  
\end{proof}

\vspace{5pt}

For later use, we note that if all input channels happen to be independent of past syndromes, $\Lambda_{\mathrm{in}|s_{1:j-1}}^{(j)}=\Lambda_{\mathrm{in}}^{(j)}$ for $j=1,\dots,D$, we can write $W_{\boldsymbol{s}}=\prod_{j=1}^D W^{(j)}_{s_{j}}$, where the syndromes $s_{j}$ are independent random variables. In this case 
\begin{align}
    \Gamma_{SALEM}^{FG}=\prod_j\mathbb{H}[(W^{(j)}_{s_j})^2].\label{Eq: Gamma_SALEM = W prod}
\end{align}

\subsection{FG-SALEM: Generalities\label{Appendix: optimal SALEM}}

\subsubsection{Expectation and shot overhead \label{Appendix: FG-SALEM gen. exp and var}}

After discussing FG-SALEM based on QPs in detail in Appendix \ref{Appendix: FG-SALEM-QP}, we now turn to a general analysis of FG-SALEM. 

Given an error-corrected circuit $C$ and logical observable $O$, an FG-SALEM protocol can be specified by a mapping of global syndromes to corresponding EM protocols and weights,
\begin{align}
    FG\text{-}SALEM: \boldsymbol{s}\mapsto (EM_{\boldsymbol{s}},w_{\boldsymbol{s}}).
\end{align}
The EM protocol $EM_{\boldsymbol{s}}$ is applied to the $N_{\boldsymbol{s}}$ out of $N$ shots in which the global syndrome $\boldsymbol{s}$ is measured, resulting in a `mitigated outcome' $o_{\boldsymbol{s}}$. The mitigated outcomes for observed syndromes ($N_{\boldsymbol{s}}>0$) are then combined to an $N$-shot SALEM estimator by averaging with the weights $w_{\boldsymbol{s}}=w_{\boldsymbol{s}}(N_{\boldsymbol{s}})$, 
\begin{align}
\overline{o}=\sum_{\boldsymbol{s}}w_{\boldsymbol{s}}o_{\boldsymbol{s}}\big /\sum_{\boldsymbol{s}}w_{\boldsymbol{s}},\label{Eq: o-bar}
\end{align} 
where $w_{\boldsymbol{s}}\geq 0$ and vanishes if $N_{\boldsymbol{s}}=0$, but  is otherwise unconstrained, for now. 

Each protocol $EM_{\boldsymbol{s}}$ involves modifications to the given circuit $C$, as well as a post-processing of the outcomes of measuring $O$ (or a related logical observable) at the end of the resulting modified circuits. Some of these circuit modifications may be  $\boldsymbol{s}$-independent, shared by all protocols $EM_{\boldsymbol{s}}$, and performed prior to circuit execution, as in standard EM and in ExtLEM. However, some circuit modifications may be $\boldsymbol{s}$-dependent, and must therefore be performed adaptively, during circuit execution, and after the relevant part of the global syndrome $\boldsymbol{s}$ is measured, in analogy with the recovery operation of EC (see Algorithm \ref{Algo: FG-SALEM-QP} for an explicit example).   

The protocol $EM_{\boldsymbol{s}}$ is designed to eliminate the effect of $\boldsymbol{\Lambda}_{L|\boldsymbol{s}}$ on the ideal expectation value $\langle O\rangle$ in the circuit $C$, in the following sense. 
\begin{definition} {\bf (Conditionally-unbiased protocols)}
The protocol $EM_{\boldsymbol{s}}$ is conditionally-unbiased if the corresponding estimator $o_{\boldsymbol{s}}$ satisfies 
\begin{align}
    \mathbb{E}[o_{\boldsymbol{s}}|N_{\boldsymbol{s}}]=\langle O\rangle.\label{Eq: un-biased per syndrome}
\end{align}
for all $N_{\boldsymbol{s}}>0$. 
\end{definition}

Setting $\boldsymbol{N}=(N_{\boldsymbol{s}})_{\boldsymbol{s}}$, we have:
\begin{corollary} {\bf (Unbiased SALEM)} \label{cor: unbiased SALEM}
    A SALEM estimator $\overline{o}$ constructed from conditionally-unbiased protocols is itself conditionally-unbiased, 
    \begin{align}
        \mathbb{E}[\overline{o}|\boldsymbol{N}]=\langle O\rangle,\label{Eq: E[o-bar|N_s]}
    \end{align}
    for any choice of weights $w_{\boldsymbol{s}}$. In particular, $\overline{o}$ is unbiased, $\mathbb{E}[\overline{o}]=\langle O\rangle$. 
\end{corollary}

\begin{proof} 
Setting  $\omega_{\boldsymbol{s}}=w_{\boldsymbol{s}}\big /\sum_{\boldsymbol{s}}w_{\boldsymbol{s}}$, 
\begin{align}
   \mathbb{E}[\overline{o}|\boldsymbol{N}]=\sum_{\boldsymbol{s}}\omega_{\boldsymbol{s}}\mathbb{E}[o_{\boldsymbol{s}}|N_{\boldsymbol{s}}]=\langle O \rangle\sum_{\boldsymbol{s}}\omega_{\boldsymbol{s}} =\langle O \rangle.
\end{align}
The law of total expectation then implies
\begin{align}
    \mathbb{E}[\overline{o}]=\mathbb{E}[\mathbb{E}[\overline{o}|\boldsymbol{N}]]=\langle O \rangle, 
\end{align}
for any choice of weights $w_{\boldsymbol{s}}$.
\end{proof}

\vspace{5pt}

The above corollary implies that the weights $w_{\boldsymbol{s}}$ can be chosen to minimize the variance $\mathbb{V}[\overline{o}]$. We assume upper bounds $\Gamma_{\boldsymbol{s}}$ for the shot overheads of the protocols $EM_{\boldsymbol{s}}$ are known, such that
\begin{align}
    \mathbb{V}[o_{\boldsymbol{s}}|N_{\boldsymbol{s}}]\leq \Gamma_{\boldsymbol{s}}/N_{\boldsymbol{s}}. \label{Eq: variance bounds}
\end{align}
With the convention $\|O\|_{\mathrm{op}}=1$, we may restrict attention to $\Gamma_{\boldsymbol{s}}\geq 1$.

\begin{theorem}\label{thm1}{\bf (Shot overhead of conditionally-unbiased FG-SALEM)} \label{thm: shot overhead of FG-SALEM} Let 
$\overline{o}$ be an $N$-shot FG-SALEM estimator (Eq.~(\ref{Eq: o-bar})) based on conditionally-unbiased protocols $EM_{\boldsymbol{s}}$ (Eq.~(\ref{Eq: un-biased per syndrome})), with shot overheads upper bounded by $\Gamma_{\boldsymbol{s}}\geq 1$ (Eq.~(\ref{Eq: variance bounds})). The corresponding upper bound on $\mathbb{V}[\overline{o}]$ is minimized by the  `inverse-variance' (IV) weights $w_{\boldsymbol{s}}=N_{\boldsymbol{s}}/\Gamma_{\boldsymbol{s}}$. The optimal bound is given by 
\begin{align}
    \mathbb{V}[\overline{o}]=&\mathbb{E}[\mathbb{V}[\overline{o}|\boldsymbol{N}]]\label{Eq: V[o-bar] bounds}\\
    \leq &\frac{\mathbb{H}[\Gamma_{\boldsymbol{s}}]}{N}+\left(\frac{\mathbb{H}[\Gamma_{\boldsymbol{s}}]}{N}\right)^2+O(N^{-3}),\nonumber 
\end{align}
where $\mathbb{H}[\Gamma_{\boldsymbol{s}}]=(\mathbb{E}[\Gamma_{\boldsymbol{s}}^{-1}])^{-1}$ is the `harmonic expectation' of $\ \Gamma_{\boldsymbol{s}}$ over the global syndrome $\boldsymbol{s}$. Thus, the shot overhead for FG-SALEM with IV weights is bounded by \begin{align}
        \Gamma_{SALEM}^{FG}=\mathbb{H}[\Gamma_{\boldsymbol{s}}],\label{Eq: Gamma_SALEM=H}
    \end{align}
up to negligible $O(1/N)$ corrections.
\end{theorem}

\begin{proof}
Using the law of total variance, 
\begin{align}
    \mathbb{V}[\overline{o}]=\mathbb{E}[\mathbb{V}[\overline{o}|\boldsymbol{N}]]+\mathbb{V}[\mathbb{E}[\overline{o}|\boldsymbol{N}]]=\mathbb{E}[\mathbb{V}[\overline{o}|\boldsymbol{N}]],
\end{align}
where the second term vanishes since $\mathbb{E}[\overline{o}|\boldsymbol{N}]=\langle O\rangle$ is independent of $\boldsymbol{N}$ (Eq.~\eqref{Eq: E[o-bar|N_s]}). Now, 
\begin{align}
    \mathbb{V}[\overline{o}|\boldsymbol{N}]=\sum_{\boldsymbol{s}}\omega_{\boldsymbol{s}}^2\mathbb{V}[o_{\boldsymbol{s}}|N_{\boldsymbol{s}}],
\end{align}
where we used the fact that, apart from the constraint $\sum_{\boldsymbol{s}}N_{\boldsymbol{s}}=N$, the different estimators $o_{\boldsymbol{s}}$ are independent random variables, since they are due to non-overlapping subsets of shots, $\mathrm{Cov}(o_{\boldsymbol{s}}o_{\boldsymbol{s'}}|\boldsymbol{N})=\delta_{\boldsymbol{s},\boldsymbol{s}'}\mathbb{V}[o_{\boldsymbol{s}}|N_{\boldsymbol{s}}]$. 

Given the upper bounds
\(\Gamma_{\boldsymbol{s}}\) in Eq.~\eqref{Eq: variance bounds}, we therefore have
\begin{align}
    \mathbb{V}[ \overline{o}]
    =\mathbb{E}\left[\sum_{\boldsymbol{s}}\omega_{\boldsymbol{s}}^2\mathbb{V}[o_{\boldsymbol{s}}|N_{\boldsymbol{s}}]\right]
    \leq \mathbb{E}\left[\sum_{\boldsymbol{s}}\omega_{\boldsymbol{s}}^2\Gamma_{\boldsymbol{s}}/N_{\boldsymbol{s}}\right].
\end{align}
Minimizing $\sum_{\boldsymbol{s}}\omega_{\boldsymbol{s}}^2\Gamma_{\boldsymbol{s}}/N_{\boldsymbol{s}}$
over the weights \(\omega_{\boldsymbol{s}}\) gives the `inverse-variance' weights,
\(\omega_{\boldsymbol{s}} \propto N_{\boldsymbol{s}}/\Gamma_{\boldsymbol{s}}\). With these
optimal weights, 
\begin{align}
    \sum_{\boldsymbol{s}}\omega_{\boldsymbol{s}}^2\Gamma_{\boldsymbol{s}}/N_{\boldsymbol{s}}=\left(\sum_{\boldsymbol{s}}N_{\boldsymbol{s}} \Gamma_{\boldsymbol{s}}^{-1}\right)^{-1}=\left(\sum_{i=1}^N \Gamma_{\boldsymbol{s}_i}^{-1}\right)^{-1}.
\end{align}
The optimal upper bound of the SALEM shot overhead is therefore given by the expectation over global syndromes $\{\boldsymbol{s}_i\}_{i=1}^N$ measured in $N$ shots, of the empirical harmonic mean of $\Gamma_{\boldsymbol{s}_i}$,
\begin{align}
    N\mathbb{V}[\overline{o}]\leq \mathbb{E}\left[\left(\frac{1}{N}\sum_{i=1}^N \Gamma_{\boldsymbol{s}_i}^{-1}\right)^{-1}\right]=:H_N[\Gamma_{\boldsymbol{s}}].
\end{align}
The expectation of the harmonic mean $H_N$ converges to the `harmonic expectation' $\mathbb{H}$,
\begin{align}
    \lim_{N\rightarrow\infty} H_N[\Gamma_{\boldsymbol{s}}]=\mathbb{ E}\left\lbrack \Gamma_{\boldsymbol{s}}^{-1} \right\rbrack^{- 1}=\mathbb{ H}\left\lbrack \Gamma_{\boldsymbol{s}} \right\rbrack.
\end{align}
The rate of convergence 
was studied in Ref.~\cite{Pakes1999}. Using Theorem 7 of this reference, we have that, assuming
bounded moments of \(\Gamma_{\boldsymbol{s}}^{- 1}\), 
\begin{align}
    H_N[\Gamma_{\mathbf{s}}]/\mathbb{H}[\Gamma_{\boldsymbol{s}}]  = 1 + c^{2}N^{- 1} + O(N^{- 2}),
\end{align}
with
\(c^{2}=\mathbb{V}[\Gamma_{\boldsymbol{s}}^{-1}]/\mathbb{E}[\Gamma_{\boldsymbol{s}}^{- 1}]^{2}\). Since $\Gamma_{\boldsymbol{s}}\geq 1$, all moments of
\(\Gamma_{\boldsymbol{s}}^{- 1}\) are indeed bounded, and moreover, $\mathbb{V}[\Gamma_{\boldsymbol{s}}^{-1}]\leq \mathbb{E}[\Gamma_{\boldsymbol{s}}^{-2}]\leq\mathbb{E}[\Gamma_{\boldsymbol{s}}^{-1}]$.
Thus, $c^2\leq 1/\mathbb{E}[\Gamma_{\boldsymbol{s}}^{-1}]=\mathbb{H}[\Gamma_{\boldsymbol{s}}]$,
and
\begin{align}
    \mathbb{V}[\overline{o}]&\leq \frac{H_N[\Gamma_{\boldsymbol{s}}]}{N}\\
    &\leq \frac{\mathbb{H}[\Gamma_{\boldsymbol{s}}]}{N}+\left(\frac{\mathbb{H}[\Gamma_{\boldsymbol{s}}]}{N}\right)^2+O(N^{-3})\nonumber.
\end{align}
\end{proof}

\vspace{5pt}

If the protocols $EM_{\boldsymbol{s}}$ are not conditionally-unbiased, the SALEM estimator $\overline{o}$ will be biased, and we therefore quantify its mean-squared-error (MSE) instead of just its variance. The following result shows that, with IV weights, the MSE contribution of the conditioned bias $b_{\boldsymbol{s}}$ in `bad syndromes', which have a large conditioned overhead $\Gamma_{\boldsymbol{s}}$, is suppressed as $\Gamma_{\boldsymbol{s}}^{-1}$. Therefore, larger conditioned biases may be tolerated for bad syndrome.  
\begin{corollary} {\bf (MSE of FG-SALEM with bounded conditional bias)}
    If the protocols $EM_{\boldsymbol{s}}$ have a bounded conditional bias 
    \begin{align}
        |\mathbb{E}[o_{\boldsymbol{s}}|N_{\boldsymbol{s}}]-\langle O 
    \rangle|\leq b_{\boldsymbol{s}},
    \end{align}
    and bounded conditional shot overheads (Eq.~\eqref{Eq: variance bounds}), the mean-squared-error $\mathrm{MSE}[\overline{o}]=\mathbb{E}\left[(\overline{o}-\langle O \rangle)^2\right]$ of the corresponding SALEM estimator with IV weights is given by 
    \begin{align}
        \mathrm{MSE}[\overline{o}]&=\mathbb{E}\left[\left(\mathbb{E}[\overline{o}|\boldsymbol{N}]-\langle O \rangle\right)^2\right]+\mathbb{E}[\mathbb{V}[\overline{o}|\boldsymbol{N}]] \nonumber
    \end{align}
    where the second term is bounded as in Eq.~\eqref{Eq: V[o-bar] bounds}, while the first is bounded as 
    \begin{align}
        \mathbb{E}\left[\left(\mathbb{E}[\overline{o}|\boldsymbol{N}]-\langle O \rangle\right)^2\right]\leq \left(\frac{\mathbb{E}[\Gamma_{\boldsymbol{s}}^{-1}b_{\boldsymbol{s}}]}{\mathbb{E}[\Gamma_{\boldsymbol{s}}^{-1}]}\right)^2+O(N^{-1}) \label{Eq: large-N bias}
    \end{align}
\end{corollary}

\begin{proof} 
We use the law of total expectation, and then split the conditioned MSE to bias and variance terms, 
\begin{align}
    \mathrm{MSE}[\overline{o}]=&\mathbb{E}\left[\mathbb{E}\left[(\overline{o}-\langle O \rangle)^2|\boldsymbol{N}\right]\right]\\
    =&\mathbb{E}\left[\left(\mathbb{E}[\overline{o}|\boldsymbol{N}]-\langle O \rangle\right)^2\right]+\mathbb{E}[\mathbb{V}[\overline{o}|\boldsymbol{N}]]\nonumber.
\end{align}
To bound the first term,
\begin{align}
    \mathbb{E}[\overline{o}|\boldsymbol{N}]-\langle O \rangle&=\sum_{\boldsymbol{s}}\omega_{\boldsymbol{s}}\left(\mathbb{E}[o_{\boldsymbol{s}}|N_{\boldsymbol{s}}]-\langle O\rangle\right)\\
    &\leq \sum_{\boldsymbol{s}}\omega_{\boldsymbol{s}}b_{\boldsymbol{s}}\nonumber\\
    &=\sum_{\boldsymbol{s}}N_{\boldsymbol{s}}\Gamma_{\boldsymbol{s}}^{-1}b_{\boldsymbol{s}}\big/\sum_{\boldsymbol{s}}N_{\boldsymbol{s}}\Gamma_{\boldsymbol{s}}^{-1}\nonumber\\
    &=\sum_{i=1}^N\Gamma_{\boldsymbol{s}_i}^{-1}b_{\boldsymbol{s}_i}\big/\sum_{i=1}^N\Gamma_{\boldsymbol{s}_i}^{-1}\nonumber,
\end{align}
so that 
\begin{align}
    \mathbb{E}\left[\left(\mathbb{E}[\overline{o}|\boldsymbol{N}]-\langle O \rangle\right)^2\right]
    \leq \mathbb{E}\left[\left(\frac{N^{-1}\sum_{i}\Gamma_{\boldsymbol{s}_i}^{-1}b_{\boldsymbol{s}_i}}{N^{-1}\sum_i\Gamma_{\boldsymbol{s}_i}^{-1}}\right)^2\right].
\end{align}
Expanding around $N=\infty$ then gives Eq.~\eqref{Eq: large-N bias} \cite{Freedman2008Ratio}. 
\end{proof}

\subsubsection{Single-shot protocols \label{Appendix: single shot}}

Our general description of FG-SALEM in Appendix \ref{Appendix: FG-SALEM gen. exp and var} allowed for protocols $EM_{\boldsymbol{s}}$ that jointly post-process final measurements (and syndromes) from all $N_{\boldsymbol{s}}$ shots in which a particular syndrome was observed. However, the detailed QP example in Appendix \ref{Appendix: FG-SALEM-QP} individually post-processed each shot (Algorithms \ref{Algo: FG-SALEM-QP}-\ref{Algo: FG-SALEM-QP N-shot}). We make the connection explicit here, by discussing the notion of `single-shot EM protocols'.

\begin{definition} {\bf (Single-shot protocols)}
$EM_{\boldsymbol{s}}$ is a single-shot EM protocol if 
\begin{align}
    o_{\boldsymbol{s}}=N_{\boldsymbol{s}}^{-1}\sum_{1\leq i \leq N;\ \boldsymbol{s}_i=\boldsymbol{s}}o_i, \label{Eq: single -> multi}
\end{align}
is an average of $N_{\boldsymbol{s}}$ out of $N$ i.i.d. estimators $o_i\sim o$, obtained from the shots $i\in\{1,\dots,N\}$ in which a particular global syndrome
$\boldsymbol{s}$ was observed. We say that $o$ is conditionally unbiased if 
\begin{align}
    \mathbb{E}[o|\boldsymbol{s}]=\langle O 
    \rangle . 
\end{align}
\end{definition}
Clearly, when $N_{\boldsymbol{s}}=1$, the protocol $EM_{\boldsymbol{s}}$ must reduce to a single-shot protocol, but this need not be the case when $N_{\boldsymbol{s}}>1$. Within FG-SALEM we condition on individual global syndromes $\boldsymbol{s}$, making the value $N_{\boldsymbol{s}}=1$ common. In contrast, within CG-SALEM, using an appropriate partition to subsets and a possible rejection of very rare subsets, we can ensure $N_{\boldsymbol{k}}
\gg1$ with high probability, for every (accepted) global subset index $\boldsymbol{k}$.

\begin{corollary} {\bf (FG-SALEM with single-shot protocols)} \label{cor: single-shot}
    For single-shot protocols, the SALEM estimator Eq.~\eqref{Eq: o-bar} can be written as 
\begin{align}
    \overline{o}=\sum_{\boldsymbol{s}}w_{\boldsymbol{s}}o_{\boldsymbol{s}}\big /\sum_{\boldsymbol{s}}w_{\boldsymbol{s}}=\sum_{i=1}^Nw_i o_i \big/ \sum_{i=1}^N w_i,
\end{align}
where $w_i=w_{\boldsymbol{s}_i}/N_{\boldsymbol{s}_i}$. The expectations of $o_{\boldsymbol{s}}$ and $o_i\sim o$ are related by 
\begin{align}
    \mathbb{E}[o_{\boldsymbol{s}}|N_{\boldsymbol{s}}]=\mathbb{E}[o|\boldsymbol{s}],
\end{align}
so if $o$ is conditionally unbiased, so is $o_{\boldsymbol{s}}$. The conditioned variances are related by 
\begin{align}
    \mathbb{V}[o_{\boldsymbol{s}}|N_{\boldsymbol{s}}]=\mathbb{V}[o|\boldsymbol{s}]/N_{\boldsymbol{s}},
\end{align}
so a bound $\mathbb{V}[o|\boldsymbol{s}]\leq \Gamma_{\boldsymbol{s}}$ implies $\mathbb{V}[o_{\boldsymbol{s}}|N_{\boldsymbol{s}}]\leq \Gamma_{\boldsymbol{s}}/N_{\boldsymbol{s}}$. More generally, $\mathrm{Cov}[o_{\boldsymbol{s}}o_{\boldsymbol{s}'}|\boldsymbol{N}]]=\delta_{\boldsymbol{s},\boldsymbol{s}'}\mathbb{V}[o|\boldsymbol{s}]/N_{\boldsymbol{s}}$. The optimal IV weights are given by 
\begin{align}
    w_i=\Gamma_{\boldsymbol{s}_i}^{-1}.
\end{align}  
\end{corollary}

\subsubsection{Advantage of FG-SALEM over ExtLEM \label{Appendix: advantage of SALEM over ExtLEM}}

We now compare the shot overhead of FG-SALEM to that of ExtLEM. To make a `fair' comparison, we assume that both ExtLEM and FG-SALEM are based on the same underlying EM protocol $EM:\boldsymbol{\Lambda}\mapsto o$, viewed as a mapping of a global logical error channel to be mitigated $\boldsymbol{\Lambda}$ to a resulting estimator  $o$. Note that this notation highlights the dependence of the estimator $o$ on the error channel $\boldsymbol{\Lambda}$, irrespective of whether or not a characterization of $\boldsymbol{\Lambda}$ is used to construct $o$, but suppresses the dependence of $o$ on the number of shots $N$, the circuit $C$ and observable $O$. Thus, the ExtLEM estimator is 
\begin{align}
    o_{ExtLEM}=EM(\boldsymbol{\Lambda}_L),
\end{align}
and the per-syndrome SALEM estimators are 
\begin{align}
    o_{\boldsymbol{s}}=EM(\boldsymbol{\Lambda}_{L|\boldsymbol{s}}).
\end{align}
Given the underlying protocol $EM$, we have a corresponding mapping  $f:\boldsymbol{\Lambda}\mapsto \Gamma$ of the mitigated global channel to a resulting bound on the shot overhead $N\mathbb{V}[o]\leq\Gamma$, such that 
\begin{align}
    \Gamma_{ExtLEM}=&f(\boldsymbol{\Lambda}_L),\\
    \Gamma_{SALEM}^{FG}=&\mathbb{H}[\Gamma_{\boldsymbol{s}}]=\mathbb{H}[f(\boldsymbol{\Lambda}_{L|\boldsymbol{s}})],\nonumber
\end{align}
or 
\begin{align}
1/\Gamma_{ExtLEM}=&1/f(\boldsymbol{\Lambda}_L),\label{Eq: 1/Gamma}\\
1/\Gamma_{SALEM}=&\mathbb{E}[1/f(\boldsymbol{\Lambda}_{L|\boldsymbol{s}})].\nonumber
\end{align}
Below we identify several conditions under which $\Gamma_{SALEM}^{FG}\leq \Gamma_{ExtLEM}$. These involve different  averaging relations of the form ``$\boldsymbol{\Lambda}_L=\mathbb{E}[\boldsymbol{\Lambda}_{L|\boldsymbol{s}}]$'' (which does not generally hold as written), and corresponding convexity assumptions on the reciprocal function $1/f$. The basic intuition for expecting a convex $1/f$ is that common expressions $\Gamma=f(\epsilon)$ for EM shot overheads as a function of an error rate $\epsilon$ (such as $f(\epsilon)=e^{\lambda\epsilon V}$ discussed in the main text and in Example \ref{Example: exp})  are monotonically increasing and convex, implying that $1/f(\epsilon)$ is convex. However, convexity of $1/f(\epsilon)$ is a weaker and highly generalizable requirement. 

\vspace{5pt}

The basic relation between $\boldsymbol{\Lambda}_{L|\boldsymbol{s}}$ and $\boldsymbol{\Lambda}_L$ follows from Eq.~\eqref{Eq: syndrome averaging of channels 2}, and involves `prefix products' of error channels and ideal layers, 
\begin{align}
    &\Lambda_{L}^{(j)}g_j\cdots\Lambda_{L}^{(1)}g_1 \label{Eq: explicit prefix averaging}\\
    =&\sum_{s_{1:j}}(p_{s_j|s_{1:j-1}}^{(j)}\cdots p_{s_1}^{(1)})\Lambda_{L|s_{1:j}}^{(j)}g_j\cdots  \Lambda_{L|s_1}^{(1)}g_1.\nonumber
\end{align}
We can express this in terms of `prefix error channels', 
$\Lambda_{L}^{(1:j)}=\Lambda_{L}^{(j)}g_j\cdots\Lambda_{L}^{(1)}g_1(g_j\cdots g_1)^{-1}$,
such that 
\begin{align}
    \Lambda_{L}^{(1:j)}=\mathbb{E}_{s_{1:j}}[\Lambda_{L|s_{1:j}}^{(1:j)}]=\mathbb{E}_{\boldsymbol{s}}[\Lambda_{L|s_{1:j}}^{(1:j)}]
\end{align}
is a convex combination of its conditioned versions $\Lambda_{L|s_{1:j}}^{(1:j)}=\Lambda_{L|s_{1:j}}^{(j)}g_j\cdots  \Lambda_{L|s_1}^{(1)}g_1(g_j\cdots g_1)^{-1}$. The list of prefixes $\boldsymbol{\Lambda}_{L|\boldsymbol{s}}^{(:)}=(\Lambda_{L|s_{1:j}}^{(1:j)})_{j=1}^D$, which parametrizes $\boldsymbol{\Lambda}_{L|\boldsymbol{s}}$, therefore satisfies 
\begin{align}
    \boldsymbol{\Lambda}_{L}^{(:)}=\mathbb{E}[\boldsymbol{\Lambda}_{L|\boldsymbol{s}}^{(:)}]. \label{Eq: prefix averaging}
\end{align}

\begin{theorem} {\bf (Advantage of SALEM over ExtLEM in shot overhead: Prefix channels)} \label{thm: prefix}
    Let \(\Gamma_{ExtLEM} = f(\boldsymbol{\Lambda}_{L}^{(:)})\) be
a bound on the shot overhead of ExtLEM with a given EM protocol, as a function of the prefix error channels \(\boldsymbol{\Lambda}_{L}^{(:)}\). Assume that the reciprocal function 
\begin{align}
    1/f:\mathrm{Hull}(\{\boldsymbol{\Lambda}_{L|\boldsymbol{s}}^{(:)}\}_{\boldsymbol{s}})\rightarrow\mathbb{R}
\end{align}
is strictly convex on the convex hull of syndrome-conditioned prefix error channels $\boldsymbol{\Lambda}_{L|\boldsymbol{s}}^{(:)}$. 
Consider FG-SALEM based on the same underlying EM protocol, such that
\(\Gamma_{\boldsymbol{s}} = f(\boldsymbol{\Lambda}_{L|\boldsymbol{s}}^{(:)})\), with IV
weights. Then the shot overhead of FG-SALEM can only be lower than
that of ExtLEM, 
\begin{align}
    \Gamma_{SALEM}^{FG} \leq \Gamma_{ExtLEM}.
\end{align}
with an equality
if and only if \(\boldsymbol{\Lambda}_{L|\boldsymbol{s}}^{(:)}=\boldsymbol{\Lambda}_L^{(:)}\)
for all $\boldsymbol{s}$.
\end{theorem}

\begin{proof} 
Using Eq.~\eqref{Eq: prefix averaging}, Jensen's inequality and Eq.~\eqref{Eq: prefix averaging}
\begin{align}
    1/\Gamma_{SALEM}&=\mathbb{E}[1/f(\boldsymbol{\Lambda}_{L|\boldsymbol{s}}^{(:)})]\\
    &\geq1/f(\mathbb{E}[\boldsymbol{\Lambda}_{L|\boldsymbol{s}}^{(:)}])\nonumber\\
    &=1/f(\boldsymbol{\Lambda}_L^{(:)})=1/\Gamma_{ExtLEM}.\nonumber
\end{align}
\end{proof}

\vspace{5pt}

An immediate application of the above result is the case where $f$ only depends on the final prefix $\Lambda_{L|s_{1:D}}^{(1:D)}$. This is the case for SALEM based on `global inversion', where informationally-complete measurement bases are sampled at the end of the circuit, and $\Lambda_{L|s_{1:D}}^{(1:D)}$ is inverted in (extensive) post-processing \cite{filippov2023scalable, qiskit_pna}. Beyond this example, note that the change of variables from $\boldsymbol{\Lambda}_{L|\boldsymbol{s}}$ to $\boldsymbol{\Lambda}_{L|\boldsymbol{s}}^{(:)}$ is not affine, and therefore convexity in the variables $\boldsymbol{\Lambda}_{L|\boldsymbol{s}}^{(:)}$ is not equivalent to convexity in $\boldsymbol{\Lambda}_{L|\boldsymbol{s}}$, which are the more natural variables from the perspective of SALEM. A simple case in which we can make a more natural convexity assumption is the history-independent case. 

{
\renewcommand{\thetheorem}{\arabic{theorem}$'$}
\addtocounter{theorem}{-1}
\begin{theorem} {\bf (Advantage of SALEM over ExtLEM in shot overhead: History independence)}
Let \(\Gamma_{ExtLEM} = f(\boldsymbol{\Lambda}_{L})\) be
a bound on the shot overhead of ExtLEM with a given EM protocol as a function of the global error channel  \(\boldsymbol{\Lambda}_{L}\). Assume the input channel to each layer is independent of past syndromes, such that \(\boldsymbol{\Lambda}_{L}=(\Lambda_{L|s_j}^{(j)})_{j=1}^D\), and that the  reciprocal function 
\begin{align}
    1/f:\mathrm{Hull}(\{\Lambda_{L|s_1}^{(1)}\}_{s_1})\times \cdots \times \mathrm{Hull}(\{\Lambda_{L|s_D}^{(D)}\}_{s_D})\rightarrow\mathbb{R}
\end{align}
is strictly convex as a function of each of its $D$ arguments, taking values in the convex hulls of syndrome-conditioned net error channels $\Lambda_{L|s_j}^{(j)}$. Note that $1/f$ need not be jointly convex. Consider FG-SALEM based on the same underlying EM protocol, such that
\(\Gamma_{\boldsymbol{s}} = f(\boldsymbol{\Lambda}_{L|\boldsymbol{s}})\), with IV
weights. Then the shot overhead of FG-SALEM can only be lower than
that of ExtLEM, 
\begin{align}
    \Gamma_{SALEM} \leq \Gamma_{ExtLEM},
\end{align}
with an equality
if and only if \(\boldsymbol{\Lambda}_{L|\boldsymbol{s}}=\boldsymbol{\Lambda}_L\)
for all $\boldsymbol{s}$.   
\end{theorem}
}

\begin{proof}
Appendix \ref{Appendix: syndrome-conditioned exact channels} shows that if the input channel $\Lambda_{\mathrm{in}|s_{1:j-1}}^{(j)}=\Lambda_{\mathrm{in}}^{(j)}$ is independent of past syndromes, we have $p_{s_j|s_{1:j-1}}^{(j)}=p_{s_j}^{(j)}$, and $\Lambda_{L|s_{1:j}}^{(j)}=\Lambda_{L|s_{j}}^{(j)}$, such that Eq.~\eqref{Eq: explicit prefix averaging} reduces to 
\begin{align}
    \Lambda_{L}^{(j)}=\mathbb{E}[\Lambda_{L|s_j}^{(j)}],
\end{align}
for all $j=1,\dots,D$. Using Jensen's inequality iteratively over layer syndromes $s_j$ then shows 
\begin{align}
    1/\Gamma_{SALEM}=&\mathbb{E}[1/f(\boldsymbol{\Lambda}_{L|\boldsymbol{s}})]\\
    =&\mathbb{E}_{s_D}\cdots \mathbb{E}_{s_1}[1/f(\Lambda_{L|s_1}^{(1)},\dots,\Lambda_{L|s_D}^{(D)})]\nonumber \\
    \geq&1/f(\mathbb{E}_{s_D}[\Lambda_{L|s_1}^{(1)}],\dots,\mathbb{E}_{s_D}[\Lambda_{L|s_D}^{(D)}]) \nonumber\\
    =&1/f(\mathbb{E}[\boldsymbol{\Lambda}_{L|\boldsymbol{s}}]) \nonumber\\
    =&1/\Gamma_{ExtLEM}.\nonumber
\end{align}
\end{proof}

\vspace{5pt}

We can allow for history dependence (which is a generic feature), and work with the channels $\boldsymbol{\Lambda}_{L|\boldsymbol{s}}$ (as opposed to prefixes $\boldsymbol{\Lambda}_{L|\boldsymbol{s}}^{(:)}$), in the QP case, where the overhead is a product $\Gamma = \prod_{j=1}^Dg_j(\Lambda^{(j)})$ over logical layers, and the function $g_j$ describes the overhead due to the corresponding error channel $\Lambda^{(j)}$ (Eq.~\eqref{Eq: Gamma_SALEM QP}). As discussed in Appendix \ref{Appendix: bi-SALEM}, the relevant functions $g_j$ admit the expansion 
\begin{align}
    g_j(\Lambda)=1+\lambda \epsilon_\Lambda+O(\epsilon_\Lambda^2), \label{Eq: g expansion}
\end{align}
with `blowup rate' $\lambda>0$, and where $\epsilon_\Lambda$ is the (entanglement) infidelity of $\Lambda$, and further satisfy  
\begin{align}
    1/g_{j}(\Lambda)
    \ge
    1-\lambda\epsilon_\Lambda.
    \label{eq:straight_line_bound}
\end{align}
Note that, given Eq.~\eqref{Eq: g expansion}, the requirement in Eq.~\eqref{eq:straight_line_bound} is weaker than convexity of $g_j$, and trivially holds for $\epsilon_\Lambda>\lambda^{-1}$. The final requirement we need in this setup is the relation 
\begin{align}
    \epsilon_{L}^{(j)}=\mathbb{E}[\epsilon_{L|s_{1:j}}^{(j)}]
\end{align}
between the infidelities of the channels $\Lambda_{L}^{(j)}$ and $\Lambda_{L|s_{1:j}}^{(j)}$. This relation is satisfied by the approximate channels $\tilde{\Lambda}_{L}^{(j)}$ (and $\tilde{\Lambda}_{L|s_{1:j}}^{(j)}$) produced by Algorithm \ref{Algo: P2LC}. Since we consider approximate channels here, we can work with space-time local channels associated with individual logical gates, as opposed to logical layers (Appendix \ref{sec:approx logical channels}).

{
\renewcommand{\thetheorem}{\arabic{theorem}$''$}
\addtocounter{theorem}{-1}
\begin{theorem} {\bf (Advantage of SALEM over ExtLEM in shot overhead: Product overhead)}
\label{thm:straight_line}
Consider an error-corrected circuit with \(V\) logical gates. Let \(\Gamma_{ExtLEM} = \prod_{j=1}^Vg_j(\Lambda_{L}^{(j)})\) be
a bound on the shot overhead of ExtLEM with a given EM protocol, which is a product of local overheads $g_j> 0$ depending on corresponding logical error channels $\Lambda_L^{(j)}$. Assume that FG-SALEM (with IV weights) is based on the same EM protocol, such that 
\begin{align}
    \Gamma_{\boldsymbol{s}}
    =
    \prod_{j=1}^V g_j(\Lambda_{j|\boldsymbol{s}}^{(j)}),
\end{align}
where \(\Lambda_{L|\boldsymbol{s}}^{(j)}\) is the logical error of the $j$'th gate, conditioned on a part of the global syndrome $\boldsymbol{s}$ (contained in the past lightcone of that gate). Denote by \(\epsilon_L^{(j)}\) and $\epsilon_{L| \boldsymbol{s}}^{(j)}$ the infidelities of \(\Lambda_L^{(j)}\) and \(\Lambda_{L|\boldsymbol{s}}^{(j)}\), such that 
\begin{align}
    \epsilon_L^{(j)}=\mathbb{E}[\epsilon_{L|\boldsymbol{s}}^{(j)}].    
\end{align}
Define the gate-averaged logical error,  $\epsilon_L=V^{-1}\sum_{j} \epsilon_{L}^{(j)}$, and assume that each local overhead $g_j$ satisfies  Eq.~\eqref{Eq: g expansion} and \eqref{eq:straight_line_bound}. Define the blowup rates 
\begin{align}
    \lambda_{ExtLEM}
    =
    \lim_{\epsilon_L\to 0}
    \frac{\log \Gamma_{\rm ExtLEM}}{V\epsilon_L},\\
    \lambda_{SALEM}^{FG}
    =
    \lim_{\epsilon_L\to 0}
    \frac{\log \Gamma_{\rm SALEM}^{FG}}{V\epsilon_L},\nonumber
\end{align}
where the limit $\epsilon_L\to0$ is obtained by taking the physical error rate $\epsilon$ to zero.  Then, the blowup rate of the blowup rate of SALEM is always lower than that of ExtLEM,
\begin{align}
    \lambda_{SALEM}^{FG}
    \le
    \lambda_{ExtLEM}=\lambda.
\end{align}
In the generic case where $\lambda_{SALEM}^{FG}<\lambda$, we have that $\Gamma_{SALEM}^{FG}<\Gamma_{ExtLEM}$ for small enough $\epsilon_L V$. 
\end{theorem}
}

\begin{proof}
First, 
\begin{align}
    \log\Gamma_{ExtLEM}&=\sum_{j=1}^V\log g_j(\Lambda_L^{(j)}) \label{Eq: log Gamma_Ext}\\
    &=\sum_{j=1}^V\log (1+\lambda\epsilon_L^{(j)}+O(\epsilon_L^{(j)})^2)\nonumber\\
    &=\lambda V\epsilon_L +O(V^2\epsilon_L^2), \nonumber
\end{align}
so $\lambda_{ExtLEM}=\lambda$. To bound $\lambda_{SALEM}$, we use the inequality $\prod_j (1-x_j)\geq 1-\sum_j x_j$ for \(x_j\in[0,1]\), and the fact that $g_j> 0$, 
\begin{align}
    \left(\Gamma_{SALEM}^{FG}\right)^{-1}
    &=
    \mathbb{E}
    \left[
        \Gamma_{\boldsymbol{s}}^{-1}
    \right]
    \\
    &= \mathbb{E}\left[\prod_{j=1}^V 1/g_j(\Lambda_{L|\boldsymbol{s}}^{(j)})\right]\nonumber\\
    &\ge
    \mathbb{E}\left[1-
    \sum_{j=1}^V
     \min(\lambda\epsilon_{L|\boldsymbol{s}}^{(j)},1)
    \right]
    \nonumber\\
    &\ge
    \mathbb{E}\left[1-
    \sum_{j=1}^V
     \lambda\epsilon_{L|\boldsymbol{s}}^{(j)}
    \right]
    \nonumber\\
    &=
    1-
    \lambda\sum_{j=1}^V
    \epsilon_L^{(j)} =1-\lambda V\epsilon_L.\nonumber
\end{align} 
It follows that 
\begin{align}
    \log\Gamma_{\rm SALEM}^{FG}&\leq
    -\log(1-\lambda V\epsilon_L)\label{Eq: logGamma_SALEM bound}\\
    &=\lambda V\epsilon_L+O(V^2\epsilon_L^2), \nonumber
\end{align}
so $\lambda_{SALEM}^{FG}\leq \lambda$. 
\end{proof}

\vspace{5pt}

Though the statement of Theorem \ref{thm:straight_line} refers to the regime of small $\epsilon_L V$, we generically expect the approximately-exponential forms 
\begin{align}
\Gamma_{ExtLEM}=\exp\left(\lambda\epsilon_L V +O(\epsilon_L^2 V)\right) \label{Eq: apprx-exp-Ext}
\end{align}
and 
\begin{align}
    \Gamma_{SALEM}^{FG}=\exp\left(\lambda_{SALEM}^{FG}\epsilon_L V +O(\epsilon_L \epsilon V)\right), \label{Eq: apprx-exp SALEM}
\end{align}
such that $\lambda_{SALEM}^{FG}<\lambda$ implies $\Gamma_{SALEM}^{FG}< \Gamma_{ExtLEM}$ up to very large $\epsilon_LV\sim \epsilon^{-1}$.  Equation \eqref{Eq: apprx-exp-Ext} simply follows from Eq.~\eqref{Eq: log Gamma_Ext}, assuming each local infidelity is bounded by the gate averaged infidelity, $\epsilon_L^{(j)}=O(\epsilon_L)$.  Equation \eqref{Eq: apprx-exp SALEM} may be justified by the (formal) cumulant expansion
\begin{align}
    \log\Gamma_{SALEM}^{FG}
    &= \log\mathbb{E}\left[e^{\sum_{j=1}^V X_j}\right]\nonumber\\
    &=\sum_j\mathbb{E}\left[X_j\right]+\sum_{i,j}\mathrm{Cov}( X_i,X_j)+\cdots,
\end{align}
where $X_j=\log g_j(\Lambda_{L|\boldsymbol{s}}^{(j)})$, and the expectation is over the global syndrome $\boldsymbol{s}=(s_i)_{i=1}^V$. If $\Lambda_{L|\boldsymbol{s}}^{(j)}$ are the approximate channels described in Appendix \ref{sec:approx logical channels}, the fact that $\Lambda_{L|\boldsymbol{s}}^{(j)}$ (and the corresponding probabilities $p_{s_j|s_{i\neq j}}^{(j)}$) only depends on syndromes  $s_i$  associated with gates $i$ which are at a finite space-time range $\|i-j\|\leq r$ from $j$, implies that $\mathrm{Cov}( X_i,X_j)=0$ for $\|i-j\|>r$. The same argument applies to all higher cumulants in the expansion, implying  $\log\Gamma_{SALEM}^{FG}=O(V)$. The generic analyticity of each local overhead $g_j(\Lambda_{L|\boldsymbol{s}}^{(j)})$ in the physical error rate $\epsilon$ then allows for an expansion $\log\Gamma_{SALEM}^{FG}/V=\lambda_{SALEM}^{FG}\epsilon_L+O(\epsilon_L\epsilon)$ in $\epsilon$, whose leading term is constrained by Eq.~\eqref{Eq: logGamma_SALEM bound}.

\begin{example}\label{ex2}{\bf (Exponential shot overhead)} \label{Example: exp} As discussed in the main text, a simplified but representative expression for the shot overhead in EM protocols is given by exponential local overheads, $g_j(\Lambda)=e^{\lambda \epsilon_\Lambda}$, such that 
\begin{align}
    \Gamma_{ExtLEM}=\prod_{j=1}^V e^{\lambda \epsilon_{L}^{(j)}}= e^{\lambda V \epsilon_L},
\end{align}
and 
\begin{align}
    \Gamma_{\boldsymbol{s}}=\prod_{j=1}^V e^{\lambda \epsilon_{L|\boldsymbol{s}}^{(j)}}= e^{\lambda V \epsilon_{L|\boldsymbol{s}}},
\end{align}
where  $\epsilon_{L|\boldsymbol{s}}=V^{-1}\sum_j \epsilon_{L|\boldsymbol{s}}^{(j)}$. Note that $g_j(\Lambda)=e^{\lambda \epsilon_\Lambda}$ satisfies Eq.~\eqref{Eq: g expansion} and \eqref{eq:straight_line_bound} (and moreover $1/g_j$ is convex). In this special case we have 
\begin{align}
    \Gamma_{ExtLEM}=e^{\lambda V\mathbb{E}[\epsilon_{L|\boldsymbol{s}}]}=e^{\mathbb{E}[\log \Gamma_{\boldsymbol{s}} ]}=\mathbb{G}[\Gamma_{\boldsymbol{s}}],
    \end{align} 
    where $\mathbb{G}$ denotes the ‘geometric expectation’. The well-known HM-GM inequality then shows directly that 
\begin{align}
    \Gamma^{FG}_{SALEM}=\mathbb{H}[\Gamma_{\boldsymbol{s}}]\leq \mathbb{G}[\Gamma_{\boldsymbol{s}}]=\Gamma_{ExtLEM},
\end{align}
in accordance with Theorem \ref{thm:straight_line}. 
\end{example}

\subsection{CG-SALEM \label{Appendix: coarse graining}}

\subsubsection{Generalities}

The following corollary of Theorems \ref{thm1}-\ref{thm:straight_line} describes the shot overhead increase in SALEM due to both  coarse-graining and rejection. 

\begin{corollary}\label{cor1}{\bf (Shot overhead of coarse-grained SALEM)} 
In analogy with Theorem \ref{thm1}, the shot overhead of SALEM with the partition $\{S_{\boldsymbol{k}}\}$ of global syndromes and IV weights is given by 
$\Gamma_{SALEM}^{\{S_{\boldsymbol{k}}\}}=\mathbb{H}[\Gamma_{\boldsymbol{k}} ]$, up to negligible corrections. Under the assumptions of any of Theorems \ref{thm: prefix}-\ref{thm:straight_line}, 
\begin{align}
    \Gamma^{FG}_{SALEM}\leq \Gamma_{SALEM}^{\{S_{\boldsymbol{k}}\}}\leq \Gamma_{ExtLEM}.
\end{align}
 Moreover, rejecting a subset $S_{Rej}\in\{S_{\boldsymbol{k}}\}$ further increases the shot overhead to 
 \begin{align}
     \Gamma_{SALEM}^{\{S_{\boldsymbol{k}}\},Rej}=\mathbb{P}_{Acc}^{-1} \mathbb{H}[\Gamma_{\boldsymbol{k}}|Acc]\geq \Gamma_{SALEM}^{\{S_{\boldsymbol{k}}\}},\label{Eq: rejection shot overhead}
 \end{align}
where $\mathbb{P}_{Acc}=1-\mathbb{P}(S_{Rej})$ is the acceptance probability. An equality occurs if and only if $\Gamma_{\boldsymbol{s}}=\infty$ for all $\boldsymbol{s}\in S_{Rej}$, or $\mathbb{P}_{Acc}=1$.
\end{corollary}

Note that the increase in shot overhead due to coarse-graining would be mild if $\Gamma_{\boldsymbol{s}}$ are relatively uniform \textit{within} each subset $S_{\boldsymbol{k}}$, with significant variations only \textit{between} subsets. The shot overhead increase due to rejection would be mild if $\Gamma_{\boldsymbol{k}}\gg \Gamma_{\boldsymbol{k}'}$ for all rejected $\boldsymbol{k}$ and accepted $\boldsymbol{k}'$.

\subsubsection{Toy model: Binary SALEM with history-independent QPs\label{Appendix: bi-SALEM}}

\begin{figure*}[ht!]
\begin{centering}
\includegraphics[width=1\textwidth]{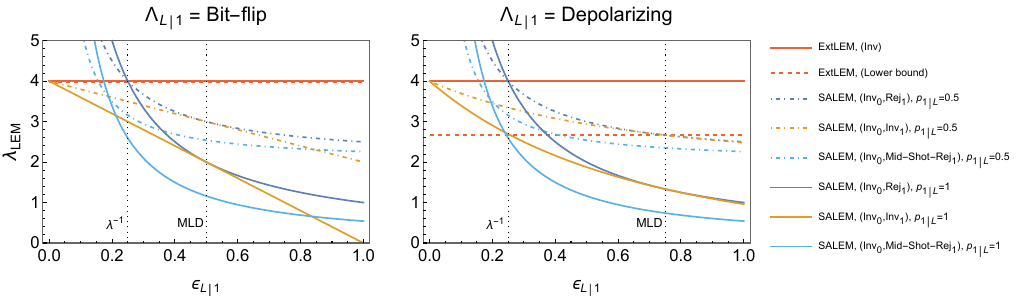}
\par\end{centering}
\caption{Blowup rate \(\lambda_{LEM}\) for three versions of binary
SALEM, compared to that of ExtLEM (red). All protocols are based on a local inversion with QPs (`Inv'), and we include lower bounds that apply to any ExtLEM protocol (red, dashed), but can be violated by SALEM (see Appendix~\ref{Appendix: lower bounds}). Binary SALEM is
defined by a partition $S=S_0\cup S_1$ of local syndromes. We
parameterize the blowup rates with 
\(\epsilon_{L|1}=\mathbb{P} (\text{logical error}|\ S_{1})\) ($x$-axis) and
\(p_{1|L}=\mathbb{P}(S_{1}|\ \text{logical error})\) (dot-dashed vs. solid SALEM lines). The `good' subset \(S_{0}\) is inverted, while the `bad' subset $S_1$ is rejected ($\text{Rej}_1$, blue), inverted ($\text{Inv}_1$, orange), or rejected mid-shot ($\text{Mid-Shot-Rej}_1$, cyan). For mid-shot rejection, we assume a circuit volume $V=\epsilon_L^{-1}$, see Appendix \ref{Appendix: mid-shot reject}. Left panel: $\Lambda_{L|1}$ is a bit-flip channel. Right: $\Lambda_{L|1}$ is a (single-qubit) depolarizing channel. ExtLEM and binary SALEM with rejection or mid-shot rejection are identical between left and right panels. Only the ExtLEM lower bound and $\text{Inv}_1$ depend on the details of $\Lambda_{L|1}$, beyond its infidelity $\epsilon_{L|1}$. Dotted vertical lines: The $\text{Rej}_1$ version of binary SALEM improves over ExtLEM only if $\epsilon_{L|1}>\lambda^{-1}=1/4$, and agrees with $\text{Inv}_1$ when $\Lambda_{L|1}$ is singular. For the bit-flip and depolarizing channels, this happens at the maximal $\epsilon_{L|1}$ possible under ML decoding, at 1/2 and 3/4, respectively. Larger values of $\epsilon_{L|1}$ correspond to decoding errors on $S_1$, which $\text{Inv}_1$ implicitly corrects without a shot overhead.  This figure can be used to understand the numerical blowup rates in Fig.~\ref{Fig: Bi-SALEM} in the main text, in light of the numerical values of $\epsilon_{L|1}$ and $p_{1|L}$ in that figure.\label{fig: blowup rate 2}}
\end{figure*}

As discussed in the main text, binary partitions are useful within CG-SALEM since they reduce and even eliminate the implementation challenges of finer partitions. Moreover, the binary case allows for an analytic computation of the SALEM blowup rate as a function of simple quantities,  providing intuition for the results presented in Appendices \ref{Appendix: FG-SALEM-QP}-\ref{Appendix: optimal SALEM} and in the main text.

To analytically compute the blowup-rate for CG-SALEM with a binary partition (`binary SALEM'), we restrict attention to the history-independent case, and start from Eq.~\eqref{Eq: Gamma_SALEM = W prod}. We further assume (i) a uniform circuit, where all logical gates are identical, and (ii) a binary partition of local syndromes $S=S_0\cup S_1$, such that 
\begin{align}
    \Gamma_{SALEM}^{\{S_0,S_1\}}=(\mathbb{H}[\Gamma_{k}])^V.
\end{align}
Here, $k\in \{0,1\}$ and $\Gamma_k=W^2(\Lambda_{L|k})$ is the shot overhead for mitigating the channel $\Lambda_{L|k}$, where $W(\Lambda_{L|k})=\sum_b|q_{b|k}|$ is the norm of a QP distribution $q_{b|k}$ satisfying $\Lambda_{L|k}^{-1}=\sum_b q_{b|k}\beta_b$, for some set of logical basis operations $\beta_b$. 

Assuming the mitigated channel $\Lambda$ is trace-preserving, we have the following lower bound on the norm of any QP representation of $\Lambda^{-1}$ \cite{QESEMpaper}, 
\begin{align}
    W^2(\Lambda)\geq\left(\frac{1+\epsilon_\Lambda}{1-\epsilon_\Lambda}\right)^2=1+4\epsilon_\Lambda+O(\epsilon_\Lambda^2),\label{Eq: OIB}
\end{align}
which shows that the optimal blow-up rate for QPs is $\lambda=4$. This is achieved by various known QP distributions, and we restrict attention to this `blowup-optimal' case, justifying Eq.~\eqref{Eq: g expansion}. 

Assuming $t$-FT logical gates,
we define the `good' set \(S_{0}\) to include all syndromes which can be
obtained due to a fault-path with weight \(\leq t\) (and possibly some syndromes that can be obtained due to higher-weight fault-paths), and
$S_{1}$ to be the complementary `bad' set. The probability
for measuring a syndrome in \(S_{1}\) is
\(p_{1}=\mathbb{P}\left( s \in S_{1} \right) = O\left( \epsilon^{t + 1} \right)\),
where \(\epsilon\) is the physical infidelity. The assumption of
\(t\)-FT implies \(\epsilon_{L} = O(\epsilon^{t + 1})\). Assuming that
\(\epsilon_{L} = \Theta(\epsilon^{t + 1})\), so no `accidental' $(t+1)$-FT occurs, we have
\begin{align}
    p_{1}  = 1-p_0= O(\epsilon_{L}).
\end{align}
It follows that
\begin{align}
    \epsilon_{L|0}=\mathbb{P}(error|\ s \in S_{0}) \leq \epsilon_{L}/p_{0}\  = O(\epsilon_{L}),
\end{align}
and therefore 
\begin{align}
    \Gamma_0=W^2(\Lambda_{L|0})=1+4\epsilon_{L|0}+O(\epsilon_L^2).
\end{align}
Note, however, that
\(\epsilon_{L|1}=\mathbb{P}(error|\ s \in S_{1}) \leq \epsilon_{L}/p_{1}\)
need not be small, and may take any value in \(\lbrack 0,1\rbrack\).
Generically, \(\epsilon_{L|1}\) will not be small, since the
leading contribution to \(S_{1}\) comes from weight-\((t + 1)\)
fault-paths. 

It is useful to parametrize \(\lambda_{SALEM}^{\{S_0,S_1\}}\) as a function of
\(\epsilon_{L|1}\) and \(p_{1|L}=\mathbb{P}(s \in S_{1}|\ error)\),
which is the fraction of logical errors contained in \(S_{1}\). Note
that \(p_{1} = \epsilon_{L}p_{1|L}/\epsilon_{L|1}\) (Bayes' theorem),
and \(\epsilon_{L} = \mathbb{E}[\epsilon_{L|k}]=p_{0}\epsilon_{L|0} + \ p_{1}\epsilon_{L|1}\). With these relations, we expand
\begin{align}
    &p_0\Gamma_0^{-1}+p_1 \Gamma_1^{-1}\\
    =&1-4\epsilon_{L|0}-p_1(1-\Gamma_1^{-1})+O(\epsilon_L^2)\nonumber\\
    =&1-\epsilon_L\left(4(1-p_{1|L})+\frac{1-\Gamma_1^{-1}}{\epsilon_{L|1}}p_{1|L}\right)+O(\epsilon_L^2),\nonumber
\end{align}
implying 
\begin{align}
    \lambda_{SALEM}^{\{S_0,S_1\}}&=\lim_{\epsilon_L\rightarrow0}\epsilon_L^{-1}\log\mathbb{H}[\Gamma_k]\\
    &=-\lim_{\epsilon_L\rightarrow0}\epsilon_L^{-1}\log\left(p_0\Gamma_0^{-1}+p_1 \Gamma_1^{-1}\right)\nonumber\\
    &=4(1-p_{1|L})+\frac{1-\Gamma_1^{-1}}{\epsilon_{L|1}}p_{1|L}.\nonumber
\end{align}
This can be understood as a weighted average of the blowup rate $\lambda_0=\lambda=4$ for mitigating $\Lambda_{L|0}$, and the blowup rate  $\lambda_1=(1-\Gamma_1^{-1})\epsilon_{L|1}^{-1}$ for mitigating $\Lambda_{L|1}$; with weights given by $1-p_{1|L}$ and $p_{1|L}$.  As a special case of Theorem \ref{thm:straight_line}, an advantage over ExtLEM is obtained when  
\begin{align}
    1/\Gamma_1=W^{-2}(\Lambda_{L|1})>1-4\epsilon_{L|1},\label{Eq: W1^-2>1-4eps1}
\end{align}
such that $\lambda_1<4$. The relation $W^{-2}(\Lambda)>1-4\epsilon_\Lambda$ holds for all (blowup-optimal) QP distributions we are aware of (see examples below), implying that $\lambda_{SALEM}^{\{S_0,S_1\}}$ is lower than $\lambda_{ExtLEM}=\lambda=4$. 

The blowup rate in case $S_1$ is rejected is obtained in the limit $\Gamma_1\rightarrow\infty$,
\begin{align}
    \lambda_{SALEM}^{\{S_0,S_1\},Rej}
    &=4(1-p_{1|L})+\epsilon_{L|1}^{-1}p_{1|L},
\end{align}
which again can be understood as a weighted average of blowup rates, where now $\lambda_1=\epsilon_{L|1}^{-1}$ is the blowup due to rejection. In accordance with Corollary \ref{cor1}, rejection always increases the blowup rate, unless $\Gamma_1=\infty$. An improvement over ExtLEM is not guaranteed, and requires $\epsilon_{L|1}>\lambda^{-1}=1/4$. 

\begin{example} {\bf (QP distributions for Pauli channels)} \label{Example: specific QP distributions}
The squared QP norm
\begin{align}
    W^{2}(\Lambda_{L|1}) = \left( 1 - 2\epsilon_{L|1}\right)^{- 2}
\end{align}
gives a very simple expression, 
\begin{align}
    \lambda_{SALEM}^{\{S_0,S_1\}} = 4(1 - p_{1|L}\epsilon_{L|1}).
\end{align}
This QP norm is obtained from a series expansion 
\begin{align}
    \Lambda_{L|1}^{-1}=\sum_{k=0}^{\infty}(I-\Lambda_{L|1})^k
\end{align}
which can be used to efficiently invert any Pauli
channel \(\Lambda_{L|1}\) with \(\epsilon_{L|1}< 1/2\). Moreover, the
same QP norm is
valid for all \(\epsilon_{L|1} \in [0,1]\) if
\(\Lambda_{L|1}= \left( 1 - \epsilon_{L|1}\right)I + \epsilon_{L|1}\sigma\)
is a Pauli channel with a single logical Pauli error \(\sigma\), since in this
case we may use the QP distribution
\begin{align}
    \Lambda^{- 1}_{L|1} = \frac{1}{1 - 2\epsilon_{L|1}}\left( \left( 1 - \epsilon_{L|1} \right)I - \epsilon_{L|1}\sigma \right)
\end{align}
At $\epsilon_{L|1}=1/2$ the single-Pauli error channel is not invertible, and the QP norm diverges. Note that \(W^{2} = \left( 1 - 2\epsilon_{L|1} \right)^{- 2}\) is convex and monotonically increasing on $[0,1/2)$, but not on $[0,1]$. However, $W^{-2}$ is convex on $[0,1]$, and in particular satisfies the weaker requirement Eq.~\eqref{Eq: W1^-2>1-4eps1}.

For \(\epsilon_{L|1} > 1/2\) this example corresponds to a `decoding
error' for dominant syndromes $s\in S_1$, since, assuming the optimal ML decoder, the identity
operator must have the highest probability in each 
\(\Lambda_{L|s}\), and therefore also in \(\Lambda_{L|1}=\sum_{s\in S_1} p_{s|1}\Lambda_{L|s}\). Accordingly, the overhead for inverting  \(\Lambda_{L|1}\) decreases in the range
\(1/2 < \epsilon_{L|1} \leq 1\), reducing to $W^2(\Lambda_{L|1})=1$ at $\epsilon_{L|1}$, where \(\Lambda_{L|1}^{- 1} = \sigma\) reduces to a deterministic recovery. Figure \ref{fig: blowup rate 2} (left panel)
summarizes this discussion. Generically, a large
\(\epsilon_{L|1}\) corresponds to a non-deterministic decoding error, and the QP
norm will be large in this regime, see Figure \ref{fig: blowup rate 2} (right panel) for an example based on the single-qubit depolarizing channel  \(\Lambda_{L|1}= \left( 1 - \epsilon_{L|1}\right)I + \epsilon_{L|1}/3(X+Y+Z)\), using the QP distribution
\begin{align}
   \Lambda^{- 1}_{L|1} = \frac{1}{1 - 4\epsilon_{L|1}/3}\left( \left( 1 - \frac{\epsilon_{L|1}}{3} \right)I - \frac{\epsilon_{L|1}}{3}(X+Y+Z)\right).
\end{align}
The squared QP norm is in this case 
\begin{align}
    W^2=\left(\frac{1+2\epsilon_{L|1}/3}{1-4\epsilon_{L|1}/3}\right)^2.
\end{align}
Similarly to the single-Pauli error case, $W^{-2}$ is convex on the entire range $\epsilon_{L|1}\in[0,1]$ (despite the singularity of $W^2$ at the "ML threshold" $\epsilon_{L|1}=3/4$), and in particular satisfies Eq.~\eqref{Eq: W1^-2>1-4eps1}.
\end{example}

\subsubsection{Mid-shot rejection\label{Appendix: mid-shot reject}}

Mid-shot (MS) rejection corresponds to the simple termination of
shots once the first rejected syndrome is measured, without waiting
for the shot to end. Assuming the shot time is monotonically increasing
in the circuit depth, this simple modification reduces the average
time per shot, and therefore improves the QPU time overhead relative to (post-shot) rejection, though the shot overhead is unchanged. Thus, even though rejection always increases the shot overhead within SALEM (Eq.~\eqref{Eq: rejection shot overhead}), MS rejection within SALEM can sometimes reduce the QPU time. 

For a quantitative analysis, we assume that the shot time is given
by $t_{s}=t_{g}t$, where $t_{g}$ is the time per logical layer, and
$t$ is the number of logical layers performed before the shot is rejected. Each logical layer contains $w$ logical gates (the `circuit width'), and the circuit volume is given by $V=wD$, where $D$ is the circuit depth (number of layers). Denoting the acceptance probability for each logical gate by $p_{acc}$, the average time per shot (in units of $t_{g}$) with MS rejection is 
\begin{align}
    \mathbb{E}[t]=Dp_{acc}^{wD}+(1-p_{acc}^w)\sum_{t=1}^{D}tp_{acc}^{w(t-1)}=\frac{1-p_{acc}^{wD}}{1-p_{acc}^w},
\end{align}
where $D$ is the circuit depth. 

The expected QPU time for an $N$-shot experiment is $T_{MS}=N\mathbb{E}[t]$,
which should be compared with the total time $T=ND$ without MS rejection. We therefore replace the shot overhead $\Gamma_{rej}=\mathbb{P}_{acc}^{-1}=p_{acc}^{-wD}$ for rejection, with the QPU time overhead 
\begin{align}
    \Gamma_{MS\ rej}=\mathbb{P}_{acc}^{-1}\mathbb{E}[t]/D=\frac{1-p_{acc}^{wD}}{Dp_{acc}^{wD}(1-p_{acc}^w)}
\end{align}
for MS rejection. To quantify
this in terms of blowup rates, we compute 

\begin{align}
\lambda_{MS\ rej} &= \frac{1}{V\epsilon_L p_{rej|L}}\log \Gamma_{MS\ rej}\label{Eq: lambda_MS}\\
&=\frac{1}{V\epsilon_{L}p_{rej|L}}\log\left(\frac{1-p_{acc}^{wD}}{Dp_{acc}^{wD}(1-p_{acc}^w)}\right)\nonumber\\
 &=\frac{1}{v p_{rej|L}}\log\left(\frac{\epsilon_{L|rej}}{p_{rej|L}v}\left(e^{\frac{p_{rej|L}v}{\epsilon_{L|rej}}}-1\right)\right)+O(\epsilon_L),\nonumber
\end{align}
where, in the last line, we set $D=v/w\epsilon_{L}$ and $p_{rej}=1-p_{acc}=p_{1|L}\epsilon_{L}/\epsilon_{L|1}$, before expanding in $\epsilon_L$.  Note that $v=V\epsilon_L$ can be thought of as a total logical infidelity, or as a normalized volume, and must be $O(1)$ for LEM to produce high accuracy estimators in reasonable time.

\begin{figure}[!th]
\begin{centering}
\includegraphics[width=1\columnwidth]{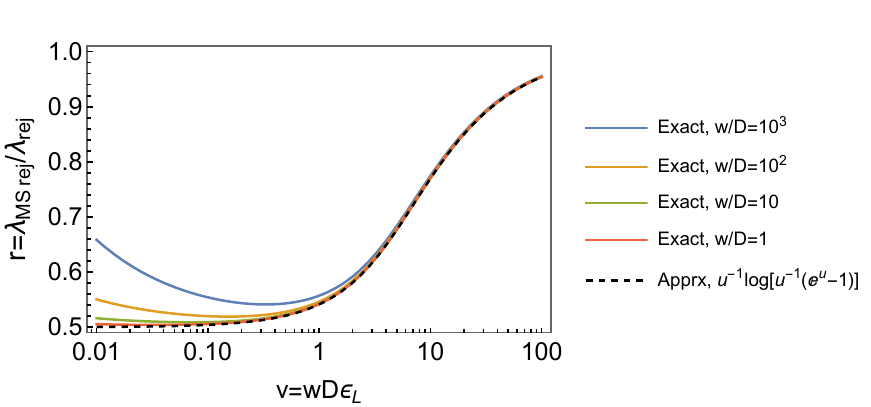}
\par\end{centering}
\caption{Ratio of the blow-up rates $\lambda_{MS\ rej}$ and $\lambda_{rej}$ for mid-shot rejection and (post-shot) rejection, as a function of the total infidelity $v=V\epsilon_L$, based on Eq.~\eqref{Eq: lambda_MS}-\eqref{Eq: r}. We set $\epsilon_L=10^{-6}, f_{rej|L}=\epsilon_{L|rej}=1/2$, though the qualitative behavior is unchanged as long as $\epsilon_L\ll1$ and $f_{rej|L}\sim\epsilon_{L|rej}$. The different colors indicate different aspect ratios $w/D$ (width/depth), while the black dashed line indicates a one-parameter, aspect-ratio-independent, approximation. Mid-shot rejection reduces the blowup rate due to rejection by a significant factor $\approx 0.5-0.8$ for challenging but feasible $v\sim1-10$, even for very  wide and shallow circuits. Moreover, the approximate expression is highly accurate in this practically relevant regime. 
\label{Fig: MS rejection}}
\end{figure}

Comparing $\lambda_{MS\ rej}$ to $\lambda_{Rej}=1/\epsilon_{L|1}$,
we define the ratio 
\begin{align}
r=\lambda_{MS\ rej}/\lambda_{rej}=u^{-1}\log\left[u^{-1}\left(e^{u}-1\right)\right]+O(\epsilon_L),\label{Eq: r}
\end{align}
which, at leading order in $\epsilon_{L}$, is a function of a single 
parameter $u=v p_{rej|L}/\epsilon_{L|rej}$. Since useful partitions of syndromes satisfy $\epsilon_{L|rej}\sim p_{rej|L}$ (see Fig.~\ref{Fig: Bi-SALEM}), one can think of $u$ as a proxy for $v$: $u\sim v$. The function in Eq.~\eqref{Eq: r} monotonically
increases from 1/2 to 1, as $u$ goes from $0$ to $\infty$, see
Fig.~\ref{Fig: MS rejection}. Thus, MS rejection can reduce the blowup rate due to rejection by a factor of up to $2$, depending on the value of $u$. 

One may be concerned that the leading order ratio $r(u)$ in Eq.~\eqref{Eq: r} is independent of the aspect ratio of the circuit, $w/D$, though it's clear that MS rejection cannot give an advantage over post-shot rejection if the aspect ratio is too large (for $D=1$, we clearly have $\Gamma_{MS\ rej}=\Gamma_{rej}$). Nevertheless, Fig.~\ref{Fig: MS rejection} shows that the leading order expression accurately captures the behavior of $r$ in the practically relevant regime of (i) circuits with aspect ratios as large as $\sim10^3$ (most quantum algorithms have aspect ratios $\ll 1$), and (ii) circuits with volume $v\sim1-10$, where the QPU time overhead for LEM is on one hand a limiting factor (and therefore important to reduce), but on the other hand feasible.

\subsection{Lower bounds for shot overheads in ExtLEM and SALEM\label{Appendix: lower bounds}}

Known lower bounds on the shot overhead of physical EM do not hold for error-corrected circuits, which involve adaptive operations \cite{takagi2022fundamental, Takagi2023universal, Tsubouchi2023universal, Quek_Eisert}. However, it is clear that these bounds do hold for ExtLEM, which mitigates $\Lambda_{L}$ while ignoring syndrome data. 
As an example, if $\Lambda_{L}$ takes the form of a layer of single-qubit depolarizing channels, we get 
\begin{align}
    \Gamma_{ExtLEM}\geq f_{bound}(\Lambda_L),
\end{align}
where $f_{bound}(\Lambda_L)=(1-4\epsilon_L/3)^{-2D}$, and $D$ is the circuit depth \cite{Takagi2023universal, Tsubouchi2023universal}.
Generalization to more general logical error channels is done by replacing $(1-4\epsilon/3)^2$ by an appropriate `contraction factor', see Appendix I in \cite{Takagi2023universal} and the parameter $\gamma^2$ in \cite{Tsubouchi2023universal}. For a layer of bit-flip channels, $f_{bound}(\Lambda)=(1-2\epsilon)^{-2D}$. For a general layer of single-qubit Pauli channels $\Lambda=(1-\epsilon)I+\epsilon (q_X X+q_Y Y+q_Z Z)$, $f_{bound}(\Lambda)=|1-2\epsilon\min_{i\neq j} (q_i+q_j)|^{-2D}$.

SALEM's use of syndrome data implies that the same lower bound does not hold for the SALEM estimator $\overline{o}$ (Eq.~\eqref{Eq: o-bar}). But it does hold for each $o_{\boldsymbol{s}}|N_{\boldsymbol{s}}$, giving $\Gamma_{\boldsymbol{s}}\geq f_{bound}(\Lambda_{L|\boldsymbol{s}})$. This implies a lower bound 
\begin{align}
    \Gamma_{SALEM}=\mathbb{H}[\Gamma_{\boldsymbol{s}}]\geq \mathbb{H}[f_{bound}(\Lambda_{L|\boldsymbol{s}})].\label{Eq: SALEM bound}
\end{align}
As in Appendix \ref{Appendix: advantage of SALEM over ExtLEM}, we can ensure that the SALEM bound is less restrictive than that of ExtLEM when $1/f_{bound}$ is convex:
\begin{align}
    \mathbb{H}[f_{bound}(\Lambda_{L|\boldsymbol{s}})]\leq f_{bound}(\Lambda_L).
\end{align}
This happens e.g., in the depolarizing and bit-flip cases (and more generally when some $q_i=0$). For layers of general single-qubit Pauli channels, $1/f_{bound}$ is convex if $\epsilon<3/4$, and violates convexity very mildly otherwise. We note that the use of ML decoding ensures that the probability for the identity is larger (or equal) to that of non-identity Paulis, which implies $\epsilon<3/4$. 

The same approach can be applied to obtain tighter lower bounds for SALEM and ExtLEM based on specific EM protocols, such as Eq.~\eqref{Eq: OIB} (which satisfies our convexity requirement). 

To summarize, we find that SALEM can break lower bounds that hold for physical EM and ExtLEM with any EM protocol. However, these lower bounds can be used to derive weaker lower bounds (Eq.~\eqref{Eq: SALEM bound}) that do hold for SALEM. 

An interesting open question is whether SALEM is the optimal LEM scheme based on syndrome data. E.g., does the lower bound $\mathbb{H}[f_{bound}(\epsilon_{L|\boldsymbol{s}})]$ hold for any LEM protocol that makes use of syndrome data?

\subsection{FT pseudo thresholds for ExtLEM and SALEM \label{Appendix: thresholds}}

We explain here the behavior of the FT (pseudo) thresholds for ExtLEM and SALEM described in the main text (Fig.~\ref{fig: performance}(c)). The standard FT (pseudo) threshold is defined as the solution to $\epsilon_L(\epsilon)=\epsilon$. In contrast, the ExtLEM threshold is defined by equating the QPU time overheads of (physical) EM and ExtLEM:
\begin{align}
    V_{EC}e^{\lambda \epsilon_L(\epsilon)V}=\Gamma_{ExtLEM}=\Gamma_{EM}=e^{\lambda \epsilon V},
\end{align}
where $V_{EC}$ is the space-time overhead for EC, discussed in Appendix~\ref{Appendix: surface code simulations}, and we adopt an exponential form for the shot overhead, for simplicity. This can be written as 
\begin{align}
    \frac{\log V_{EC}}{\lambda V}+ \epsilon_L(\epsilon)=\epsilon,
\end{align}
which is identical to the defining equation of the FT threshold, but with a $V$-dependent addition to the logical error. It is therefore clear that the ExtLEM threshold is always lower than the FT threshold, and converges to it as $V\rightarrow \infty$. Similarly, the equation defining the SALEM threshold is
\begin{align}
    V_{EC}e^{\lambda_{SALEM} \epsilon_L(\epsilon)V}=\Gamma_{SALEM}=\Gamma_{EM}=e^{\lambda \epsilon V},
\end{align}
such that 
\begin{align}
    \frac{\log V_{EC}}{ \lambda V}+ \frac{\lambda_{SALEM}}{\lambda}\epsilon_L(\epsilon)=\epsilon.
\end{align}
Since $\lambda_{SALEM}<\lambda$, it's clear that the SALEM threshold is always higher than the ExtLEM threshold and that, as $V\rightarrow\infty$, we obtain a SALEM threshold which is higher than the FT threshold. On the other hand, at small volume, the first term dominates. To identify the transition, parametrize $V=v/\epsilon_L(\epsilon)$. The SALEM threshold is higher than the FT threshold if $v>v_0=\log V_{EC}/(\lambda-\lambda_{SALEM})$. As discussed in Appendix \ref{Appendix: surface code simulations}, for surface codes of distance $d$ we have $V_{EC}=5(2d^2-1)$.
Since $\log(V_{EC})$ is, say, $<10$ up to very high code distances $\approx60$, and considering values of $\lambda-\lambda_{SALEM}>1$ presented in the main text and in Appendix \ref{Appendix: surface code simulations}, it is expected that volumes $V>v_0/\epsilon_L$, where the SALEM threshold is higher than the standard threshold, are feasible for SALEM with reasonable QPU time overhead. Note that in Fig.~\ref{fig: performance}(c) we used the convention $V=v/\epsilon$ as opposed to $V=v/\epsilon_L(\epsilon)$.

\subsection{Numerical simulations\label{Appendix: numerical simulations}}

\subsubsection{Steane code \label{Sec: color code simulations}}

We use Steane's \( [\![7,1,3]\!] \) code to fully simulate the methods discussed above for deep memory circuits on a single logical qubit.

For syndrome extraction, we employ the 1-FT scheme of Ref.~\cite{Chao_2018}. 
This scheme uses one ancilla qubit for syndrome measurement and an additional ancilla qubit to flag hook errors.
The measurement procedure is adaptive and consists of flagged and unflagged rounds. 
In the first round, syndromes are measured sequentially using flagged gadgets. 
If a flag or nontrivial syndrome is detected, the round is terminated and \emph{all} syndromes are subsequently measured using unflagged gadgets. 
If no such event occurs, the second round is skipped. 
Recovery is performed using a lookup table (LUT) that applies minimal-weight decoding conditioned on which gadget triggered the flag.
Since there are six stabilizer generators \(|\mathcal{P}_7/\mathcal S|=4^7/2^6=256\).
The flag register takes 7 possible values: one value indicating that no flag was raised, and six values specifying which stabilizer gadget raised the flag. Thus, there are \(7\cdot 2^6 = 448\) possible flagged syndrome outcomes.
Let us introduce some useful notation. 
Denote by
\(\mathcal{R}_{\mathcal{S}}\) a set of representatives for the cosets
\(\mathcal{P}_7/\mathcal{S}\) of the stabilizer group. In the notation of
Appendix~\ref{Appendix: setup}, we have
\(\mathcal{R}_{\mathcal{S}}=\{\sigma_{\mathrm{corr}}^a\sigma_{\mathrm{log}}^b\}_{a,b}\).
To avoid cluttering the equations, throughout this section we will denote the coset representatives by \(\sigma\).
We will choose the coset representative of \(\mathcal{S}\) to be \(I \in \mathcal{P}_7\).
Further, denote by \(G\) the faulty error-correction cycle. 

Our circuit-level noise model consists of a two-qubit depolarizing channel applied after each CNOT 
with probability \(\epsilon\), an \(X\) error after each reset operation with probability \(\epsilon/2\), and an identical \(X\) error before each ancilla measurement. 
We assume the initial logical state preparation and final logical measurement in the circuit are fault-free.

For a small number of syndrome subsets, P2LC can generally be efficiently implemented by sampling fault-paths, as described in \cite{qedma_logical_errors_2024} and done in Appendix \ref{Appendix: surface code simulations}. 
However, the present case allows for a brute-force enumeration of error-fault combinations (EFCs), 
due to the low FT level, \(t=1\), the relatively small volume of the syndrome-extraction gadgets, and the small number of stabilizer group cosets and total number of syndromes. 
Here an EFC is a pair \((\sigma_{\mathrm{in}},f)\), where \(\sigma_{\mathrm{in}}\) labels the input error coset and \(f\) is a set of internal faults. The probability assigned to the EFC is the probability \(\mathbb{P}(f)\) of the internal faults; \(\sigma_{\mathrm{in}}\) is treated as a conditioned input. 
The \emph{length} of an EFC is defined as
\(|f|+\delta_{\sigma_{\mathrm{in}}\ne I},\)
namely, the number of internal faults, plus one if the conditioned input error is nontrivial.

We enumerate all EFCs of length \(\le 2\),
as described in Algorithm~\ref{alg:bf_enum_faults}, to compute the \emph{``transition tensor''}:
\begin{definition}[Transition tensor]
Let \(G=\sum_{s\in S}G_s\) be the faulty error-correction cycle written as
an instrument, where \(s\) is the measured flagged syndrome outcome. For any
\(\sigma_{\mathrm{in}}\in\mathcal{R}_{\mathcal S}\) and any code state
\(\sket{c}\), we define
\begin{align}
    G_s\sigma_{\mathrm{in}}\sket{c}
    =
    \left(
    \sum_{\sigma_{\mathrm{out}}\in\mathcal{R}_{\mathcal S}}
    \cT_{s,\sigma_{\mathrm{out}}}^{\sigma_{\mathrm{in}}}
    \sigma_{\mathrm{out}}
    \right)\sket{c}.
    \label{eq:action_of_E}
\end{align}
Thus, \(\cT_{s,\sigma_{\mathrm{out}}}^{\sigma_{\mathrm{in}}}\) is the joint
probability that the EC cycle, initialized with input error
\(\sigma_{\mathrm{in}}\), produces syndrome \(s\) and output error
\(\sigma_{\mathrm{out}}\).
\end{definition}

For every EFC, we compute the measured syndrome \(s\) and the final output
error \(\sigma_{\mathrm{out}}\), and add the probability of that EFC to
\(\cT^{\sigma_{\mathrm{in}}}_{s,\sigma_{\mathrm{out}}}\).
The omitted probability mass from EFCs of length \(>2\) is referred to below as the
missing probability.

\begin{algorithm}[H]
  \caption{Construction of the transition tensor by brute-force enumeration}\label{alg:bf_enum_faults}
  \label{Algo: transition-tensor-enumeration}
  \begin{flushleft}
    \Input A set of chosen coset representatives \(\mathcal{R}_{\mathcal S}\);
    the set of EFCs \(\mathcal{F}_{\le w}\) of length at most \(w\), 
  \end{flushleft}
  \begin{algorithmic}[1]
    \State Initialize \(\cT^{\sigma_{\mathrm{in}}}_{s,\sigma_{\mathrm{out}}} \leftarrow 0\).
    \For{each error-fault combination \((\sigma_{\mathrm{in}}, f) \in \mathcal{F}_{\le w}\)}
      \State Propagate \((\sigma_{\mathrm{in}}, f)\) through the circuit to obtain \(s\) and \(\sigma_{\mathrm{out}}\).
      \State Update \(\cT\) via
      \[
        \cT^{\sigma_{\mathrm{in}}}_{s,\sigma_{\mathrm{out}}}
        \leftarrow
        \cT^{\sigma_{\mathrm{in}}}_{s,\sigma_{\mathrm{out}}}
        +
        \mathbb{P}(f).
      \]
    \EndFor
    \State \Return \(\cT\).
  \end{algorithmic}
\end{algorithm}

We first show how to compute the channels \(\tilde{\Lambda}^{(j)}_{L}\) which approximate the exact channels $\Lambda^{(j)}_L$.
Following the P2LC procedure in Appendix~\ref{Sec: logical channels}, we proceed as follows. 
In the memory case we have $G_j = G$ and $g_j = I$, for $1\leq j \leq V$. Moreover, in our setup, there are no state preparation and encoding errors so the first cycle does not have input errors. 

We represent the channels
\(\Lambda^{(j)}_{\mathrm{in}} : \mathcal{K} \to \mathcal{H}\),
\(\Lambda^{(j)}_{\mathrm{out}} : \mathcal{K} \to \mathcal{H}\), and
\(\Lambda^{(j)}_{L} : \mathcal{K} \to \mathcal{K}\)
as the sums
\(\sum_{\sigma \in \mathcal{R}_{\mathcal{S}}} p^{(j)}_{\mathrm{in}, \sigma} \sigma\),
\(\sum_{\sigma \in \mathcal{R}_{\mathcal{S}}} p^{(j)}_{\mathrm{out}, \sigma} \sigma\), and
\(\sum_{\sigma \in \mathcal{R}_{\mathcal{S}}} p^{(j)}_{L, \sigma} \sigma\), respectively.
For logical channels, the coefficients \(p_{L,\sigma}\) are nonzero only for
representatives whose correctable component is trivial.
Since the channels act on the code space \(\mathcal{K}\), this representation is independent of the choice of
\(\mathcal{R}_{\mathcal{S}}\).

We thus have, as \(\Lambda^{(1)}_{\mathrm{in}} = I\), 
\begin{align}\label{Eq: prev out}
\Lambda^{(1)}_{\mathrm{out}}\sket{c} = G \sket{c},
\end{align}
which gives by Eq.~\eqref{eq:action_of_E},
\begin{align}\label{Eq: memory I->out}
\Lambda^{(1)}_{\mathrm{out}} \sket{c}= 
\sum_{s,\sigma}  \cT^{I}_{s,\sigma } \sigma\sket{c}  \quad ;
\quad  p^{(1)}_{\mathrm{out},\sigma} = \sum_{s}  \cT^{I}_{s,\sigma}.
\end{align}
We now extract from \(\Lambda^{(1)}_{\mathrm{out}} \) a logical error channel by applying the noiseless version of $G$,
\begin{align}
    \Lambda^{(1)}_L \sket{c} =& G_\mathrm{ideal} \Lambda^{(1)}_{\mathrm{out}} \sket{c} 
    \quad ; \\
    p^{(1)}_{L,\sigma} =& \sum_{\sigma': G_{\mathrm{ideal}}\sigma'\sket{c}= \sigma\sket{c}} p^{(1)}_{\mathrm{out},\sigma'}.
\end{align}
To proceed and compute \(\tilde{\Lambda}^{(2)}_L\) we need to compute the input channel to the second gate via a convolution
\begin{align}
    \Lambda^{(2)}_{\mathrm{in}} \sket{c} = &
     \Lambda^{(1)}_{\mathrm{out}} (\Lambda^{(1)}_L)^{-1} \sket{c}
     \quad
     ;
     \quad
     p^{(2)}_{\mathrm{in},\sigma} = \sum_{\sigma'} p^{(1)}_{\mathrm{out}, \sigma\cdot \sigma'} q_{\sigma'},
\end{align}
where \( (\Lambda^{(1)}_{L} )^{-1}= \sum_{\sigma} q_{\sigma} \sigma \). 
Hence we obtain
\begin{align}
    \Lambda^{(2)}_{\mathrm{out}} \sket{c} = G \Lambda^{(2)}_{\mathrm{in}} \sket{c}
     \quad
     ;
     \quad
     p^{(2)}_{\mathrm{out},\sigma} = \sum_{s,\sigma'} \cT^{\sigma'}_{s,\sigma} p^{(2)}_{\mathrm{in}, \sigma'}.
\end{align}
Finally, we extract \(\Lambda^{(2)}_{L}\), and use the time-local approximation as described in Algorithm~\ref{Algo: P2LC} to get
\begin{align}\label{eq:ch_ext_lem_memory}
\tilde{\Lambda}^{(j)}_{L}=
\begin{cases}
\Lambda^{(1)}_L \text{ if } j=1, \\
\Lambda^{(2)}_L \text{ if } j>1.
\end{cases}
\end{align}

\begin{figure}[ht!]
\begin{centering}
\includegraphics[width=0.45\textwidth]{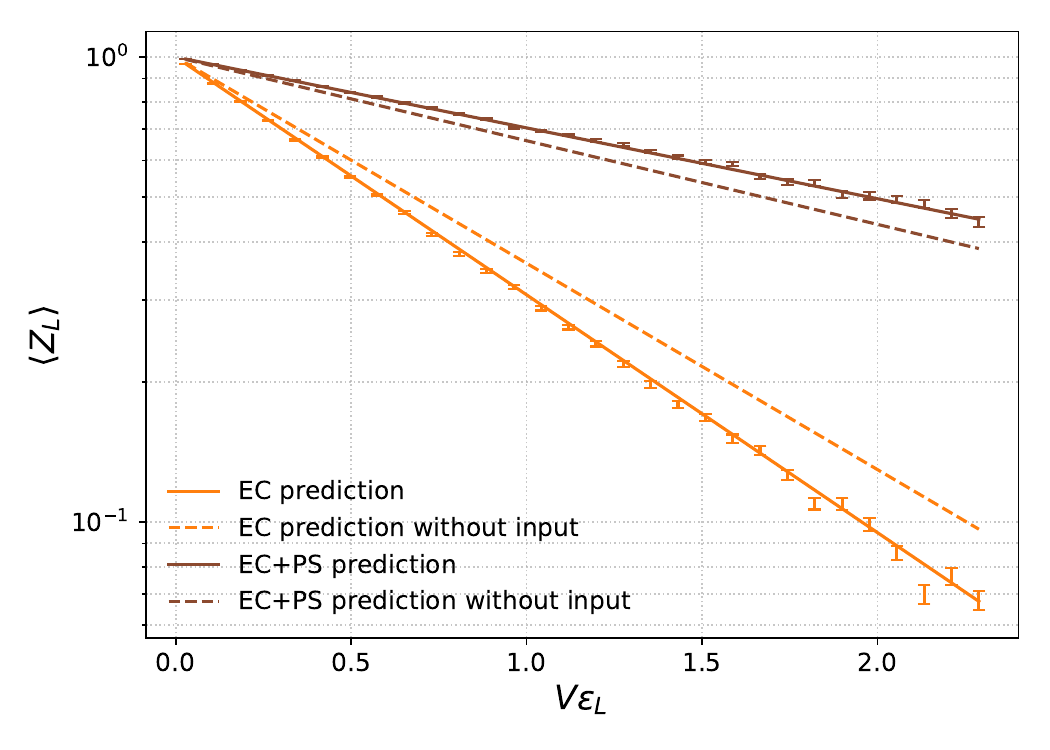}
\par\end{centering}
\caption{
Here we demonstrate the precision of the time-local characterization by comparing simulation results with our characterization predictions. 
Error bars indicate the simulated values; 
solid curves show the predictions using our logical channels: Eq.~\eqref{eq:ch_ext_lem_memory} for EC and Eq.~\eqref{eq:ch_ps_memory} for EC+PS; 
dashed curves show the predictions when neglecting the input error, that is taking \(\Lambda^{(1)}_{L}\) for EC and \( \Lambda_{L|0} \) for EC+PS.
The dash–dot curve illustrates how neglecting input errors leads to a substantial over- or under-estimation of the decay rates.
}\label{fig:decays_steane}
\end{figure}

For the binary SALEM results in this section, we will use a binary syndrome partitioning \(S = S_0 \cup S_1 \) with
\[
S_1 = \{\, s : \epsilon_{L|s} > \tau \,\}.
\]
where here \(\epsilon_{L|s}\) is the logical infidelity for \(G\) applied to a code state. 
We consider three variants of binary SALEM.
The first, which we call
\((\mathrm{Inv}_0,\mathrm{Rej}_1)\), rejects shots if any syndrome lies in \(S_1\) and inverts the logical noise channels in the remaining shots.
In the second variant, \((\mathrm{Inv}_0,\mathrm{Inv}_1)\), we do not reject any shots and instead invert all logical channels conditioned on the corresponding global syndrome.
The third variant, \(\invmsrej\), uses the same characterized accepted channels as
\(\invrej\), the only difference is that shots are terminated as soon as a syndrome in \(S_1\) is observed.

We consider syndrome histories of length at most \(\ 2\) to obtain time-local approximate channels \(\tilde{\Lambda}^{(j)}_{L|\boldsymbol{k}}\). 
Algorithm~\ref{Algo: syndrome-history-memory} shows how to compute the logical channel corresponding to an arbitrary syndrome history  \( (k_1, \dots, k_l) \) where the state before \(k_1\) is in the code space, which we denote by \(\Lambda_{L|k_1,\dots,k_l}\).

\begin{algorithm}[H]
  \caption{Characterization of a syndrome history for memory}
  \label{Algo: syndrome-history-memory}

  \begingroup
  \small
  \linespread{0.9}\selectfont
  \setlength{\abovedisplayskip}{2pt}
  \setlength{\belowdisplayskip}{2pt}
  \setlength{\abovedisplayshortskip}{1pt}
  \setlength{\belowdisplayshortskip}{1pt}

  \begin{algorithmic}[1]
    \Require Transition tensor
    \(\cT^{\sigma_{\mathrm{in}}}_{s,\sigma_{\mathrm{out}}}\);
    subset history \((k_1,\ldots,k_l)\); syndrome subsets \(S_k\).
    \Ensure Logical channel
    \(\Lambda_{L|k_1,\ldots,k_l}=\sum_{\sigma}p_{L,\sigma}\sigma\)
    of the final EC cycle, conditioned on \((k_1,\ldots,k_l)\).

    \State Set \(p_{\mathrm{in},\sigma}=\delta_{\sigma,I}\).

    \For{\(j=1,\ldots,l\)}
      \State Compute
      \[
        P(s,\sigma_{\mathrm{out}})
        =
        \sum_{\sigma}
        p_{\mathrm{in},\sigma}
        \cT^{\sigma}_{s,\sigma_{\mathrm{out}}}.
      \]

      \State Condition on \(s\in S_{k_j}\):
      \[
        p=\sum_{s\in S_{k_j},\sigma}P(s,\sigma),
        \qquad
        p_{\mathrm{out},\sigma}
        =
        p^{-1}\sum_{s\in S_{k_j}}P(s,\sigma).
      \]

      \State Map the output channel to a logical channel:
      \[
        p_{L,\sigma}
        =
        \sum_{\sigma':\,G_{\mathrm{ideal}}\sigma'\sket{c}
        =
        \sigma\sket{c}}
        p_{\mathrm{out},\sigma'}.
      \]

      \If{\(j<l\)}
        \State Invert \(\Lambda_L=\sum_{\sigma}p_{L,\sigma}\sigma\),
        writing \(\Lambda_L^{-1}=\sum_{\sigma}q_{\sigma}\sigma\), and update
        \[
          p_{\mathrm{in},\sigma}
          =
          \sum_{\sigma'}
          p_{\mathrm{out},\sigma\cdot\sigma'}q_{\sigma'}.
        \]
      \EndIf
    \EndFor

    \State \Return \((p_{L,\sigma})_{\sigma\in\mathcal{R}_{\mathcal S}}\).
  \end{algorithmic}

  \endgroup
\end{algorithm}

\begin{figure}[ht!]
\begin{centering}
\includegraphics[width=1\columnwidth]{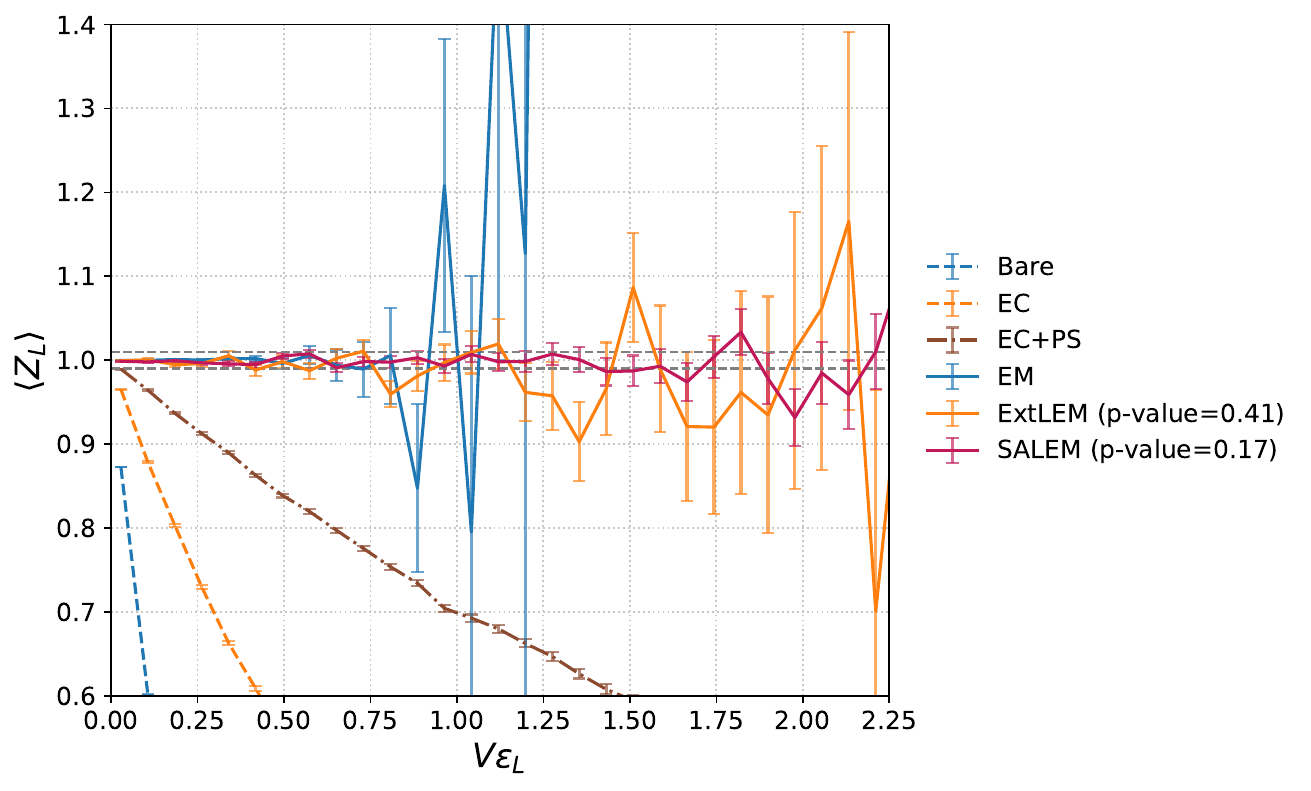}
\par\end{centering}
\caption{
Analog of Fig.~\ref{fig: performance}(a) for Steane's code. We use
\(\epsilon=4\cdot10^{-4}\), corresponding to
\(\epsilon_L=1.12\cdot10^{-4}\), and consider memory circuits with lengths
\(V\) between \(256\) and \(20480\).
We simulate $10^5$ shots for each value of $V$.
The ``SALEM'' curve corresponds to the $\invrej$ variant. It exhibits a significantly lower error than ExtLEM and remains unbiased within statistical uncertainty.
The rest of the curves are computed in the same way as in Fig.~\ref{fig: performance}(a). 
}\label{fig: <z> vs vol eps steane}
\end{figure}

For the $(\mathrm{Inv}_0,\mathrm{Rej}_1)$ variant we apply Algorithm~\ref{Algo: syndrome-history-memory} for the syndrome histories \((0)\) and \((0,0)\) to obtain \( \Lambda_{L|0} \) and \( \Lambda_{L|0,0} \),
and use
\begin{align}\label{eq:ch_ps_memory}
\tilde{\Lambda}^{(j)}_{L|\boldsymbol{k}}=
\begin{cases}
\Lambda_{L|0} \text{ if } j=1, \\
\Lambda_{L|0,0 } \text{ if } j>1.
\end{cases}
\end{align}

For $(\mathrm{Inv}_0,\mathrm{Inv}_1)$ variant we also consider history of length up to 2, and apply Algorithm~\ref{Algo: syndrome-history-memory} for the syndrome histories \((0), (1), (0,0), (0,1), (1,0), (1,1)\).
\begin{align}\label{eq:ch_hist_len_2_memory}
\tilde{\Lambda}^{(j)}_{L|\boldsymbol{k}}=
\begin{cases}
\Lambda_{L|k_1} \text{ if } j=1, \\
\Lambda_{L|k_{j-1},k_{j} } \text{ if } j>1.
\end{cases}
\end{align}

\begin{figure*}[ht!]
\begin{centering}
\includegraphics[width=0.9\textwidth]{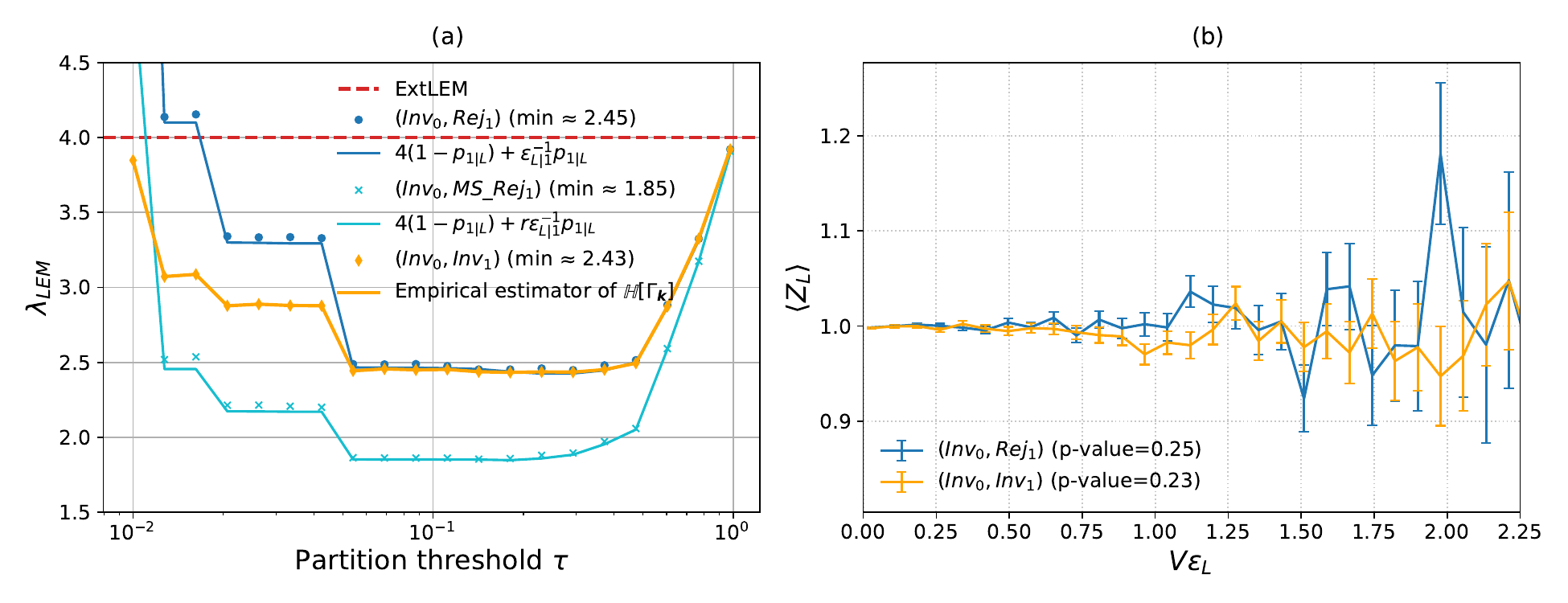}
\par\end{centering}
\caption{
Results for one logical qubit memory for Steane's code.
We assume that the initial logical state preparation and final logical measurement are fault-free. 
In all the simulations $\epsilon = 4\cdot 10^{-4}$ which gives $\epsilon_L=1.12\cdot 10^{-4}$.
(a) Empirical blowup rates for three SALEM variants. These are measured by initializing the logical qubit in the state \(\ket{+_L}\).
We find a good agreement between the dashed lines, which correspond to the measured values, and the analytical solid curves. 
Here $r$ is the factor defined in Eq.~\eqref{Eq: r}.
We find optimal blowup rates at $\tau \approx 0.2$ for all variants, with a significant improvement from mid-shot rejection, further reducing the optimal blowup rate from 2.43 to 1.86.
(b)
Mitigation results at the non-optimal threshold of \(\tau = 0.03\).
We consider an initial logical state $|0_L \rangle$.
At high values of \(\epsilon_L V\), the blowup rates are 2.84 and 3.28 for $\invinv$ and $\invrej$, respectively, consistent with the values in (a).
\label{fig: lamlem and miti non opt}
}
\end{figure*}

Fig.~\ref{fig:decays_steane} compares the decay of $\langle Z_L \rangle$ predicted by the characterized 
with full simulations of faulty error-corrected memory circuits. We show the decay for the EC computation which we predict using the channels in Eq.~\eqref{eq:ch_ext_lem_memory}, and the decay for the post-selection on all \(0\)s using the channels in Eq.~\eqref{eq:ch_ps_memory}. 
As one can see, the prediction agrees with the simulation within statistical error. 
The importance of taking into account the input errors to EC cycles is illustrated by the dashed lines, 
which show the corresponding predictions when the input to the EC cycles is ignored.

We have implemented binary SALEM using our P2LC channels for the variants $(\mathrm{Inv}_0,\mathrm{Rej}_1)$, $(\mathrm{Inv}_0,\mathrm{Inv}_1)$ and $(\mathrm{Inv}_0,\mathrm{MS\_Rej}_1)$ with a full stabilizer simulation of the error-corrected computation (using \texttt{Stim} \cite{stim}).
In all runs we used \(\epsilon = 4\cdot 10^{-4}\) which gives \(\epsilon_L = 1.12 \cdot 10^{-4}\) corresponding to the channel \(\Lambda^{(2)}_{L}\) in Eq.~\eqref{eq:ch_ext_lem_memory}.

\begin{figure*}[ht!]
\begin{centering}
\includegraphics[width=1\textwidth]{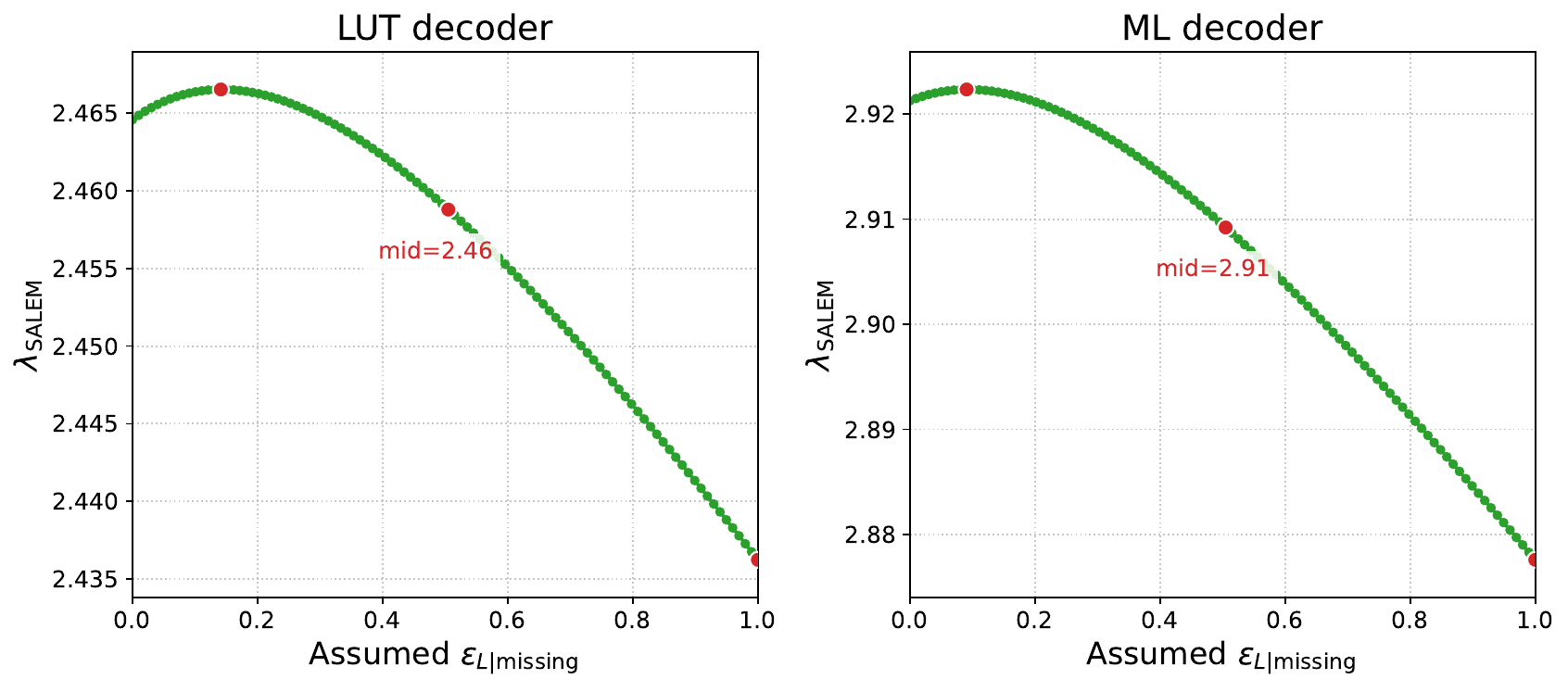}
\par\end{centering}
\caption{Estimation of $\lambda_{SALEM}^{FG}$ for Steane's code.
We use a brute-force enumeration of length $\leq 2$ EFCs, and treat the missing probability, due to higher-length EFCs, as corresponding to a single additional syndrome, with unknown conditioned logical error $\epsilon_{L|missing}\in[0,1]$. 
Red dots mark the minimal and maximal values obtained for $\lambda_{SALEM}^{FG}$, as well as the values obtained for $\epsilon_{L|missing}=1/2$, assuming either the LUT decoder, or the ML decoder obtained by taking the logical error channel obtained for the LUT decoder, and applying the most probable logical error. 
The small changes in $\lambda_{SALEM}^{FG}$ as a function of $\epsilon_{L|missing}$ (note the scales of $y$-axes) indicate that higher-length EFCs have a negligible effect.\label{fig: lam eps miss steane}
}
\end{figure*}

Fig.~\ref{fig: lamlem and miti non opt}(a) shows the empirical blowup rates of these variants as a function of the partition threshold \(\tau\).
The measured optimal blowup rate for $\invrej$ is $2.45$, and for $\invinv$ is $2.43$. Therefore, in this setup, 
inversion of logical error channels corresponding to histories with a syndrome in \(S_1\) gives only a negligible reduction in the optimal blowup rate, compared to rejection.
Mid-shot rejection gives a significant reduction of the blowup rate to $1.86$ for $\invmsrej$.
We observe good agreement between the measured 
blowup rates and the blowup rates predicted from our P2LC channels.
Fig.~\ref{fig: lamlem and miti non opt}(b) shows mitigation results at the non-optimal partition threshold of \(\tau = 0.03\), for which $(\mathrm{Inv}_0,\mathrm{Inv}_1)$ has a lower blowup rate than $(\mathrm{Inv}_0,\mathrm{Rej}_1)$.
The p-values indicate that no statistically significant bias is resolved for
\(\invinv\).
This shows that P2LC can produce channels that are accurate enough for binary SALEM, 
including channels for syndrome histories involving \(S_1\) syndromes.

Fig.~\ref{fig: <z> vs vol eps steane} shows results analogous to Fig.~\ref{fig: performance}(a) but with a full stabilizer simulation of the error-corrected circuit (using \texttt{Stim}). 
The ``SALEM'' curve corresponds to \(\invrej\) with syndrome partition threshold of $\tau=0.2$ as found from the simulation in Fig.~\ref{fig: lamlem and miti non opt}(a).
The estimated values are consistent with the ideal value of \(\langle Z_L\rangle=1\) and exhibit a significantly lower error than ExtLEM.

The optimal blowup rate of 2.43 achieved by $(\mathrm{Inv}_0,\mathrm{Inv}_1)$ is consistent with the predicted optimal value obtained for FG-SALEM with LUT decoder as depicted on the left in Fig.~\ref{fig: lam eps miss steane}. 
Here we get a range of blowup rates corresponding to the unknown logical errors for the missing probability.
The small variation of about 0.03 indicates that the effect of the missing EFCs is negligible.

\subsubsection{Surface codes\label{Appendix: surface code simulations}}

To study the interplay of SALEM with decoding, as well as its scaling to
larger-distance codes, we consider logical memory circuits for the family
of distance-$d$ (rotated) surface codes, for which several decoders are
publicly available. Specifically, we use the fault-tolerant memory circuits
implemented in \texttt{Stim} \cite{stim}, with $d$ rounds of syndrome
measurement, together with the compatible minimum-weight perfect matching
(MWPM) decoder provided by \texttt{PyMatching}~\cite{pymatching}, and the circuit-level 
tensor-network (TN) decoder constructed in
Ref.~\cite{piveteau2023tensornetworkdecoding}.

We consider circuit-level noise models throughout, where each physical single-qubit (two-qubit) gate is followed by a single-qubit (two-qubit) depolarizing channel, and each physical $Z$-basis reset (measurement) operation is followed (preceded) by a bit-flip channel, where all channels have the same physical error rate $\epsilon$. 

In contrast to our detailed simulations of the Steane code presented in
Appendix~\ref{Sec: color code simulations}, the surface-code circuits
considered here involve two simplifying assumptions: (i) Only flips of a
single logical operator ($Z_L$) are measured, reducing the logical error channel to a logical bit-flip channel, and (ii) a final ideal round
of syndrome measurement is performed, the outcomes of which are fed to the decoder. This latter assumption removes correctable errors
that would otherwise serve as input errors to subsequent rounds of EC. These simplifying  assumptions do not allow for  simulations of deep memory circuits with repeated error correction, but they suffice for estimating the shot overheads in various versions of SALEM, as well as the overhead and bias in EC and EC+PS.

\begin{figure*}[ht!]
\begin{centering}
\includegraphics[width=1\textwidth]{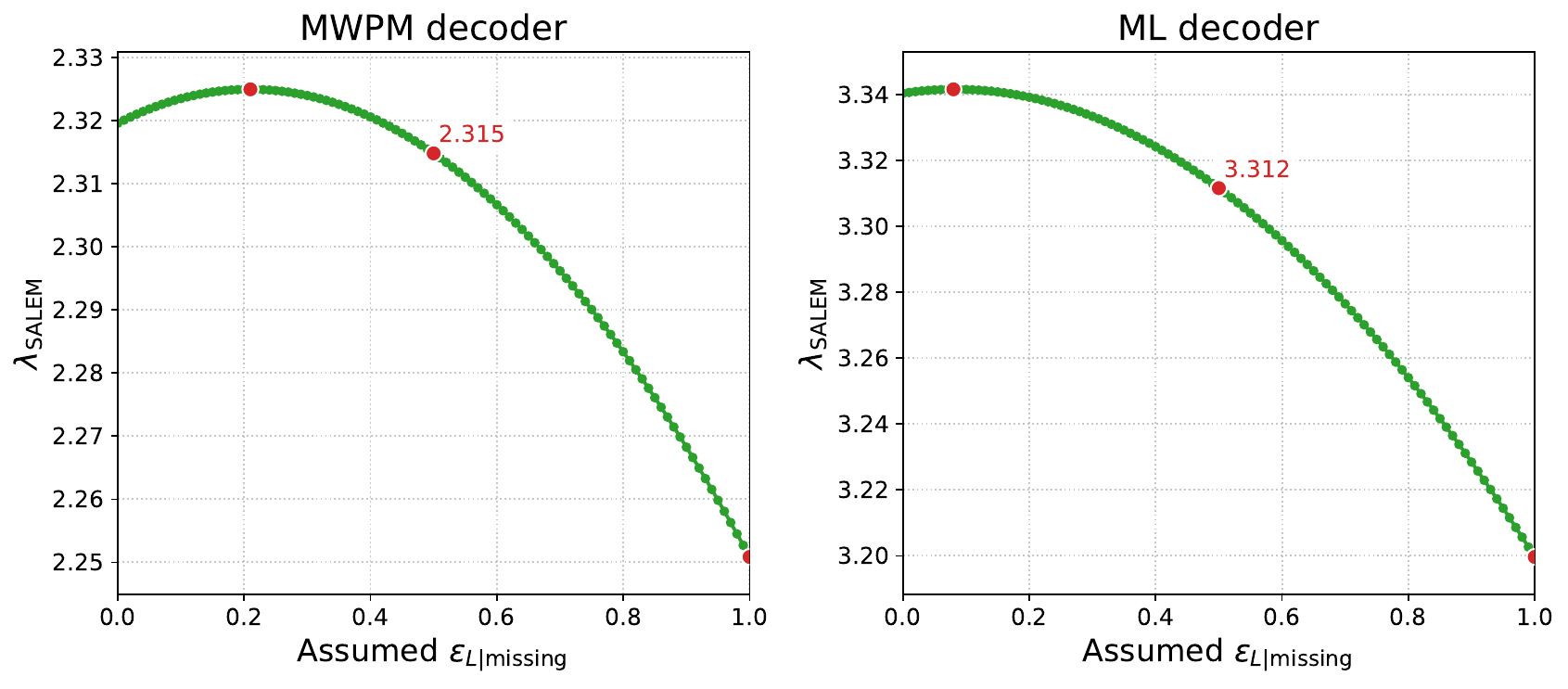}
\par\end{centering}
\caption{Estimating $\lambda_{SALEM}^{FG}$ for the $d=3$ surface code. We use a brute-force enumeration of weight $\leq 3$ fault-paths, and treat the missing probability, due to higher-weight fault-paths, as corresponding to a single additional syndrome, with unknown conditioned logical error $\epsilon_{L|missing}\in[0,1]$. Red dots mark the minimal and maximal values obtained for $\lambda_{SALEM}^{FG}$, as well as the values obtained for $\epsilon_{L|missing}=1/2$, assuming either the MWPM decoder (left), or an approximation of the optimal ML decoder based on the same brute-force enumeration (right). The small changes in $\lambda_{SALEM}^{FG}$ as a function of $\epsilon_{L|missing}$ (note the scales of $y$-axes) indicate that higher-weight fault-paths have a negligible effect, and we use the values of $\lambda^{FG}_{SALEM}$ obtained for $\epsilon_{L|missing}=1/2$ for Tab.~\ref{Tab: per-syndrome lambda} in the main text.\label{Fig: lambda as a function of eps missing}}
\end{figure*}

As discussed in Sec.~\ref{Sec: fine-grained SALEM and ML decoding}, the syndrome-conditioned logical characterization needed for FG-SALEM is a hard computational problem.  
For the distance-3 surface code, we perform a brute-force syndrome-conditioned logical characterization, by enumerating all weight $\leq 3$ combinations of independent physical errors (or faults). This allows for an accurate approximation of ML decoding, and of the blowup rate for FG-SALEM, see Fig.~\ref{Fig: lambda as a function of eps missing}. Brute force enumeration quickly becomes intractable as the code distance grows, and we were not able to perform it to a high-enough accuracy for $d>3$. 

As discussed in Sec.~\ref{Sec: coarse-grained SALEM main text}, for binary SALEM the syndrome-conditioned characterization splits into two tractable computational problems - classification, and subset-conditioned characterization. We perform classification using two soft-output decoders, based on the MWPM and TN decoders. A simplified physical-to-logical characterization (P2LC) of the subset-conditioned channels $\Lambda_{L|0}$ and $\Lambda_{L|1}$ (and the subset probabilities $\mathbb{P}(S_0),\mathbb{P}(S_1)$) is then performed by sampling and simulating fault-paths (using \texttt{Stim}), decoding with MWPM, and collecting a joint distribution over subsets and logical errors. 

For the MWPM decoder, we follow Ref.~\cite{smith2024mitigating, meister2024efficientsoftoutputdecoderssurface, gidney2023yokedsurfacecodes} to obtain a weight gap $\Delta w_s=|w_{0,s}-w_{1,s}|$, where $w_{0,s}$ ($w_{1,s}$) is the minimal weight of a fault-path in the `no logical error' ('logical error') sector, given the syndrome $s$. These minimal weights are obtained by adding the measured value $l=0,1$ of the logical operator as an additional bit in the syndrome, $s'=s\cup\{l\}$, and forcing $l$ to either 0 or 1. The logical recovery operation is given by $R_s=Z_L^{l_s}$, where $l_s=\text{argmin}_{l}(w_l)$, while the gap $\Delta w_s$ indicates the confidence of the decoder in this recovery, with a larger gap indicating higher confidence. We partition based on the condition 
\begin{align}
    e^{-\Delta w_s}>\tau_{MWPM},
\end{align}
which defines the `bad' set of syndromes $S_1$. The quantity $e^{-\Delta w_s}$ takes values in $[0,1]$, and can be understood as the ratio of probabilities between the most probable fault-path (identified by MWPM) given the syndrome, and the most probable fault-path given the syndrome in the opposite logical sector.

The soft-output of the TN decoder directly corresponds to an estimate of $\epsilon_{L|s}$. In Ref.~\cite{piveteau2023tensornetworkdecoding}, a `difference TN' $Z_{-}$, which estimates $\mathbb{P}(l=1,s)-\mathbb{P}(l=0,s)$, is used for decoding, such that $l_s^{TN}=\text{Heaviside}( Z_{-})$ defines the logical recovery operation. We use the TN decoder as a classifier, assuming decoding is performed with the weaker but faster MWPM. This is done by additionally computing a TN $Z_{+}$ which estimates $\mathbb{P}(s)=\mathbb{P}(l=1, s)+\mathbb{P}(l=0,s)$. We then obtain an estimate  $\tilde{\epsilon}_{L|s}=(1+(-1)^{l_s}Z_{-}/Z_{+})/2$ of $\epsilon_{L|s}$. Note that $l_s$ corresponds to the recovery operation of MWPM, and not of the TN decoder. Binary partition is then performed based on the criterion 
\begin{align}
\tilde{\epsilon}_{L|s}>\tau_{TN},
\end{align}
defining the `bad' set $S_1$. 

Note that we needed two TNs to perform classification, as opposed to just one needed for decoding. However, in the realistic case with $4^k$ logical Paulis $\sigma$ for $k$ logical qubits, classification is easier than decoding. The latter requires $4^k-1$ TNs, corresponding to all non-zero Fourier components (with respect to $\sigma$) of $\mathbb{P}(\sigma,s)$, while the latter requires only two TNs (independent of $k$), corresponding to $\mathbb{P}(s)$ and $\epsilon_{L, s}=\mathbb{P}(\sigma\neq I,s)$.

\begin{figure*}[ht!]
\begin{centering}
\includegraphics[width=1\textwidth]{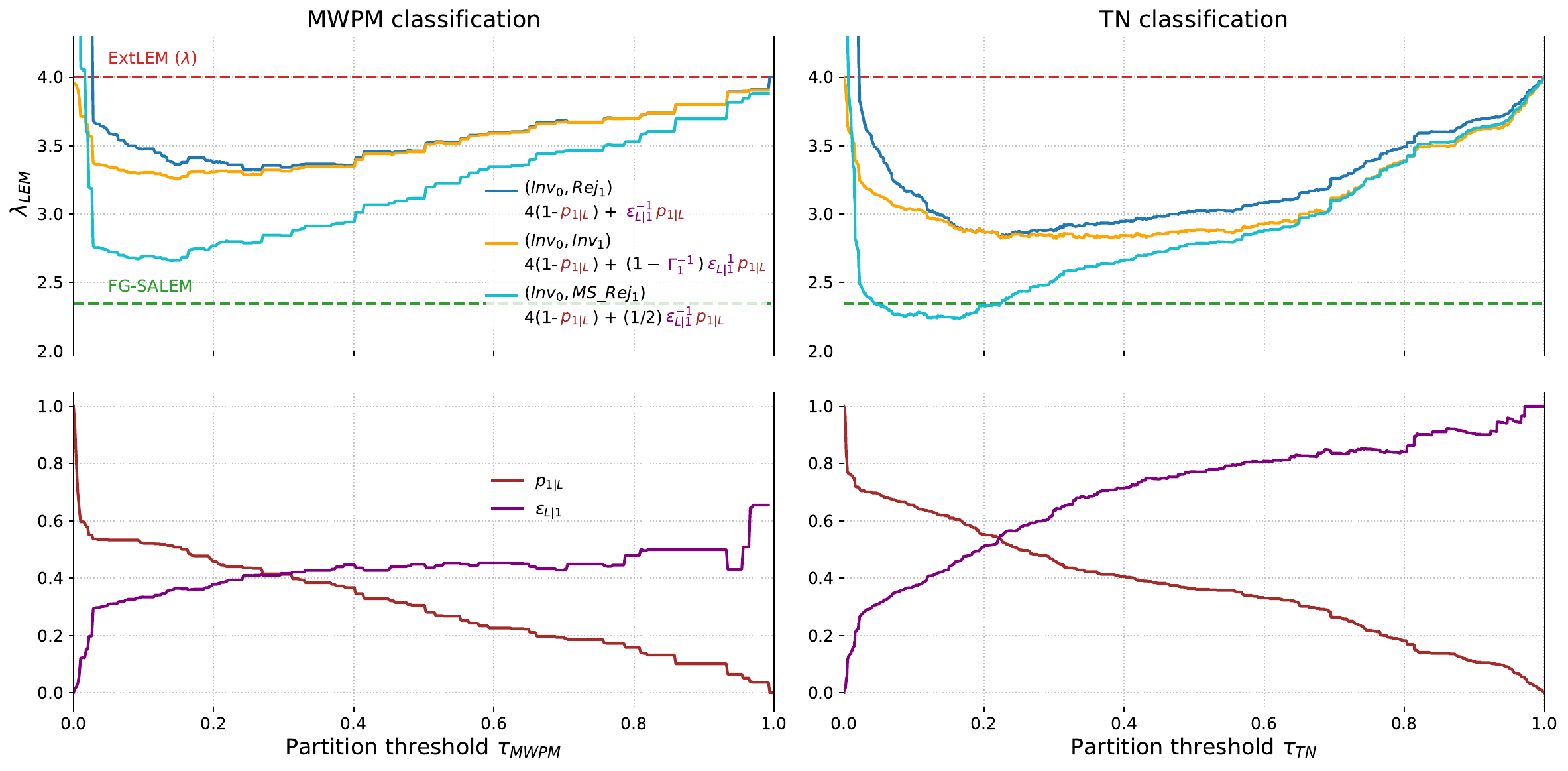}
\par\end{centering}
\caption{Binary SALEM for the $d=3$ surface code memory circuit. See Fig.~\ref{Fig: Bi-SALEM} for details. Note that Mid-shot rejection can even improve the QPU time overhead (but not the shot overhead) over FG-SALEM. 
\label{Fig: Bi-SALEM d=3}
 }
\end{figure*}

\begin{figure*}[ht!]
\begin{centering}
\includegraphics[width=1\textwidth]{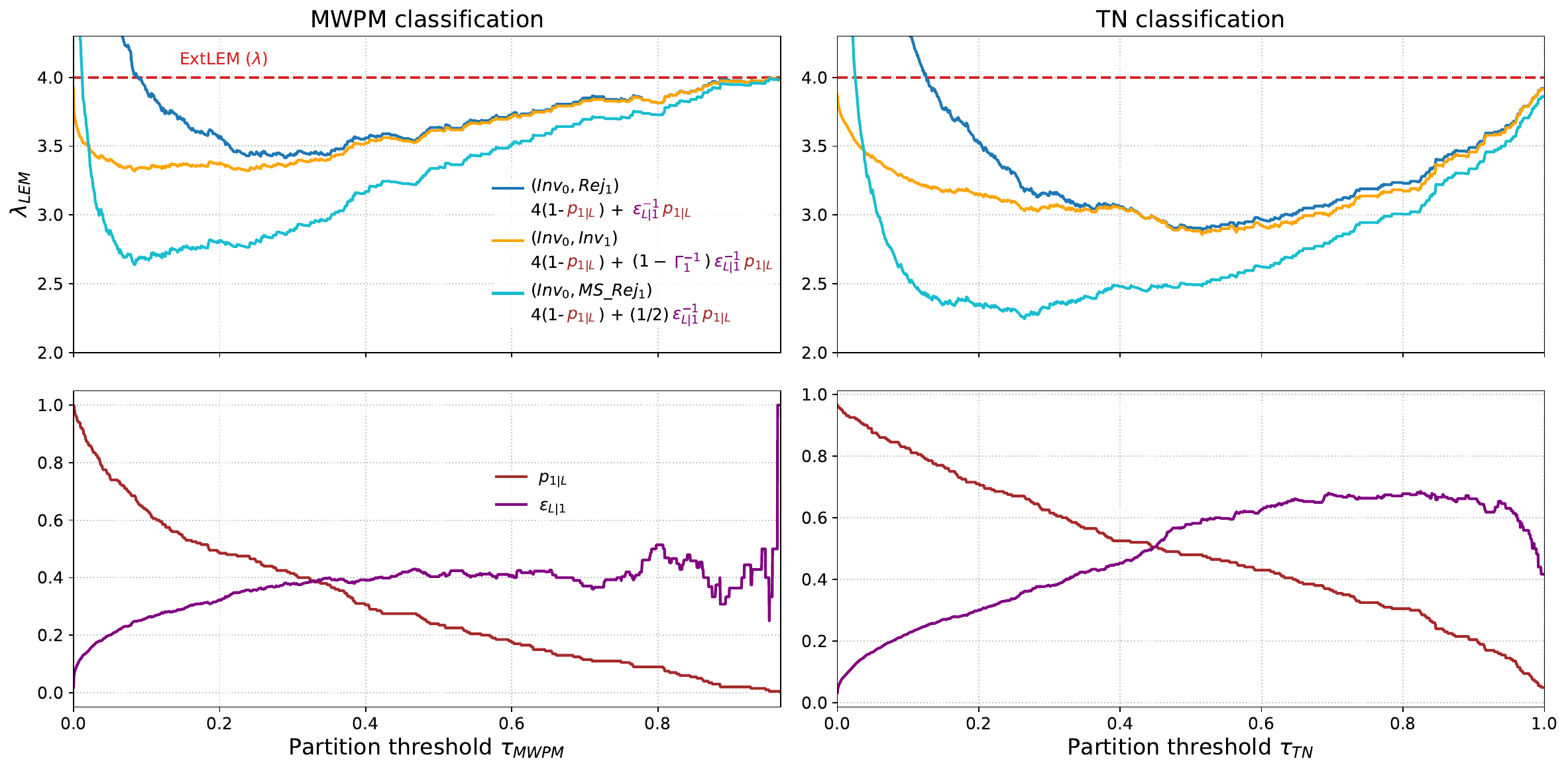}
\par\end{centering}
\caption{Binary SALEM for the $d=5$ surface code memory circuit. See Fig.~\ref{Fig: Bi-SALEM} for details. The drop in $\epsilon_{L|1}$ for $\tau_{TN}>0.8$ shows that the TN classifier we use struggles to approximate $\epsilon_{L|s}$ for the `worst' syndromes $s$ where this quantity is close to 1. Nevertheless, the TN classifier approximates $\epsilon_{L|s}$ well enough to produce strong partitions, which significantly improve the blowup rate for binary SALEM relative to the MWPM classifier. Optimal blowup rates with both classifiers and all three variants of binary SALEM are close to those observed for $d=3$. 
\label{Fig: Bi-SALEM d=5}
 }
\end{figure*}

The parameters we use to construct and contract TNs for $d=3,4,5$ are shown in Fig.~\ref{Tab: TN_params}, along with the physical error rates, numbers of sampled fault-paths, and the number of distinct observed syndromes, which are decoded once to reduce runtime. As noted in Ref.~\cite{piveteau2023tensornetworkdecoding}, the  authors did not focus on optimizing TN decoding runtime, which is excessive, with $\sim 15$ hours of parallelized computation on a machine with 90 CPUs needed for $d=5$. Results for $d=4$ are shown in Fig.~\ref{Fig: Bi-SALEM}, while results for $d=3,5$ are shown in Fig.~\ref{Fig: Bi-SALEM d=3}-\ref{Fig: Bi-SALEM d=5}. We observe an even-odd effect with respect to code distance, where binary SALEM blowup rates are smaller for even $d$ than for odd $d$. This is rooted in the well-known ability of distance $d=2t+2$ codes to correct $t$ errors and additionally detect $t+1$ errors (implying a better performance of EC+PS), which suggests the existence of partitions with simultaneously larger $\epsilon_{L|1}$ and $p_{1|L}$ (see bottom panels in Fig.~\ref{Fig: Bi-SALEM}, \ref{Fig: Bi-SALEM d=3},\ref{Fig: Bi-SALEM d=5}).

\renewcommand{\arraystretch}{1.2}
\begin{table}[t]
\caption{Simulation parameters used to generate Fig.~\ref{Fig: Bi-SALEM}, \ref{Fig: Bi-SALEM d=3} and \ref{Fig: Bi-SALEM d=5}. The second group of parameters is relevant for TN classification, see Ref.~\cite{piveteau2023tensornetworkdecoding} for details.}
\label{Tab: TN_params}
\centering
\begin{ruledtabular}
\begin{tabular}{lccc}
Parameter & $d=3$ & $d=4$ & $d=5$ \\
\hline
Physical error & $0.001$ & $0.002$ & $0.007$ \\
Sampled fault-paths & $10^6$ & $10^5$ & $10^4$ \\
Distinct syndromes & $3,769$ & $14,081$ & $9,861$ \\
\hline
Truncation dimension & 8 & 8 & 8 \\
Bond dimension  & 20 & 12 & 12 \\
MPS bond dimension & 128 & 64 & 64 \\
Split dimension & 20 & 12 & 12 \\
SVD cutoff & $10^{-12}$ & $10^{-12}$ & $10^{-12}$ \\
\end{tabular}
\end{ruledtabular}
\end{table}

To produce Fig.~\ref{fig: performance} in the main text, we use a particular binary SALEM method for $d=4$, defined by the optimal partition for $(\text{Inv}_0,\text{MS\_Rej}_1)$ with the MWPM classifier (minimal point on the cyan line in Fig.~\ref{Fig: Bi-SALEM}, Top Left). We then perform three types of parameter scans, generating an estimate for the bias and standard deviation of Bare, EM, EC, ExtLEM, EC+PS and SALEM, for each parameter value. We refer to the sum of the bias and statistical error as a `total estimation error'. For Fig.~\ref{fig: performance}(a), we scan over circuit volume $V$, and plot the estimated bias and standard deviation for each method as a function of $V$ (data points and error bars are simply obtained by sampling from corresponding normal distributions). For Fig.~\ref{fig: performance}(b), we scan over the required accuracy $1-\delta$, and identify the maximal volume for each method such that the total estimation error is $\leq \delta$. For Fig.~\ref{fig: performance}(c), we scan over physical error rates $\epsilon$ around the FT (pseudo) threshold, and plot the total estimation error as a function of $\epsilon$. For  Fig.~\ref{fig: performance}(a)-(b), we allocate a space-time volume corresponding to a total of $10^4$ error-corrected shots, for all error reduction methods. For Fig.~\ref{fig: performance}(c) we allocate a total of $10^8$ error-corrected shots.

We set the bias to 0 for EM, ExtLEM and SALEM. The bias for Bare is $1-(1-4\epsilon/3)^V$, corresponding to a physical depolarizing channel with parameter $\epsilon$. The shot overhead for EM is $\Gamma=(1-4\epsilon/3)^{-2V}/V_{EC}$, corresponding to mitigating only the $X$ and $Y$ errors which affect the measured $Z$, and we divide by the space-time overhead of error-correction $V_{EC}$, to account for the reduced time of physical shots compared to error-corrected shots, as well as EM's ability to parallelize shots using  multiple qubit patches, corresponding to the (net) rate of the code used. For the surface code, the net rate is $1/(2d^2-1)$, and each round of stabilizer measurements involves four layers of physical two-qubit gates and one layer of measurement and reset operations. We therefore take the space-time overhead per round to be $V_{EC}=5(2d^2-1)=155$ for $d=4$. All other biases and standard deviations are estimated using the simplified P2LC described above, where all quantities are computed per-round. In particular, the bias for EC is $1-(1-2\epsilon_L/d)^V$, and the bias for EC+PS is $1-(1-2\epsilon_{L|0}/d)^{V}$, both corresponding to logical bit-flip channels.

\clearpage

\noappendixtoc         
% \nocite{apsrev41Control}
% \bibliography{SALEM.bib,revtex-custom}
\input{SALEM.bbl}

\end{document}

%% file: fig_decomposition_of_G_j3.tikz.tex
\begin{tikzpicture}[
  font=\small,
  >=Stealth,
  block/.style={draw, rounded corners=3pt, thick, minimum height=16mm, align=center},
  big/.style={block, minimum width=10mm, fill=pink!60},
  logical/.style={block, minimum width=10mm, fill=white},
  inchan/.style={
    block,
    minimum width=10mm,
    shade,
    left color=white,
    right color=pink!60,
    shading=axis
  },
  braceLabel/.style={midway, yshift=10pt}
]

%%%%%%%%%%%%%%%%%%%%%%%%%%%
\node[big] (G1) {$G_1$};
\node[big, right=1mm of G1] (G2top) {$G_2$};
\node[big, right=1mm of G2top] (G3top) {$G_3$};
\node[right=1mm of G3top] (dotstop) {$\cdots$};

\draw[thick] ([yshift=2pt,xshift=2pt]dotstop.east) -- ([yshift=-2pt,xshift=2pt]dotstop.east);
\draw[->, thick] ([xshift=2pt] dotstop.east) -- ++(5mm,0);

% next line
\node[logical, yshift=-2.5cm] (g1a) {$g_1$};
\node[logical, right=1mm of g1a] (lambda1a) {$\Lambda_L^{(1)}$};
\node[inchan, right=1mm of lambda1a] (lambda_in_2) {$\Lambda_{\mathrm{in}}^{(2)}$};
\node[big, right=1mm of lambda_in_2] (G2mid) {$G_2$};
\node[big, right=1mm of G2mid] (G3mid) {$G_3$};
\node[right=1mm of G3mid] (dotsmid) {$\cdots$};

\draw[decorate, line width=1.3pt, decoration={brace, amplitude=5pt}]
  (lambda1a.north west) -- (lambda_in_2.north east)
  node[braceLabel, yshift=5pt] {$\Lambda_{\mathrm{out}}^{(1)}$};

\draw[thick] ([yshift=2pt,xshift=2pt]dotsmid.east) -- ([yshift=-2pt,xshift=2pt]dotsmid.east);
\draw[->, thick] ([xshift=2pt] dotsmid.east) -- ++(5mm,0);

% next line
\node[logical, yshift=-5cm] (g1b) {$g_1$};
\node[logical, right=1mm of g1b] (lambda1b) {$\Lambda_L^{(1)}$};
\node[logical, right=1mm of lambda1b] (g2b) {$g_2$};
\node[logical, right=1mm of g2b] (lambda2b) {$\Lambda_L^{(2)}$};
\node[inchan, right=1mm of lambda2b] (lambda_in_3) {$\Lambda_{\mathrm{in}}^{(3)}$};
\node[big, right=1mm of lambda_in_3] (G3bot) {$G_3$};
\node[right=1mm of G3bot] (dotsbot) {$\cdots$};

\draw[decorate, line width=1.3pt, decoration={brace, amplitude=5pt}]
  (lambda2b.north west) -- (lambda_in_3.north east)
  node[braceLabel, yshift=5pt] {$\Lambda_{\mathrm{out}}^{(2)}$};

\end{tikzpicture}

%% file: fig_proof_prop_1.tikz.tex
\begin{tikzpicture} [
  font=\small,
  >=Stealth,
  block/.style={draw, rounded corners=3pt, thick, minimum height=16mm, align=center},
big/.style={
  block,
  minimum width=10mm,
  shade,
  shading=axis,
  left color=pink!100,
  right color=white,
  shading angle=135
},    
  mid/.style={block, minimum width=10mm},
  dash/.style={block, dashed, minimum width=10mm},
  braceLabel/.style={midway, yshift=10pt}
]

\tikzset{
  grayx/.style={
    draw=none,
    minimum size=2mm,
    path picture={
      \draw[gray, line width=0.8pt]
        (path picture bounding box.south west) -- (path picture bounding box.north east)
        (path picture bounding box.north west) -- (path picture bounding box.south east);
    }
  }
}

\begin{scope}
\node[big] (gate1) {$G_{j-2}$};
\node at ([yshift=3mm] gate1.north west) {(a)};
\node[big, right=1mm of gate1] (gate2) {$G_{j-1}$};
\node[big, right=1mm of gate2] (gate3) {$G_{j}$};   

\node[grayx] at ([xshift=-2mm,yshift=5mm] gate1.east) {};
\node[grayx] at (1.5,0.3) {};
\node[grayx] at (2.4,-0.5) {};
\end{scope}
%%%%%%%%%%%%%%%
\begin{scope}[xshift=5cm]
\node[big] (gate1) {$G_{j-2}$};
\node at ([yshift=3mm] gate1.north west) {(b)};
\node[big, right=1mm of gate1] (gate2) {$G_{j-1}$};
\node[big, right=1mm of gate2] (gate3) {$G_{j}$};

\node[grayx] at ([xshift=-2mm,yshift=5mm] gate1.east) {};
\node[grayx] at (1.5,0.3) {};
\node[grayx] at (2.4,-0.5) {};
\node[grayx] at (2.6,0.5) {};
\end{scope}

%%%%%%%%%%%%%%%
\begin{scope}[yshift=-2.5cm]
\node[big] (gate1) {$G_{j-2}$};
\node at ([yshift=3mm] gate1.north west) {(c)};

\node[big, right=1mm of gate1] (gate2) {$G_{j-1}$};

\node[big, right=1mm of gate2] (gate3) {$G_{j}$};

\node[grayx] at ([xshift=-2mm,yshift=5mm] gate1.east) {};
\node[grayx] at (1.5,0.3) {};
\node[grayx] at (1.2,-0.5) {};

\end{scope}

%%%%%%%%%%%%%%%
\begin{scope}[xshift=5cm, yshift=-2.5cm]

\node[big] (gate1) {$G_{j-2}$};
\node at ([yshift=3mm] gate1.north west) {(d)};

\node[big, right=1mm of gate1] (gate2) {$G_{j-1}$};

\node[big, right=1mm of gate2] (gate3) {$G_{j}$};

\node[grayx] at (1.5,0.3) {};
\node[grayx] at (1.4,-0.5) {};
\node[grayx] at (2.6,0.5) {};

\end{scope}
\end{tikzpicture}

%% file: fig_logical_channel_construction.tikz.tex
\begin{tikzpicture}[
  font=\small,
  >=Stealth,
  block/.style={draw, rounded corners=3pt, thick, minimum height=16mm, align=center},
big/.style={
  block,
  minimum width=10mm,
  shade,
  shading=axis,
  left color=pink!100,
  right color=white,
  shading angle=135
},  
  mid/.style={block, minimum width=10mm},
  dash/.style={block, dashed, minimum width=10mm},
  braceLabel/.style={midway, yshift=10pt}
]

\node [big] (Gj) {$G_{j-1}$};

\node at ([xshift=-8mm] Gj.west) (I2) {$I$};
\draw[->,thick] (I2.east) -- ++(5mm,0);

\draw[->,thick] (Gj.east) -- ++(10mm,0) node [midway, above] {${\tilde{\Lambda}}_c^{(j-1)}$};
\node [big] at ([xshift=15mm] Gj.east)(Gjj) {$G_j$};

\draw[->,thick] (Gjj.east) -- ++(10mm,0) node [midway, above] {${\tilde{\Lambda}}_{\mathrm{out}}^{ (j)}$};

\node [big] at ([xshift=18mm] Gjj.east)(Gjjj) {$G_{j+1,\mathrm{ideal}}$};

\draw[->,thick] (Gjjj.east) -- ++(5mm,0) node [pos=1.7] {$\tilde{\Lambda}_{L}^{(j)}$};

\end{tikzpicture}

%% file: fig_one_step_past_light_cone.tikz.tex
\begin{tikzpicture} [
  font=\small,
  >=Stealth,
  block/.style={draw, rounded corners=3pt, thick, minimum height=16mm, align=center},
  % big/.style={block, minimum width=15mm},
big/.style={
  block,
  minimum width=10mm,
  shade,
  shading=axis,
  left color=pink!100,
  right color=white,
  shading angle=135
},    
  mid/.style={block, minimum width=15mm},
  dash/.style={block, dashed, minimum width=10mm},
  braceLabel/.style={midway, yshift=10pt}
]

% \tikzset{
%   grayx/.style={
%     draw=none,
%     minimum size=2mm,
%     path picture={
%       \draw[gray, line width=0.8pt]
%         (path picture bounding box.south west) -- (path picture bounding box.north east)
%         (path picture bounding box.north west) -- (path picture bounding box.south east);
%     }
%   }
% }

\begin{scope}
% \node[box] (gate1) {$G_{j-2}$};
% \node [right=of gate1] (text1) {$e_{j-2}$};
\node[big] (gate1) {$G_{j-1, \{1,2\}}$};
% \node [right=of gate2] (text2) {$e_{j-1}$};
\node[big] at ([xshift=8mm,yshift=9mm] gate1.east) (gate2) {$G_{j, \{0,1\}}$};   
\node[big, above=0.5mm of gate1, minimum height=8mm, minimum width = 16mm] (gate3) {$G_{j-1, \{0\}}$};
\end{scope}

\begin{scope} [xshift=3.6cm]
\node[big] (gate1) {$G_{j+1, \{1,2\}, \mathrm{ideal}}$};
\node[big, above=0.5mm of gate1, minimum height=8mm, minimum width = 23mm] (gate3) {$G_{j+1, \{0\},\mathrm{ideal}}$};
\end{scope}

% \draw[->] (gate1.east) -- (gate2.west);% node [midway, above] {$e_{j-2}$};

% \draw[->] (gate2.east) -- (gate3.west);% node [midway, above] {$e_{j-1}$};

% \node[grayx] at ([xshift=-2mm,yshift=1mm] gate1.east) {};
% \node[grayx] at ([xshift=-2mm,yshift=1mm] gate2.east) {};
% \node[grayx] at (4,-0.5) {};

\draw [->] (-1,-1.5)--(5,-1.5) node [above] {time};
% \end{scope}

\end{tikzpicture}